\documentclass[usenatbib]{mn2e}
\usepackage{amssymb,amsmath,graphicx,natbib,setspace,pifont}
\newcommand{\HI}{\hbox{{\sc H}{\sc i}} }
\newcommand{\HII}{\hbox{{\sc H}{\sc ii}} }

\newcommand{\Hm}{\rm H_{2} }
\newcommand{\NHI}{{N_{\rm HI}}}

\newcommand{\cmsq}{\,{\rm cm^{-2}}}
\newcommand{\cms}{\,{\rm cm^{2}}}
\newcommand{\cmcb}{\,{\rm cm^{-3}}}

\newcommand{\nH}{n_{\rm _{H}} }

\newcommand{\Ob}{\Omega_{\rm b} }
\newcommand{\Om}{\Omega_{\rm m} }

\newcommand{\Ol}{\Omega_{\Lambda} }
\newcommand{\ns}{n_{\rm s} }
\newcommand{\sigeight}{\sigma_{\rm 8} }

\newcommand{\Msun}{{\rm M_{\odot}} }
\newcommand{\Msunh}{h^{-1} {\rm M_{\odot}} }
\newcommand{\Mpch}{h^{-1} {\rm Mpc} }
\newcommand{\kpch}{h^{-1} {\rm kpc} }

\newcommand{\Gadget}{{\small GADGET-3} }

\newcommand{\SUBFIND}{{\small SUBFIND} }
\newcommand{\TRAPHIC}{{\small TRAPHIC}~}

\newcommand{\apjl}{{ApJ}}
\newcommand{\aj}{{AJ}}
\newcommand{\apj}{{ApJ}}
\newcommand{\apjs}{{ApJS}}
\newcommand{\aap}{{A\&A}}
\newcommand{\mnras}{{MNRAS}}
\newcommand{\araa}{{ARA\&A}}

\begin{document}
\title[Strong $\HI$ absorbers and galaxies at $z=3$]{Predictions for the relation between strong $\HI$ absorbers and galaxies at redshift 3}

\author[A.~Rahmati \& J.~Schaye]
  {Alireza~Rahmati$^{1,2}$\thanks{rahmati@mpa-garching.mpg.de} and Joop Schaye$^1$\\
  $^1$Leiden Observatory, Leiden University, P.O. Box 9513, 2300 RA Leiden, The Netherlands\\
  $^2$Max-Planck Institute for Astrophysics, Karl-Schwarzschild-Strasse 1, 85748 Garching, Germany}

\maketitle

\begin{abstract} 
We combine cosmological, hydrodynamical simulations with accurate radiative transfer corrections to investigate the relation between strong $\HI$ absorbers ($\NHI \gtrsim 10^{17}\cmsq$) and galaxies at redshift $z = 3$. We find a strong anti-correlation between the column density and the impact parameter that connects the absorber to the nearest galaxy. The median impact parameters for Lyman Limit (LL) and Damped Lyman-$\alpha$ (DLA) systems are $\sim 10$ and $\sim 1$ proper kpc, respectively. If normalized to the size of the halo of the nearest central galaxy, the median impact parameters for LL and DLA systems become $\sim 1$ and $\sim 10^{-1}$ virial radii, respectively. At a given $\HI$ column density, the impact parameter increases with the mass of the closest galaxy, in agreement with observations. We predict most strong $\HI$ absorbers to be most closely associated with extremely low-mass galaxies, $\rm{M_{\star}} < 10^{8}~\Msun$ and star formation rate $<10^{-1}~\Msun~\rm{yr^{-1}}$. We also find a correlation between the column density of absorbers and the mass of the nearest galaxy. This correlation is most pronounced for DLAs with $\NHI > 10^{21} \cmsq$ which are typically close to galaxies with $\rm{M_{\star}} \gtrsim 10^{9}~\Msun$. Similar correlations exist between column density and other properties of the associated galaxies such as their star formation rates, halo masses and $\HI$ content. The galaxies nearest to $\HI$ absorbers are typically far too faint to be detectable with current instrumentation, which is consistent with the high rate of (often unpublished) non-detections in observational searches for the galaxy counterparts of strong $\HI$ absorbers. Moreover, we predict that the detected nearby galaxies are typically not the galaxies that are most closely associated with the absorbers, thus causing the impact parameters, star formation rates and stellar masses of the observed counterparts to be biased high.

\end{abstract}

\begin{keywords}
  methods: numerical -- galaxies: formation -- galaxies: high-redshift -- galaxies: absorption lines -- quasars: absorption lines -- intergalactic medium
\end{keywords}

\section{Introduction}

Studies of high-redshift galaxies are nearly always based on the light emitted by stars and/or ionized gas. This limits the observations to the small fraction of galaxies that are bright enough to be detected in emission. Given the large number density of faint galaxies, it is likely that most high-redshift galaxies are missing from observational studies. The analysis of absorption features in the spectra of background QSOs provides an alternative probe of the distribution of matter at high redshifts opening up a window to study an unbiased sample of matter that resides between us and background QSOs. 

The rare strong $\HI$ Ly$\alpha$ absorbers which are easily recognizable in the spectra of background QSOs due to their damping wings, for which they are called Damped Lyman-$\alpha$ (DLA) systems\footnote{The official column density limit of a DLA is somewhat arbitrarily defined to be $\NHI > 10^{20.3}\cmsq$.}, are of particular interest. DLAs are likely to arise in, or close to, the interstellar medium (ISM). DLAs thus provide a unique opportunity to define an absorption-selected galaxy sample and to study the ISM, particularly at the early stages of galaxy formation, and they have therefore been studied intensely since their discovery (see \citealp{Wolfe05} for a review). 

Based on the observed velocity width of metal lines associated with DLAs, it was initially suggested that large, massive galactic disks are responsible for the observed DLAs at $z \sim 3$ \citep{PW97,PW98}. However, it has been shown that (collections of) smaller systems are also capable of having high velocity dispersions as a result of infall of material during structure formation \citep{Haehnelt98} or galactic winds \citep{MM99,Schaye01a}. Nevertheless, reproducing the observed velocity width distribution remains a challenge for hydrodynamical simulations \citep[e.g.,][]{Razoumov06,Pontzen08}.

Some recent studies suggest that at $z\sim 2-3$, a large fraction of strong $\HI$ absorbers like Lyman Limit Systems (LLS; $\NHI > 10^{17}\cmsq$) and DLAs are associated with galaxies similar to Lyman-break galaxies \citep[e.g.,][]{Steidel10,Rudie12,Font-Ribera12}, which have stellar and total halo masses $\sim 10^{10}$ and $\sim 10^{12}~\Msun$, respectively. If such massive galaxies were indeed the prime hosts of strong $\HI$ absorbers, then many of the galaxy counterparts of strong absorbers should be detectable with current surveys. However, observations that aim to find galaxies close to DLAs often result in non-detections \citep[e.g.,][]{Foltz86,Smith89,Lowenthal95,Bunker99,Prochaska02,Kulkarni06,Rahmani10,Bouche12} or find galaxies that are at unexpectedly large impact parameters from DLAs \citep[e.g.,][]{Yanny90,Teplitz98,Mannucci98}. In addition, deep Ly$\alpha$ observations at $z \approx 2-3$ have revealed a population of faint Ly$\alpha$ emitters whose number density is large enough to account for most LLSs and DLAs \citep{Rauch08,Rauch11}. Those findings suggest that strong $\HI$ systems such as DLAs are more closely associated to low-mass galaxies which are too faint to be observable with the detection thresholds of the current studies.

Because observational studies are limited by the small number of known DLAs and are missing low-mass galaxies, we resort to cosmological simulations to help us understand the link between DLAs and galaxies. Many studies have used simulations to investigate the nature of strong $\HI$ absorbers and particularly DLAs \citep[e.g.,][]{Gardner97,Gardner01,Haehnelt98,Nagamine04,Razoumov06,Pontzen08,Tescari09,Fumagalli11,Cen12,Voort12a,Altay13}. To maximize the numerical resolution required for accurate modeling of the high $\HI$ column densities, most previous studies have used small simulation boxes or zoomed simulations. Those studies often try to compensate for the lack of a full cosmological distribution of absorbers by combining the results from their small-scale simulations with analytic halo mass functions to predict the properties of the DLA population \citep[e.g.,][]{Gardner97,Gardner01} or to determine what kinds of galaxies dominate the cosmic DLA distribution \citep[e.g.,][]{Pontzen08}. This approach requires some preconceptions about the types of environments that can give rise to DLA absorbers and cannot easily account for the large scatter in the distribution of absorbers in halos with similar properties. As a result, the statistical properties found using zoomed simulations may be biased. Finally, the impact of finite detection thresholds on the observed relation between strong $\HI$ absorbers and galaxies cannot be studied with simulations that do not contain a representative sample of $\HI$ absorbers and galaxies.

In this work, we use cosmological hydrodynamical simulations that contain a representative sample of the full distribution of strong $\HI$ systems \citep{Rahmati13a}. Similar to what is done observationally, we connect each absorber to its nearest galaxy. A significant improvement in this work is the use of photoionization corrections that are based on accurate radiative transfer simulations and that account for both the uniform ultraviolet background (UVB) radiation and recombination radiation \citep{Rahmati13a}. In addition, we show that our main conclusions are insensitive to the inclusion of local sources and to variations in the subgrid physics (see Appendix \ref{ap:feedback} and Appendix \ref{ap:LSR}).

We predict correlations between the column density of strong $\HI$ absorbers, their impact parameters, and the properties of the associated galaxies at $z = 3$. While the fraction of $\HI$ absorbers that are linked to relatively massive galaxies increases with $\HI$ column density, most LLS and DLAs are closely associated with very low-mass galaxies (i.e., $\rm{M_{\star}} \lesssim 10^{8}~\Msun$), that are generally undetectable with current instruments. We show that our predictions are nevertheless in good agreement with existing observations, including those of \citet{Rudie12} who found that a large fraction of strong $\HI$ absorbers at $z \sim 3$ are within $300$ proper kpc radius from massive Lyman-break galaxies.

The structure of this paper is as follows. In $\S$\ref{secDLA:ingredients} we discuss our numerical simulations and ionization calculations for obtaining the $\HI$ column densities and describe our methods for connecting $\HI$ systems to their host galaxies. We present our results in $\S$\ref{secDLA:results} and compare them with observations. In this section we also investigate how the distribution of $\HI$ absorbers varies with the properties of their host galaxies. Finally, we conclude in $\S$\ref{secDLA:conclusions}.

\section{Simulation techniques}
\label{secDLA:ingredients}

In this section we briefly describe the hydrodynamical simulations that are post-processed to get the $\HI$ distribution by accounting for various ionization processes ($\S$\ref{secDLA:hydro}). Then we explain our halo finding method ($\S$\ref{secDLA:galfind}), our $\HI$ column density calculations ($\S$\ref{secDLA:HI_method}), and the procedures we use to connect $\HI$ absorbers to their host halos ($\S$\ref{secDLA:matching}).

\subsection{Hydrodynamical simulations}
\label{secDLA:hydro}
We use cosmological simulations performed using a significantly modified and extended version of the smoothed particle hydrodynamics (SPH) code \Gadget (last described in \citealp{Springel05}). The simulation code was used for the Overwhelmingly Large Simulations (OWLS) described in \citet{Schaye10}. Our reference model is identical to the OWLS reference model except for the choice of cosmology. We use the subgrid pressure-dependent star formation prescription of \citet{Schaye08} which reproduces the observed Kennicutt-Schmidt law. The chemodynamics is described in \citet{Wiersma09b} and follows the abundances of eleven elements assuming a \citet{Chabrier03} initial mass function. These abundances are used for calculating radiative cooling/heating rates, element-by-element and in the presence of the uniform cosmic microwave background and the \citet{HM01} UVB model \citep{Wiersma09a}. Galactic winds driven by star formation are modeled using a kinetic feedback recipe that assumes that $40\%$ of the kinetic energy generated by Type II SNe is injected as outflows with initial velocity of $600~{\rm{kms^{-1}}}$ and with a mass loading factor $\eta = 2$ \citep{DallaVecchia08}. To bracket the impact of feedback, we also consider simulations with different feedback and sub-grid models. We found that our results are insensitive to the variations in feedback and sub-grid physics (see Appendix \ref{ap:feedback}).

We adopt cosmological parameters consistent with the WMAP year 7 results: $\{\Om=0.272,\ \Ob=0.0455,\ \Ol=0.728,\ \sigeight=0.81,\ \ns=0.967,\ h=0.704\} $ \citep{Komatsu11}. Our reference simulation has a periodic box of $L = 25$ comoving $\Mpch$ and contains $512^3$ dark matter particles with mass $6.3 \times 10^6~\Msunh$ and an equal number of baryons with initial mass $1.4 \times 10^6~\Msunh$. The Plummer equivalent gravitational softening length is set to $\epsilon_{\rm{com}} = 1.95~\kpch$ and is limited to a maximum physical scale of $\epsilon_{\rm{prop}} = 0.5~\kpch$. In addition to our reference simulation explained above, we use simulations with different resolutions and box-sizes to investigate numerical effects (see Appendix \ref{ap:res}).

\subsection{Finding galaxies}
\label{secDLA:galfind}
To identify individual galaxies in our cosmological simulations, we assume that galaxies are gravitationally bound structures of baryons and dark matter. We first use the Friends-of-Friends (FoF) algorithm to identify groups of dark matter particles that are near each other (i.e., FoF halos), using a linking length of $b = 0.2$. Then, we use \SUBFIND \citep{Dolag09} to connect gravitationally bound particles as part of unique structures (halos) and to identify the center of each halo/galaxy as the position of the most bound particle in that halo. We take the radius within which the average density of a given halo reaches 200 times the mean density of the Universe at a given redshift, $R_{200}$, as the size of that halo. The most massive substructure in each halo is considered to be a \emph{central} galaxy and all the other gravitationally bound structures in that FoF halo are considered \emph{satellite} galaxies. Note that while satellites are always part of a FoF halo, we do not require them to be within the $R_{200}$ of their central galaxy.

In our analysis, we use all the simulated galaxies that have star formation rates SFR $> 4\times 10^{-3}~\rm{M_{\odot}~yr^{-1}}$. By using this SFR threshold, more than $99\%$ of our selected galaxies are resolved with $>100$ resolution elements (i.e., dark matter particles and/or baryonic particles). We test the impact of different SFR thresholds on our results, which provides useful insights for observational studies with finite detection threshold (see $\S$\ref{secDLA:results}).

\begin{figure*}
\centerline{\fbox{{\includegraphics[width=0.43\textwidth]
             {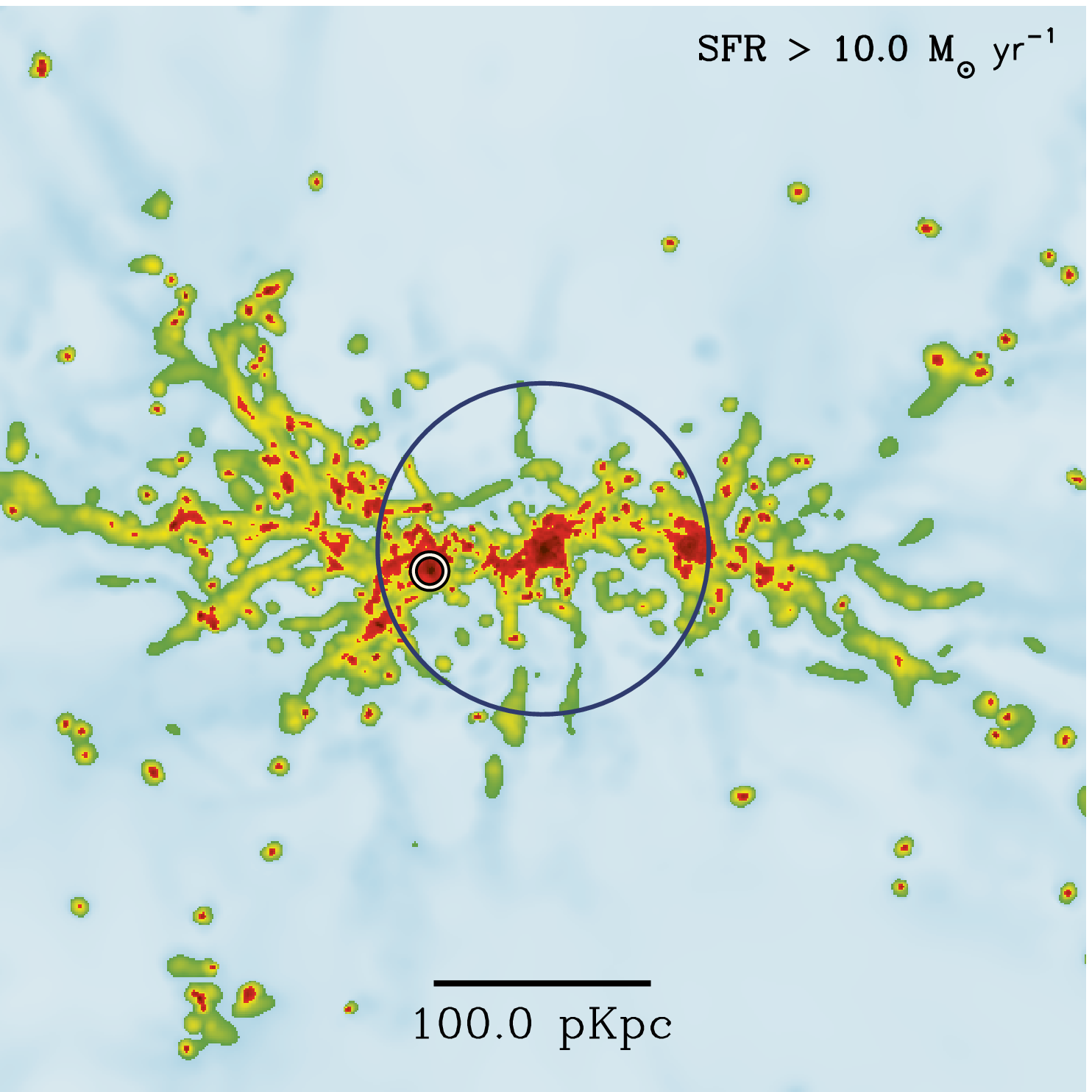}}}
             \fbox{{\includegraphics[width=0.43\textwidth]
             {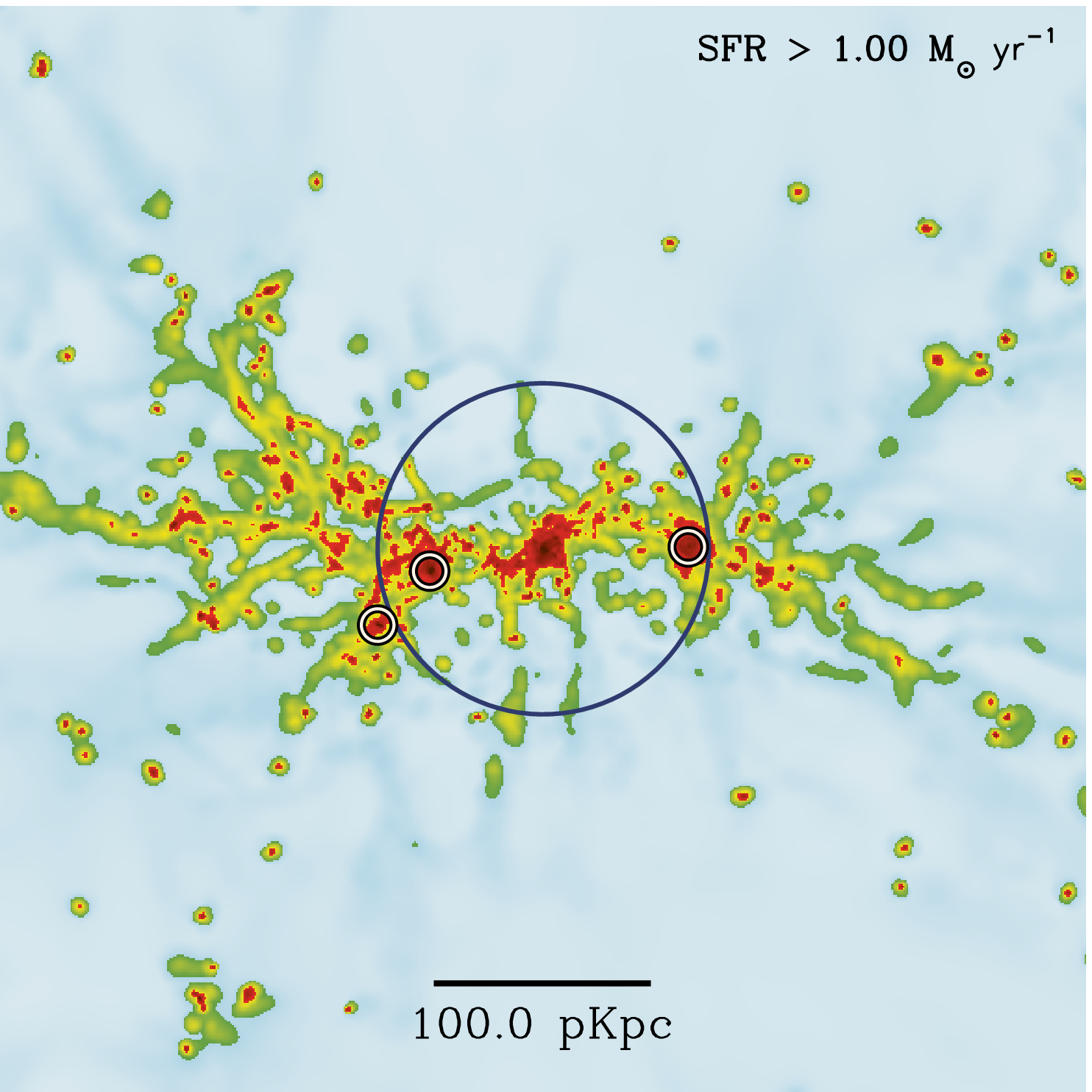}}}}
\centerline{\fbox{\includegraphics[width=0.43\textwidth]
             {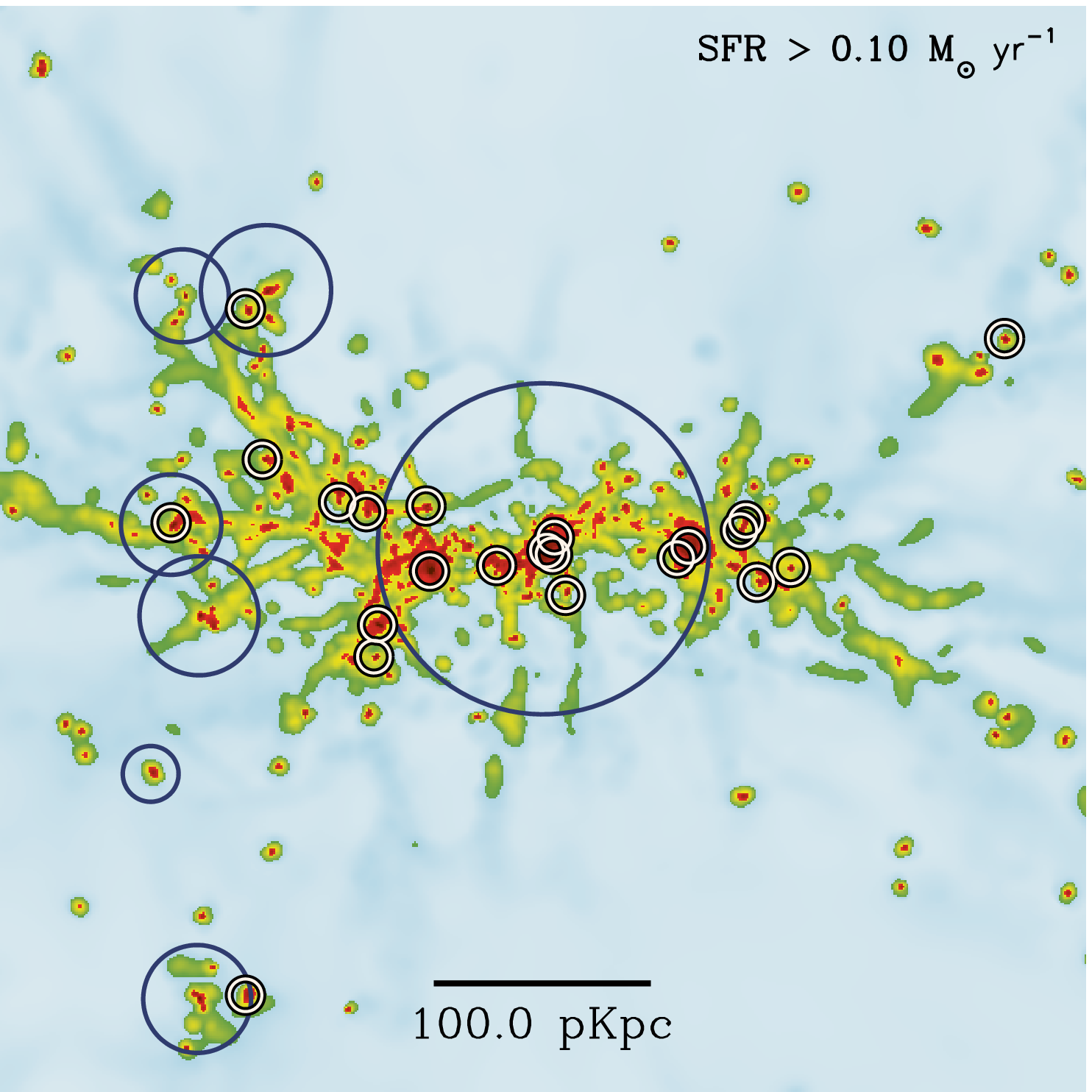}} 
             \fbox{\includegraphics[width=0.43\textwidth]
             {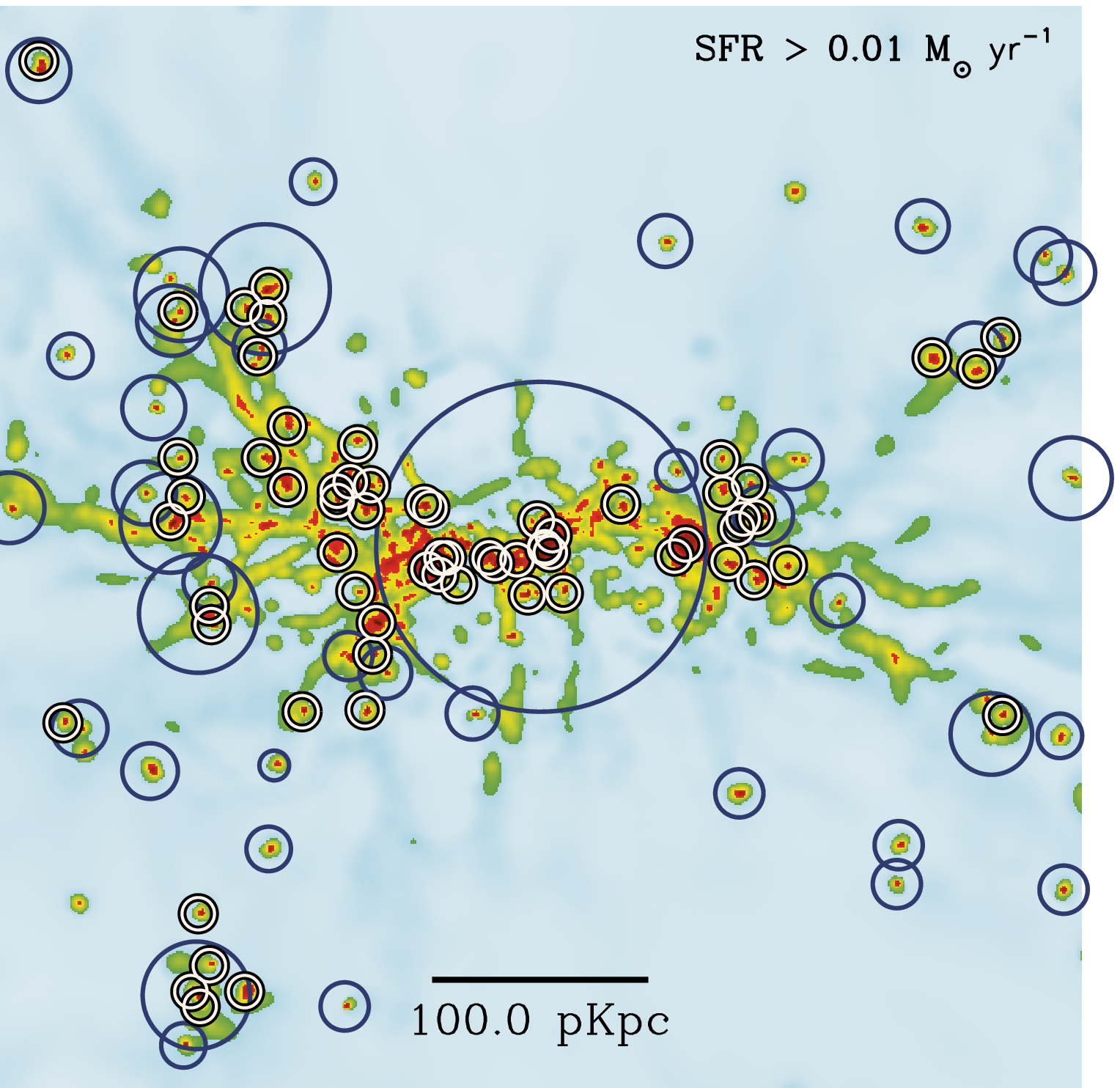}}}
\centerline{\hbox{\includegraphics[width=0.9\textwidth]
             {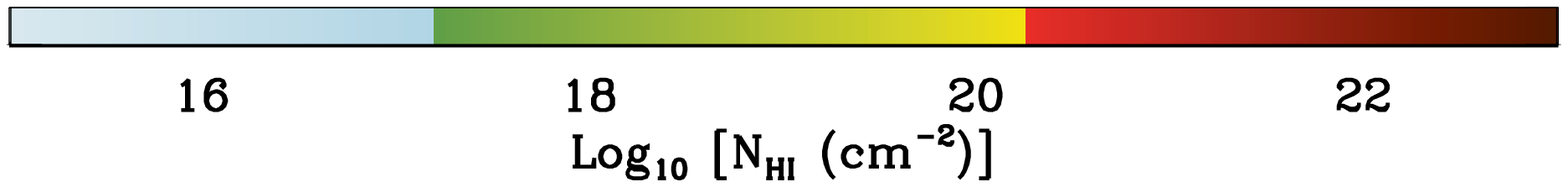}}}
\caption{The simulated $\HI$ column density distribution around a massive galaxy with $\rm{M_{\star}}= 10^{10}~\Msun$ and $\rm{SFR} = 29 ~\rm{\Msun~yr^{-1}}$ at $z = 3$. Circles indicate the positions of galaxies. The size of each dark circle indicates the virial radius of a central galaxy ($R_{200}$) while the small white circles all have the same size and indicate the locations of satellite galaxies. From top-left to bottom-right, panels show galaxies with $\rm{SFR} > 10 ~\rm{\Msun~yr^{-1}}$, $\rm{SFR} > 1 ~\rm{\Msun~yr^{-1}}$, $\rm{SFR} > 0.1 ~\rm{\Msun~yr^{-1}}$ and $\rm{SFR} > 0.01 ~\rm{\Msun~yr^{-1}}$, respectively. As the SFR threshold decreases, more galaxies are detected and the typical impact parameter between galaxies and absorbers decreases.}
\label{figDLA:stamp}
\end{figure*}

\subsection{Finding strong $\HI$ absorbers}
\label{secDLA:HI_method}
The first step in identifying $\HI$ absorbers in the simulations, is to calculate the hydrogen neutral fractions. To accomplish this, the main ionization processes that shape the distribution of neutral hydrogen must be accounted for. In this context, photoionization by the metagalactic UVB radiation is the main contributor to the bulk of hydrogen ionization at $z \gtrsim 1$ while collisional ionization becomes more important at lower redshifts \citep{Rahmati13a}. Although the photoionization from local stellar radiation is the dominant source of ionization at high $\HI$ column densities \citep[e.g.,][]{Schaye06,Rahmati13b}, our tests show that it does not have a significant impact on our conclusions (see Appendix \ref{ap:LSR}).

We use the UVB model of \citet{HM01} to account for the large-scale photoionization effect of quasars and galaxies. The same UVB model is used for calculating heating/cooling in our hydrodynamical simulations. It has been shown that this UVB model is consistent with metal absorption lines at $z\sim3$ \citep{Aguirre08} and the observed $\HI$ column density distribution function and its evolution over a wide range of redshifts \citep{Rahmati13a}. 

We characterize the UVB by its optically thin hydrogen photoionization rate, $\Gamma_{\rm{UVB}}$, and the effective hydrogen absorption cross-section, $\bar{\sigma}_{\nu_{\rm{HI}}}$ (see equations 3 and 4 and Table 2 in \citealp{Rahmati13a}). In self-shielded regions, $\Gamma_{\rm{UVB}}$ is attenuated to an effective total photoionization rate, $\Gamma_{\rm{Phot}}$, which decreases with density. In \citet{Rahmati13a} we performed radiative transfer simulations of the UVB and recombination radiation in cosmological density fields using \TRAPHIC \citep{Pawlik08,Pawlik11,Raicevic13}. We showed that at all densities the effective photoionization rate is accurately reproduced by the following fitting function:
\begin{eqnarray}
&{}&\frac{\Gamma_{\rm{Phot}}} {\Gamma_{\rm{UVB}}} = 0.98~\left[1 + \left(\frac{\nH}{n_{\rm{H, SSh}}}\right)^{1.64}  \right]^{-2.28} \nonumber \\
&{}& \qquad  \qquad  \qquad \qquad +0.02~\left[1 + \frac{\nH}{n_{\rm{H, SSh}}} \right]^{-0.84},
\label{eqDLA:Gamma-fit}
\end{eqnarray}
where $\nH$ is the hydrogen number density and $n_{\rm{H, SSh}}$ is the self-shielding density threshold predicted by the analytic model of \citet{Schaye01b}
\begin{eqnarray}
&{}& {n_{\rm{H,SSh}}} \sim~  6.73\times10^{-3} \cmcb \left(\frac{\bar{\sigma}_{\nu_{\HI}} }{2.49\times10^{-18}\cms}\right)^{-2/3} \nonumber \\
&{}& \qquad  \qquad  \times ~ \left(\frac{\Gamma_{\rm{UVB}}}{10^{-12}~{\rm{s}}^{-1}}\right)^{2/3}.
\label{eqDLA:densitySSH}
\end{eqnarray}

We use the photoionization rate from equations \eqref{eqDLA:Gamma-fit} and \eqref{eqDLA:densitySSH} together with the hydrogen number density and temperature of each SPH particle in our hydrodynamical simulations to calculate the equilibrium hydrogen neutral fraction of that particle in post-processing including also collisional ionization (see Appendix A2 in \citealp{Rahmati13a}). It is also worth noting that in our simulations, ISM gas particles (which all have densities $\nH > 0.1~\cmcb$) follow a polytropic equation of state that defines their temperatures. Since these temperatures are not physical and only measure the imposed effective pressure \citep{Schaye08}, we set the temperature of the ISM particles to $T_{\rm{ISM}} = 10^4~\rm{K}$, the typical temperature of the warm-neutral phase of the ISM.

At very high $\HI$ column densities, where the gas density and the optical depth for $\Hm$-dissociating radiation is high, hydrogen is expected to be mainly molecular. This process has been suggested as an explanation for the observed cut-off in the abundance of absorbers at $\NHI \gtrsim 10^{22}\cmsq$ \citep{Schaye01c,Krumholz09,PW09}\footnote{Note, however, that recent studies based on low-resolution spectra found that the $\HI$ column density distribution function extends beyond $10^{22}\cmsq$ (e.g., \citealp{Noterdaeme12}).}. It has been also shown that accounting for $\Hm$ formation can produce a good agreement between cosmological simulations and observations of the $\HI$ column density distribution function \citep{Altay11,Rahmati13a}. To test the impact of $\Hm$ formation on the spatial distribution of $\HI$ absorbers, we adopted the observationally inferred pressure law of \citet{BR06} to compute the $\Hm$ fractions in post-processing (see Appendix A in \citealp{Rahmati13b}). Once the $\Hm$ fractions have been calculated, we exclude the molecular hydrogen from the total neutral gas when calculating the $\HI$ column densities. We note that the adopted empirical relation for calculating the $\Hm$ fractions was calibrated using observations of low-redshift galaxies and may overestimate the $\Hm$ fraction in very low metallicity gas.

We calculate $\HI$ column densities by projecting the $\HI$ content of the simulation box along each axis onto a grid with $10,000^2$ pixels\footnote{Using $10,000^2$ cells produces converged results. The corresponding cell size is similar to the minimum smoothing length of SPH particles at $z = 3$ in our simulation.}, using SPH interpolation. While the projection could in principle merge distinct systems along the line of sight, this effect is not significant for high $\HI$ column density systems because such chance alignments are rare in the relatively small simulation boxes that we use. We tested the impact of projection effects by performing projections using multiple slices instead of the full box. We found that at $z = 3$ and for simulations with box sizes comparable to that of our simulation, the effect of projection starts to appear only at $\NHI < 10^{16} \cmcb$. Since the focus of our study is to characterize the properties of strong $\HI$ absorbers with $\NHI \gtrsim 10^{17} \cmcb$, our results are insensitive to the above mentioned projection effect.

In addition to $\HI$ column densities, we calculate the HI-weighted velocity along each line of sight (LOS), ${\left\langle V_{\rm{LOS}}\right\rangle}_{\HI}$ accounting for both Hubble and peculiar velocities. We use ${\left\langle V_{\rm{LOS}}\right\rangle}_{\HI}$ to define the position of the strongest absorber along the projection direction and verified that it is not subject to significant projection effects for the box size used here. 

\subsection{Associating $\HI$ absorbers with galaxies}
\label{secDLA:matching}
An example of the distribution of galaxies and $\HI$ absorbers in our simulation is shown in Figure \ref{figDLA:stamp}. The colored map, which is repeated in all four panels, shows the $\HI$ column density distribution in a $500\times500$ proper $\rm{kpc}^2$ region which is centered on a galaxy with $\rm{M_{\star}} = 10^{10}~\Msun$. Galaxies are shown with circles while the star formation rate threshold decreases from $\rm{SFR} > 10 ~\rm{\Msun~yr^{-1}}$ in the top-left panel to $\rm{SFR} > 0.01 ~\rm{\Msun~yr^{-1}}$ in the bottom-right panel. The dark circles have sizes proportional to the virial radii of galaxies and are centered on central galaxies. The small white circles all have identical sizes and indicate the locations of satellite galaxies. As this figure shows, LLSs and DLAs (that are shown using green and red colors, respectively) are strongly correlated with the positions of galaxies. In addition, the $\HI$ column density of absorbers tends to increase near galaxies. For a quantitative study, a well defined connection between absorbers and galaxies must first be established.
\begin{figure}
\centerline{\hbox{\includegraphics[height=0.45\textwidth]
             {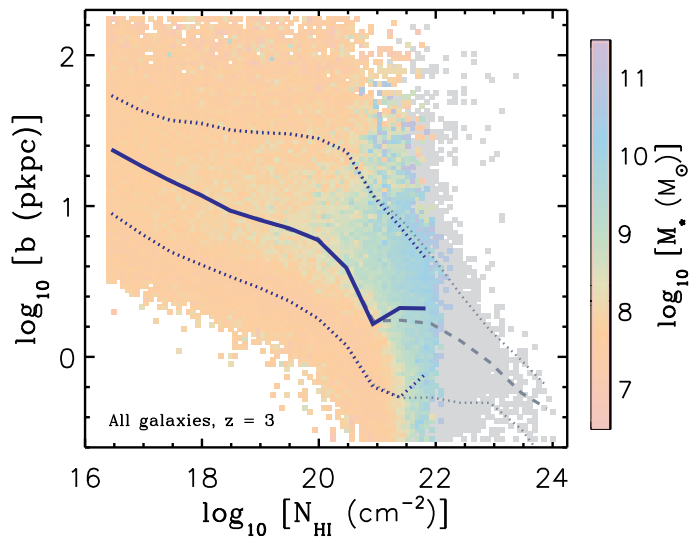}}}
\caption{The predicted impact parameters (in proper kpc) of absorbers as a function of $\HI$ column density at $z = 3$. The color of each cell (in the $b-\NHI$ plane) indicates the median stellar mass of the galaxies associated with the $\HI$ absorbers in that cell. The median impact parameter as a function of $\NHI$ is shown by the blue solid curve while the dotted curves indicate the $15\%-85\%$ percentiles. The gray cells show the region where $\Hm$ formation drains the atomic gas. The gray dashed and dotted curves show the median impact parameter and the $15\%-85\%$ percentiles as a function of $\NHI$ that we obtain if the conversion of high pressure gas into $\Hm$ is neglected.}
\label{figDLA:b-NHIall-2D}
\end{figure}

Galaxies and absorbers can be connected in two ways: by linking any given absorber to its closest galaxy (i.e., absorber-centered) or by finding absorbers that are closest to a given galaxy (i.e., galaxy-centered). 
In the present work, we use the absorber-centered matching to connect the simulated $\HI$ absorbers to neighboring galaxies. This approach is particularly efficient for associating rare strong $\HI$ absorbers to galaxies and it is the relevant approach for comparison with observational searches for the galaxy counterparts of $\HI$ absorbers.

The projected distances between $\HI$ absorbers and galaxies, together with their LOS velocity differences, can be used to associate them with each other. We use this method for a direct comparison between simulations and observational studies that employ the same approach. First, we calculate the velocity of each simulated galaxy along the LOS by adding its peculiar and Hubble velocities along the projection direction taking the periodic boundary conditions into account. Then, for every absorber we define the galaxy counterpart to be the galaxy with the shortest projected distance (i.e., the smallest impact parameter) among the galaxies with LOS velocity differences less than a chosen maximum value, $\Delta V_{\rm{LOS,~max}}$, with respect to the LOS velocity of the absorber, ${\left\langle V_{\rm{LOS}}\right\rangle}_{\HI}$. With this approach, each galaxy can be connected to more than one absorber, but each absorber is connected to one and only one galaxy.

We note that the difference between the LOS velocities of absorbers and galaxies includes not only the distance between the absorbers and galaxies along the LOS, but also their relative peculiar velocities along the LOS. Therefore, choosing values of $\Delta V_{\rm{LOS,~max}}$ that are less than the expected peculiar velocities around galaxies results in unphysical associations between $\HI$ absorbers and neighboring galaxies. We know that accretion of the gas into halos together with galactic outflows produces typical peculiar velocities of a few hundreds of kilometers per second. Similar velocity differences have been observed between the LOS velocity of absorbers and their host galaxies \citep[e.g.,][]{Fynbo99,Rakic12,Rudie12} in addition to being common in our simulations \citep{Voort12b}. For this reason, we chose $\Delta V_{\rm{LOS,~max}} = 300~\rm{km~s^{-1}}$, which is consistent with recent observations \citep{Rudie12}. However, as we show in Appendix \ref{ap:max-vel-dif}, our results are insensitive to this particular choice, provided $\Delta V_{\rm{LOS,~max}} > 100~\rm{km~s^{-1}}$, although the scatter in the $b-\NHI$ relation only converges for $\Delta V_{\rm{LOS,~max}} = 300~\rm{km~s^{-1}}$.

The simulated $\HI$ distribution and its connection to galaxies could in principle both depend on the resolution of our simulations. The $\HI$ column density distribution function is converged for LLSs and most DLAs at the resolution we use in this work \citep{Rahmati13a}, but this does necessarily imply that their distribution relative to galaxies with certain properties has converged. Indeed, we show in Appendix~\ref{ap:res} that a simulation with 8 times lower mass resolution only agrees with our fiducial model if we limit the analysis to galaxies with $\rm{SFR} > 0.4 ~\rm{\Msun~yr^{-1}}$, which suggests that our fiducial run may be converged down to the 8 times smaller value of $0.05 ~\rm{\Msun~yr^{-1}}$ (i.e.\ $\log_{10}[\rm{SFR} (\rm{\Msun~yr^{-1}})] > -1.3$). 
\begin{figure*}
\centerline{\hbox{\includegraphics[width=0.53\textwidth]
             {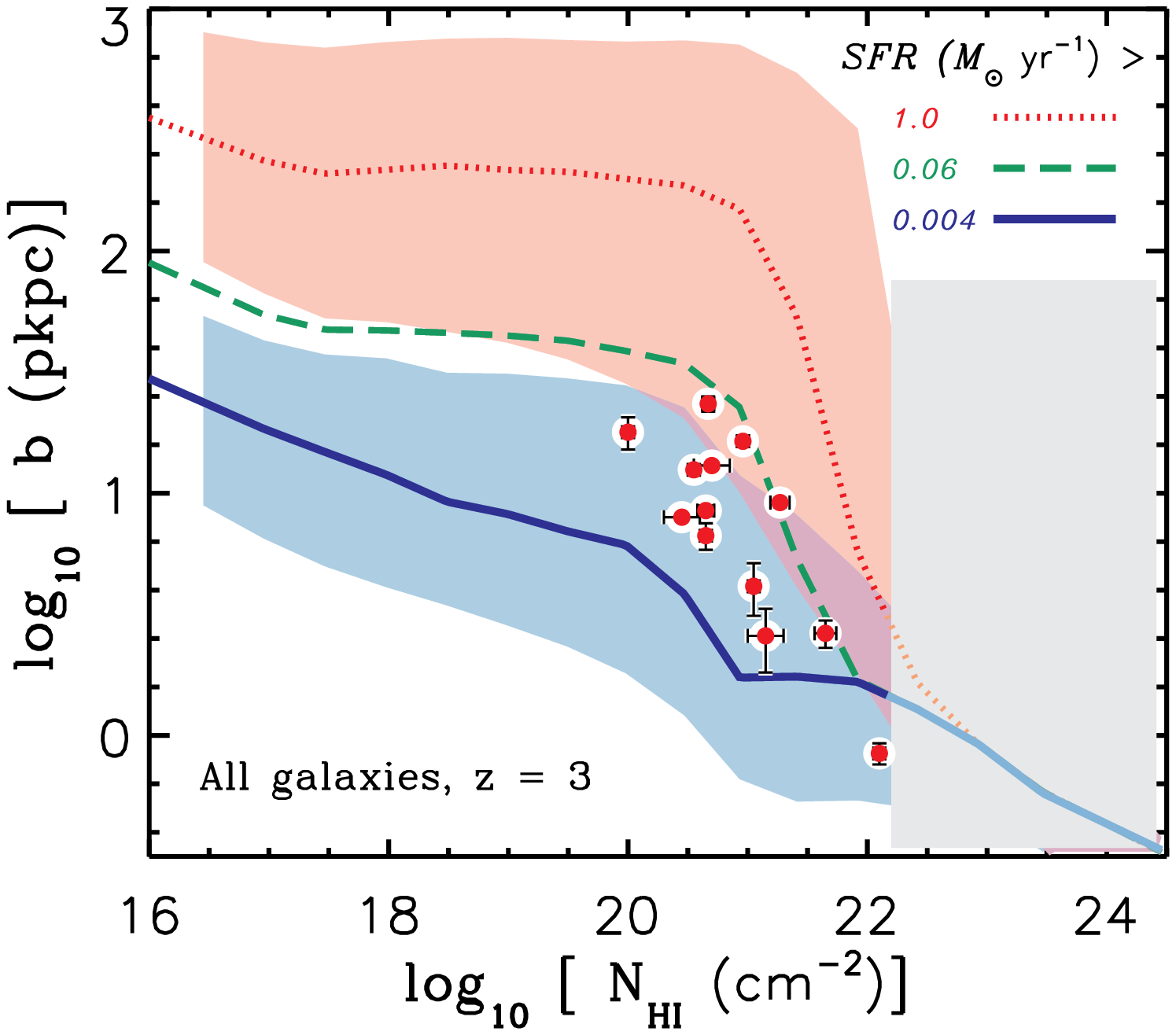}} 
             \hbox{\includegraphics[width=0.53\textwidth]
             {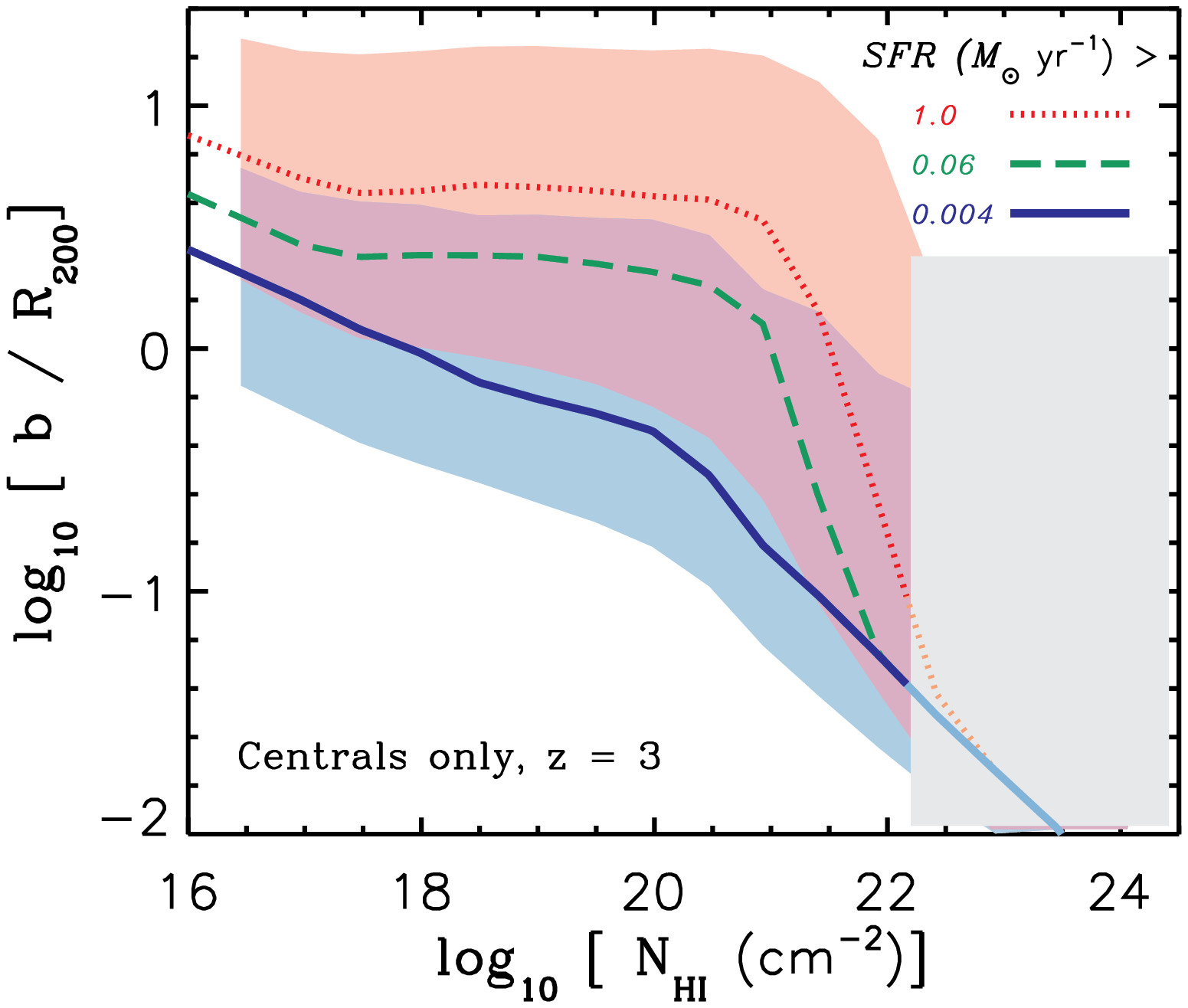}}}
\caption{Left: Predicted median impact parameter vs. $\NHI$ for different SFR thresholds at $z = 3$. Right: Median impact parameter normalized to the virial radius ($R_{200}$) as a function of $\NHI$. Since satellite galaxies reside in the halos of central galaxies, they do not have a well-defined virial radius. Therefore, the result in the right panel is based on matching $\HI$ absorbers to the central galaxy with the smallest impact parameter. In both panels, the red dotted, green dashed and blue solid curves correspond to SFR thresholds of  1, 0.06 and 0.004 $\Msun~\rm{yr^{-1}}$, respectively. The shaded areas around the blue solid and red dotted curves show the $15\%-85\%$ percentiles. Red data points in the left panel show a compilation of DLAs with observed galaxy counterparts described in Table \ref{tbl:obs}. Because of the efficient conversion of hydrogen atoms into molecules, absorbers with $\NHI \gtrsim 10^{22}\cmsq$ (indicated by the gray areas) are expected to be very rare.}
\label{figDLA:b-NHI}
\end{figure*}

We find, however, that these resolution limits are not determined by our ability to locate low-mass galaxies, but by our ability to predict their SFRs. The SFRs of the smallest galaxies in each simulation tend to decrease with increasing resolution. In Appendix~\ref{ap:res} we demonstrate that convergence is much better if we select galaxies by their cumulative number density (after sorting them by SFR and starting with the highest value) rather than by SFR. In many plots we therefore opted to show the relation between absorbers and galaxies down to $\rm{SFR} >4 \times 10^{-3} ~\rm{\Msun~yr^{-1}}$, which corresponds to a cumulative number density of $0.5$ galaxies per comoving $\rm{Mpc}^3$ ($31.5$ per proper $\rm{Mpc}^3$), but we stress that SFRs lower than $0.05 ~\rm{\Msun~yr^{-1}}$ are probably overestimated. Hence, a higher resolution simulation would likely produce nearly the same results as our fiducial run if we select galaxies with the same cumulative number density. However, for a fixed SFR threshold the predictions of the higher resolution simulation would start to differ if this threshold is lower than $0.05 ~\rm{\Msun~yr^{-1}}$ with the higher resolution simulation predicting somewhat larger impact parameters.

\section{Results and discussion}
\label{secDLA:results}
Using the procedure described in the previous section, we match $\HI$ column density systems that have $\NHI > 3\times10^{16}\cmsq$ to the galaxies with non-zero SFRs in our simulations (i.e., $\approx 2\times10^6$ strong $\HI$ absorbers and more than 10,000 galaxies for every projection). In the following we use these associations to study the connection galaxies and $\HI$ absorbers. 

\subsection{Spatial distribution of $\HI$ absorbers}

After connecting absorbers and galaxies, one can measure the typical projected distances (i.e., impact parameters, $b$) separating them. The predicted distribution of impact parameters as a function of $\HI$ column density is shown in Figure \ref{figDLA:b-NHIall-2D}. The color of each cell indicates the median stellar mass of galaxies that are associated with the absorbers in that cell (see the color bar on the right-hand side). To show how the impact parameter of typical absorbers is distributed at any given $\NHI$, we plot the median impact parameter as a function of $\NHI$ using the blue solid curve and the $15\%-85\%$ percentiles using the blue dotted curves. For reference, we note that if we randomize the positions of the galaxies, then the median impact parameter becomes $\approx 60$ proper kpc for all column densities. 

Our simulation predicts a strong anti-correlation between the $\HI$ column density of absorbers and their impact parameters. While weak LLSs with $\NHI \approx 10^{17}\cmsq$ have typical impact parameters $b\approx 20$ proper kpc, the impact parameter decreases with increasing $\HI$ column density such that strong DLAs with $\NHI > 10^{21}\cmsq$ are typically within a few proper kpc from the center of a galaxy. The increase in the impact parameter of $\HI$ absorbers with decreasing $\HI$ column density is in agreement with observations \citep{Moller98,Christensen07,Monier09,Rao11,Peroux11,Krogager12} and consistent with previous theoretical studies \citep{Gardner01,Pontzen08}. 

Despite the strong anti-correlation between the median impact parameter and $\NHI$, there is a large amount of scatter in the impact parameter at any given $\HI$ column density, as the dotted curves in Figure \ref{figDLA:b-NHIall-2D} show. Since galaxies actively exchange material with their surroundings through accretion and outflows, the $\HI$ distribution around them has a complex geometry (see the top-left panel of Figure \ref{figDLA:stamp} for an example). This complexity is a major contributor to the scatter in the impact parameters. In addition, part of the scatter is due to the fact that, at any given $\NHI$, galaxies with a range of sizes contribute to the total distribution of absorbers. This can be seen from the color gradients in Figure \ref{figDLA:b-NHIall-2D}: at any given impact parameter, the typical masses of galaxies that are linked to $\HI$ absorbers increases with $\NHI$ and at any given $\HI$ column density (particularly for DLAs), the typical masses of galaxy counterparts increases with the impact parameter of the $\HI$ absorbers. We will discuss this further in $\S$\ref{secDLA:phys-prop}.

To show the impact of $\Hm$ formation Figure \ref{figDLA:b-NHIall-2D} also shows the regions where the $\HI$ gas is converted into molecules using gray cells (see $\S$\ref{secDLA:HI_method} for details on the $\Hm$ calculation). The median impact parameters and $15\%-85\%$ percentiles that we obtain if we do not account for $\Hm$ formation are shown as gray dashed and gray dotted curves, respectively. The comparison between the colored and gray areas (and curves) in Figure \ref{figDLA:b-NHIall-2D} shows that $\Hm$ formation only strongly affects $\HI$ column densities $\NHI > 10^{22} \cmsq$. This is consistent with the sharp cut-off in the observed $\HI$ column density distribution at $\NHI > 10^{22} \cmsq$ as shown in \citet{Rahmati13a} (see also \citealp{Altay11,Erkal12}). The formation of $\Hm$ does not significantly affect our predictions for the impact parameters of $\HI$ absorbers with $\NHI < 10^{22} \cmsq$. 

%\begin{sidewaystable}
%
\begin{table*}
\caption{A compilation of confirmed $z\sim 2-3$ DLA-galaxy pairs from the literature. The columns from left to right show respectively: the ID of the background quasar, the DLA redshift, \HI column density, the impact parameter in arc seconds, the impact parameter in proper kpc, the SFR and stellar mass of the associated galaxy and finally the references from which these values are extracted. We note that the SFR estimates are often based on Ly$\alpha$ emission, which provides a lower limit for the SFR since it is difficult to account for dust extinction. SFR estimates that have been corrected for dust extinction are indicated with bold-face.} 
\begin{tabular}{lccccccc}
\hline
ID & $\rm{z_{DLA}}$ & $\log{\NHI}$ & $b$ & $b_{\rm{p}}$ & SFR &  $\rm{M_{\star}}$  & Reference\\  
    &                   &  $(\cmsq)$   & (arcsec) & (pkpc) & $(\Msun~\rm{yr^{-1})}$  &    ($10^{9}~\Msun$)  & \\ 
\hline 
 
Q2206-1958         &  1.92 & $20.65$ & $0.99$  & $8.44$    & 3       &   -   &  [1]\\ 
 
Q0151+048A        & 1.93 &  $20.45 $& $0.93$  & $7.93$   &71       &    -   & [3] \& [4]\\

PKS 0458-02        & 2.04 &  $21.65$ & $0.31$  & $2.63$   & 6        &     -  & [2]\\

Q1135-0010         &  2.21 &  $22.10$ & $0.10$ & $0.84$   & 25      &     -  &  [6]\\

Q0338-0005         &  2.22 &  $21.05$ & $0.49$ & $4.12$   & -         &     -  & [5]\\

Q2243-60             &  2.33 &  $20.67$ & $2.28$ & $23.37$ &{\bf36} &    -  & [7]\\ 

Q2222-0946        &  2.35 &  $20.65$ & $0.8$    & $6.67$  & {\bf13}&   2    & [8] \& [9]\\

Q0918+1636        &  2.58 &  $20.96$ & $2.0$   & $16.38$& {\bf27}& 12.6 &[ 10] \& [5] \& [15]\\

Q0139-0824         &  2.67 &  $20.70$ & $1.6$   & $13.01$  & -          &    -   &  [11]\\

J073149+285449 &  2.69 &  $20.55$ & $1.54$ & $12.50$  & {\bf12} &  -    & [12]\\

PKS 0528-250      & 2.81 &  $21.27$ & $1.14$ & $9.15$     & 17      &  -     &  [1]\\

2233.9+1318	   &  3.15 &  $20.00$ & $2.3$  & $17.91$   &  20      &   -   &  [13]\\

Q0953+47            &  3.40 &  $21.15$ & $0.34$   & $2.58$    & -       &    -   & [14]\\

\hline
\end{tabular}

[1]- \citet{Moller02}; [2]- \citet{Moller04}; [3]- \citet{Moller98}; [4]- \citet{Fynbo99}; [5]- \citet{Krogager12}; [6]- \citet{Noterdaeme12}; [7]- \citet{Bouche12}; [8]- \citet{Fynbo10}; [9]- \citet{Krogager13}; [10]- \citet{Fynbo11}; [11]- \citet{Wolfe08}; [12]- \citet{Fumagalli10}; [13]- \cite{Djorgovski96}; [14]- \citet{Prochaska03}; [15]- \citet{Fynbo13}
\label{tbl:obs}
\end{table*}%{sidewaystable}%

\subsection{The effect of a finite detection limit}

As seen from the colors in Figure \ref{figDLA:b-NHIall-2D}, our simulation predicts that most $\HI$ absorbers with $10^{17} <\NHI \lesssim 10^{21}\cmsq$ are closely associated with low-mass galaxies, with typical stellar masses of $\rm{M_{\star}} \lesssim 10^8~\Msun$. The typical SFR for those galaxies is $\lesssim 10^{-1} ~\rm{\Msun~yr^{-1}}$. On the other hand, the typically accessible sensitivity of observations only allows the detection of galaxies that have $\rm{SFR}\gtrsim 1-10 ~\rm{\Msun~yr^{-1}}$ at $z \approx 3$\footnote{For a Chabrier IMF and adopting the optically-thick limit for hydrogen ionizing photons, a star formation rate of $1 ~\rm{\Msun~yr^{-1}}$ corresponds to a Ly$\alpha$ luminosity of $2\times 10^{42}~{\rm erg \, s^{-1} }$ \citep{Kennicutt98}, which translates into an observed flux of $\approx 6.7\times 10^{-17}~{\rm erg \, s^{-1} \, cm^{-2}}$ and $\approx 2.5\times 10^{-17}~{\rm erg \, s^{-1} \, cm^{-2}}$ at redshifts $z = 2$ and $z =3$, respectively.} (but see \citealp{Rauch08}). Because of this relatively high detection threshold, most galaxy counterparts are not detectable and the chance of observing galaxies that host LLSs and DLAs is slim. This could be the main reason why observational surveys that are aiming to find galaxies close to DLAs, often result in non-detections \citep[e.g.,][]{Foltz86,Smith89,Lowenthal95,Bunker99,Prochaska02,Kulkarni06,Rahmani10,Bouche12}. Moreover, the relatively low sensitivity of observational surveys may bias in the measured distribution of impact parameters of absorbers by mis-associating them to the closest detectable galaxy in their vicinity, instead of their real hosts that are likely to fall below the detection limit. 

Our simulation includes galaxies down to SFRs that are much lower than the typical detection threshold of observations. Therefore, we are able to analyze the impact of varying the detection limit. Figure \ref{figDLA:stamp} shows the distribution of the $\HI$ column densities and positions of galaxies in a simulated region of size 500 proper kpc around a randomly selected massive galaxy at $z = 3$. The top-left panel of Figure \ref{figDLA:stamp} shows the distribution of $\HI$ column density and galaxies that have $\rm{SFR} > 10 ~\rm{\Msun~yr^{-1}}$. With this detection threshold, only the central galaxy (shown with the dark circle whose size is proportional to the virial radius of the central galaxy) and one of its satellites (shown with the small white circle) are detectable. Other panels in this figure show that as the SFR threshold for detecting galaxies decreases, many more galaxies show up in the field, which strongly decreases the typical impact parameter. 

The effect of the detection threshold on the impact parameters of strong $\HI$ absorbers (i.e., $\NHI \gtrsim 10^{17}\cmsq$) is shown more quantitatively in the left panel of Figure \ref{figDLA:b-NHI}. Different curves show the median impact parameter as a function of $\NHI$ assuming different SFR detection thresholds. The blue solid curve assumes a SFR detection threshold identical to that of Figure \ref{figDLA:b-NHIall-2D}, where all galaxies with $\rm{SFR} > 4 \times 10^{-3} ~\rm{\Msun~yr^{-1}}$ are considered as potential galaxy counterparts. The green dashed and red dotted curves, which respectively correspond to SFR thresholds of $ > 6 \times 10^{-2} ~\rm{\Msun~yr^{-1}}$ and $ > 1 ~\rm{\Msun~yr^{-1}}$, show the effect of increasing the SFR threshold on the impact parameter distribution. The shaded areas around the blue solid and red dotted curves show the $15\%-85\%$ percentiles, and the overlap region between the two shaded areas is shown in purple. The gray area at $\NHI > 10^{22} \cmsq$ shows the region affected by the formation of $\Hm$. 

The comparison between the three curves in the left panel of Figure \ref{figDLA:b-NHI} shows that the impact parameters increase strongly with the detection threshold. If we lower the SFR threshold from 1 to $4\times 10^{-3} ~\rm{\Msun~yr^{-1}}$, then nearly every $\HI$ absorber becomes associated to a different, fainter galaxy. Moreover, the anti-correlation between the impact parameter and $\NHI$ is also sensitive to the detection threshold. As the green dashed and red dotted curves show, for detection threshold $\rm{SFR} \gtrsim 10^{-1} ~\rm{\Msun~yr^{-1}}$ the strong anti-correlation between the impact parameter and $\NHI$ becomes insignificant at $\NHI \lesssim 10^{20}\cmsq$. The main reason for this is that most galaxy counterparts that are detectable with relatively low sensitivities (i.e., high detection thresholds) are not physically related to the strong $\HI$ absorbers. As a result, the probability distribution of the impact parameters for those systems is controlled by the average projected distribution of the detectable galaxies. As the red dotted curve in the left panel of Figure \ref{figDLA:b-NHI} shows, our simulation predicts that with a detection threshold of $\rm{SFR} > 1 ~\rm{\Msun~yr^{-1}}$ the typical impact parameters between strong $\HI$ absorbers and their nearest galaxies vary from several tens of kpc to a few hundred kpc. This result is in excellent agreement with the measured impact parameters between DLAs and galaxies in observational surveys that used similar detection thresholds at $z \approx 2-3$ \citep[not shown in this plot but see e.g.,][]{Teplitz98,Mannucci98}. It is also worth noting that the anti-correlation between $b$ and $\NHI$ remains in place for strong DLAs (i.e., $\NHI > 10^{21}\cmsq$), even for a relatively high detection threshold like $\rm{SFR} > 1 ~\rm{\Msun~yr^{-1}}$. However, the impact parameters of those systems are increasingly over-estimated as the galaxy detection threshold is increased. Nevertheless, the existence of an anti-correlation implies that a significant fraction of strong DLAs, but not LLSs and weak DLAs, is physically related to a galaxy with a relatively high SFR.

We show the observed impact parameters of a compilation of confirmed DLA-galaxy pairs using red symbols with error-bars in the left panel of Figure \ref{figDLA:b-NHI} (see Table \ref{tbl:obs} for more details). The observed impact parameters are generally above the blue solid curve in the left panel of Figure \ref{figDLA:b-NHI}, which corresponds to SFR $> 4\times 10^{-3} ~\rm{\Msun~yr^{-1}}$. This is not surprising, as we have already noted that such faint galaxies cannot be detected. Indeed, the observed galaxies that have been associated with DLAs have SFR $\sim 10 ~\rm{\Msun~yr^{-1}}$ (see Table \ref{tbl:obs}). On the other hand, comparison of the data points with the red curve and the red shaded area shows that impact parameters as low as those of the observed galaxies should be rare for a detection limit of $10 ~\rm{\Msun~yr^{-1}}$. We can, however, understand this apparent discrepancy by noting that there are many more non-detections of galaxy counterparts of DLAs in the literature than there are detections \citep[e.g.,][]{Foltz86,Smith89,Lowenthal95,Bunker99,Prochaska02,Kulkarni06,Rahmani10,Bouche12} and that there are most likely many more unpublished non-detections. For these non-detections the nearest galaxy with SFR $> 10 ~\rm{\Msun~yr^{-1}}$ will typically have a much larger impact parameter, sufficiently large to be either outside the survey area or to be considered unrelated by the observers. Therefore, the existing observational sample is unrepresentative and the impact parameters are biased low. However, this does not mean that the existing observed pairs are not real associations. Indeed, as we will show in $\S$\ref{secDLA:phys-prop}, the subset of DLAs for which the nearest galaxy is bright enough to be detectable do have impact parameters that are similar to the observed values, which means that the simulation does in fact agree with the observations.

\subsection{Distribution of $\HI$ absorbers relative to halos}

It is useful to compare the impact parameters that connect absorbers to their neighboring galaxies with the size of the halos hosting those galaxies. We therefore define the normalized impact parameter of absorbers as the ratio of the impact parameter and the virial radius (${R_{200}}$) of the host galaxy. Since the virial radius is only well defined for central galaxies, we only consider those objects when associating absorbers to galaxies. 

As shown by the blue solid curve in the right panel of Figure \ref{figDLA:b-NHI}, stronger $\HI$ absorbers tend to be located closer to the centers of halos. Most LLSs are found within the virial radius of the halo hosting the nearest (in projection) central galaxy and the majority of DLAs is found within a few tenths of the viral radius. There is, however, a large scatter in the normalized impact parameters at given $\NHI$ which is shown by the shaded area around the blue solid and red dotted curves in the right panel of Figure \ref{figDLA:b-NHI}. A non-negligible fraction of DLAs have impact parameters comparable to, or even larger than the virial radius of their associated central galaxies. This is in part due to the neglect of satellite galaxies in matching absorbers and galaxies (because they do not have a well defined virial radius) which effectively associates the strong $\HI$ absorbers that are near satellites to their closest central galaxies\footnote{Note that satellite galaxies (and hence their associated absorbers) are not necessarily within the virial radius of their host galaxy, due to the non-spherical distribution of FoF structures. This is shown in Figure \ref{figDLA:stamp}.}. The complex and highly structured distribution of strong $\HI$ absorbers which often extends to distances beyond the virial radius of central galaxies (see Figure \ref{figDLA:stamp}) also contributes to the large scatter around the median normalized impact parameter at a given $\NHI$.

The anti-correlation between the normalized impact parameter of absorbers and their $\HI$ column density is steeper than that between the absolute impact parameter and $\NHI$ (compare the panels of Figure \ref{figDLA:b-NHI}). This difference is most pronounced at $\NHI \gtrsim 10^{21}\cmsq$ where the impact parameter flattens with increasing $\NHI$ while the normalized impact parameter still decreases steeply with $\NHI$. This trend can be explained by the contribution of massive (and hence large) galaxies becoming increasingly more dominant at very high $\HI$ column densities (see Figure \ref{figDLA:b-NHIall-2D} and $\S$\ref{secDLA:phys-prop}). 

The trends we discussed earlier for the effect of varying the SFR threshold on the impact parameter of absorbers also hold qualitatively for the normalized impact parameters. As the green dashed and red dotted curves in the right panel of Figure \ref{figDLA:b-NHI} show, increasing the SFR threshold for the galaxies that are considered when matching absorbers to galaxies, results in larger normalized impact parameters. Despite the qualitatively similar trends, the differences between the normalized impact parameters for different SFR thresholds are smaller than the differences between the absolute impact parameters. This difference is due to the fact that galaxies with higher SFRs tend to be hosted by more massive galaxies and hence larger halos. For $10^{17} < \NHI < 10^{21} \cmsq$ both the green dashed and red dotted curves in the right panel of Figure \ref{figDLA:b-NHI} are nearly flat. Such absorbers are several virial radii away from the closest central galaxies with $\rm{SFR} > 6 \times 10^{-2} ~\rm{\Msun~yr^{-1}}$ and $\rm{SFR} > 1 ~\rm{\Msun~yr^{-1}}$, respectively, regardless of $\NHI$. The absence of an anti-correlation between $b/{{R_{200}}}$ and $\NHI$ supports our earlier statement that most of the associations between $\HI$ absorbers with $\NHI < 10^{21} \cmsq$ and galaxies are not physical unless galaxies with $\rm{SFR} \ll 10^{-1} ~\rm{\Msun~yr^{-1}}$ are considered.
\begin{figure*}
\centerline{\hbox{\includegraphics[width=0.53\textwidth]
             {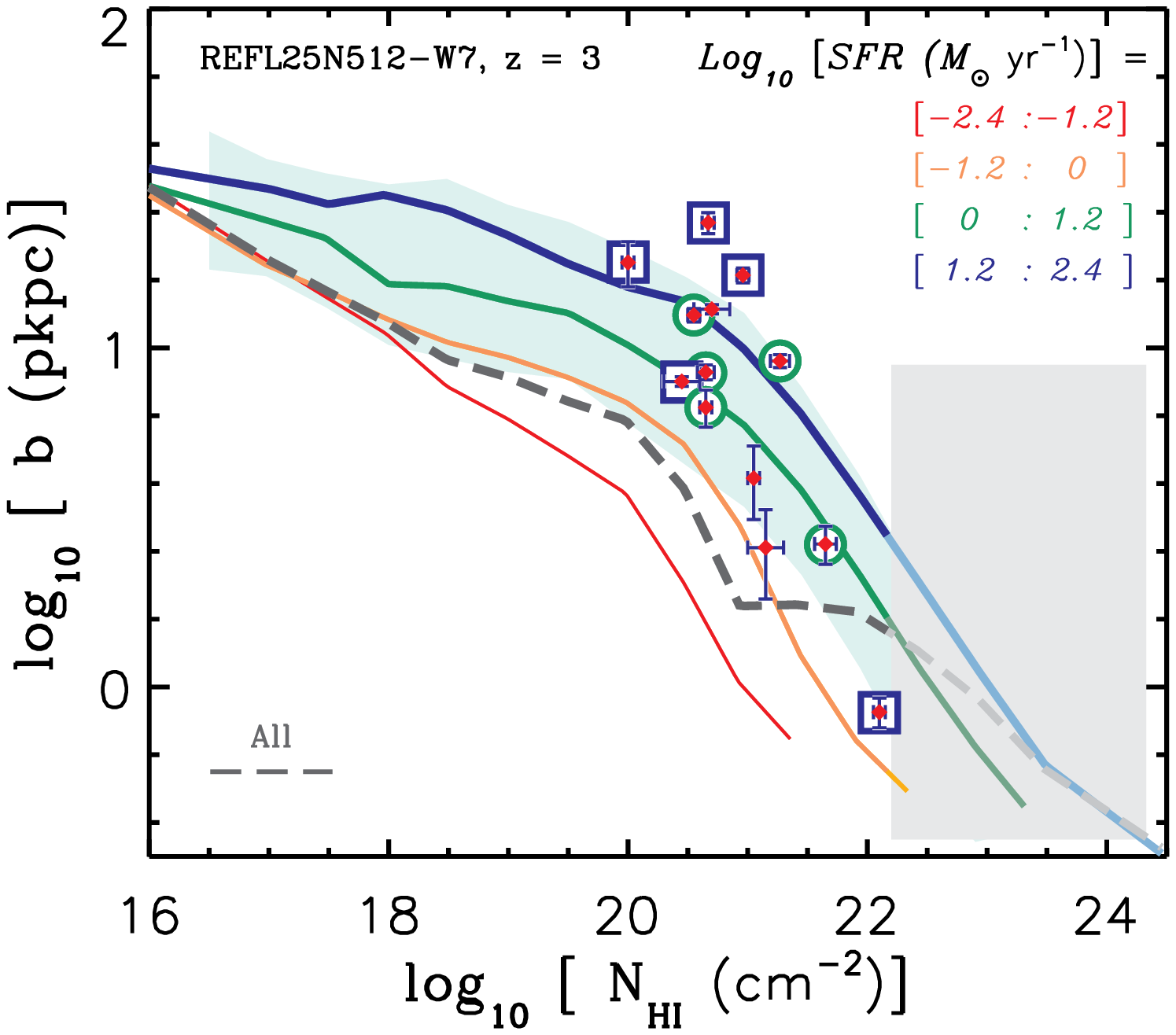}} 
             \hbox{\includegraphics[width=0.53\textwidth]
             {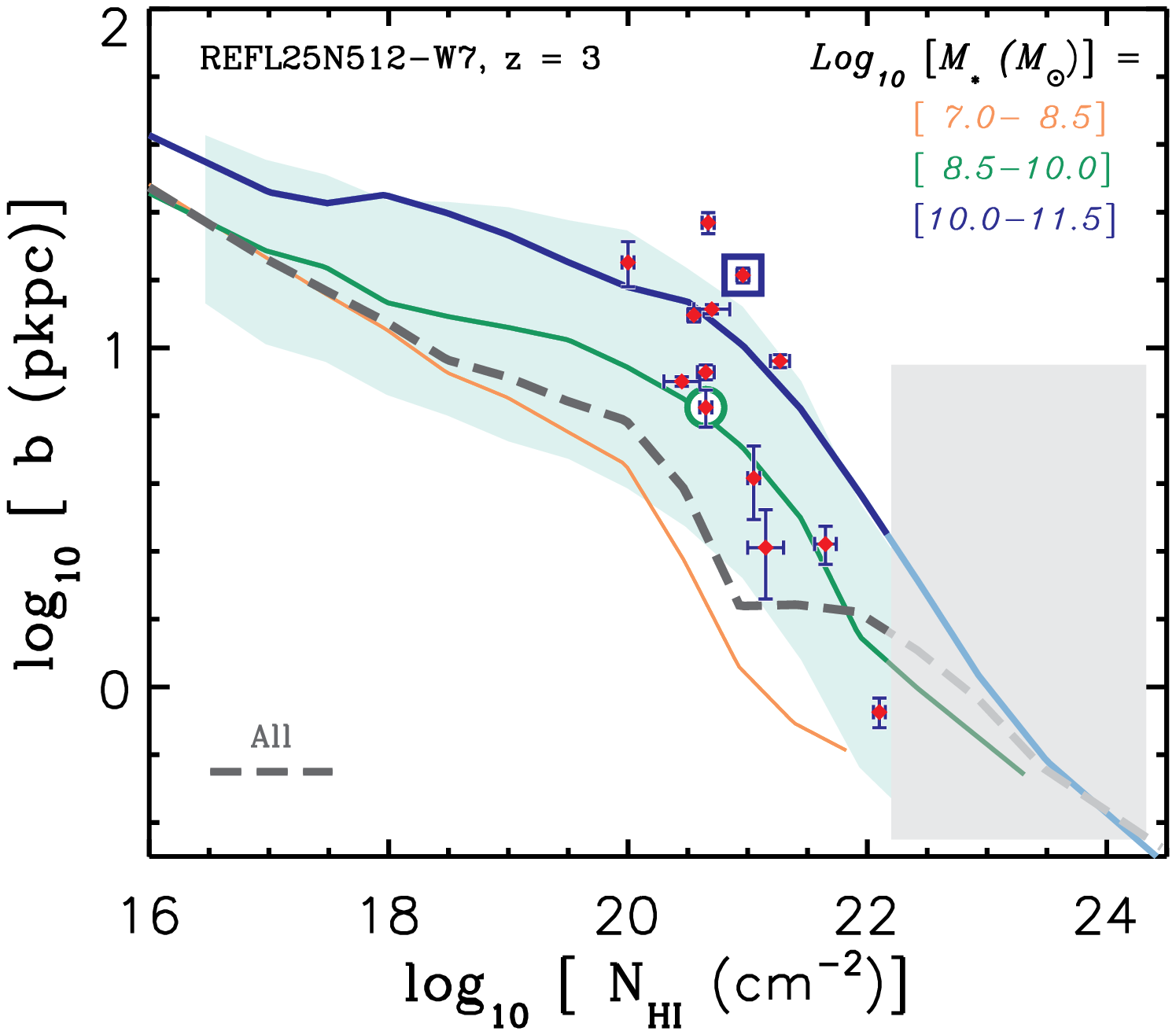}}}
\caption{The predicted median impact parameters for subsets of $\HI$ absorbers associated with galaxies in different star formation rate bins (left) and stellar mass bins (right) as a function of $\NHI$, at $z = 3$. The shaded area around the solid green curves in the left (right) panel shows the $15\%-85\%$ percentile in the distribution of absorbers that are linked to galaxies with $1 < \rm{SFR} < 16 ~\rm{\Msun~yr^{-1}}$ ($3\times 10^{8}< \rm{M_{\star}} < 10^{10}~\Msun$). The dashed curves in both panels show the median impact parameter of all absorbers as a function of $\NHI$. The data points show the observed impact parameters for confirmed observed DLA-galaxy pairs (see Table \ref{tbl:obs}). The colored circles and squares around the data points indicate the SFR/mass bin to which they belong. Note that the squares and circles, which show the two bins with the highest values (of SFR or mass) respectively, are in agreement with our results. Because of a very efficient conversion of hydrogen atoms into molecules, absorbers with $\NHI \gtrsim 10^{22}\cmsq$ (indicated with the gray areas) are expected to be very rare.}
\label{figDLA:prof-Ms-sfrI}
\end{figure*}
\begin{figure*}
\centerline{\hbox{\includegraphics[width=0.53\textwidth]
             {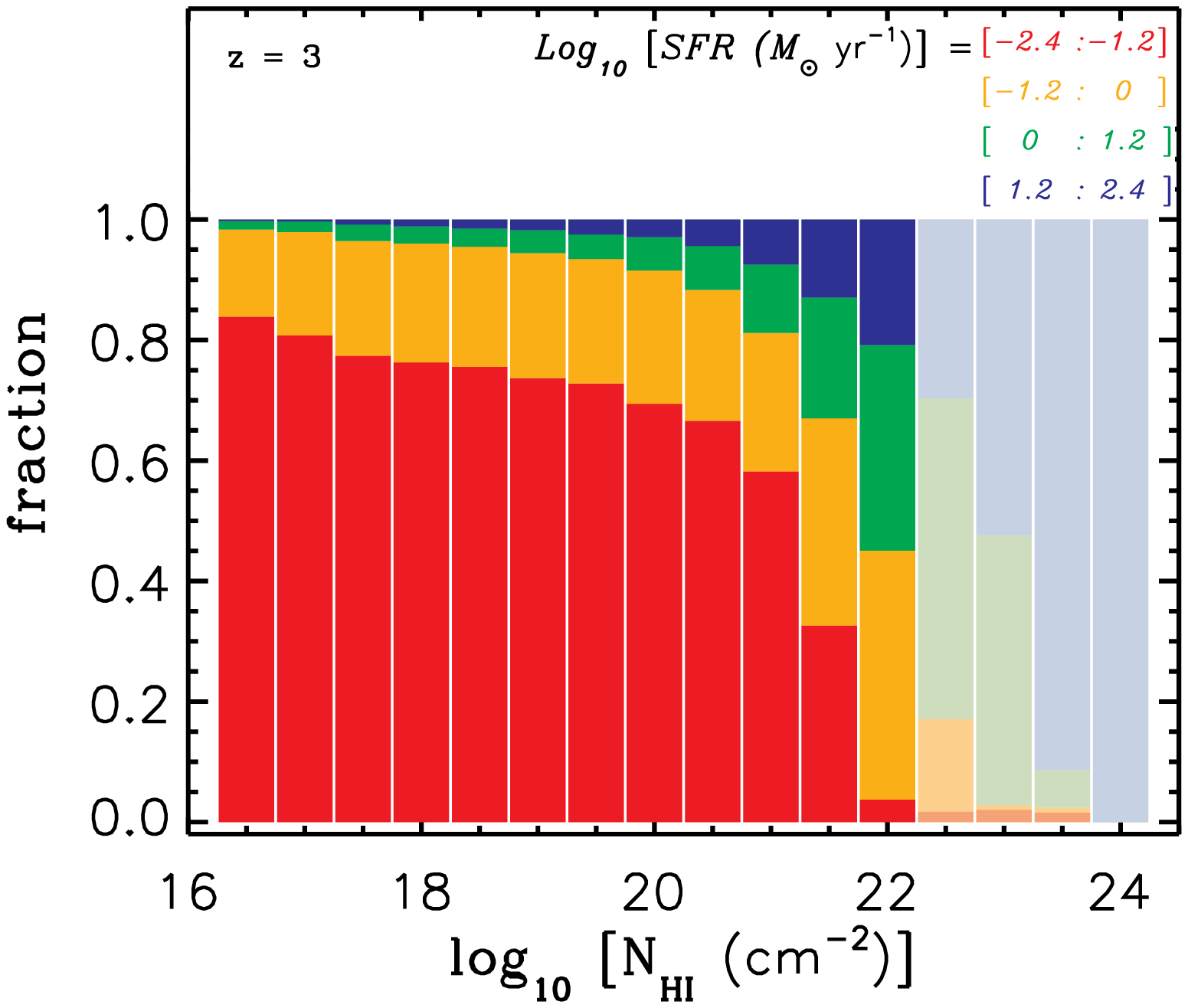}} 
             \hbox{\includegraphics[width=0.53\textwidth]
             {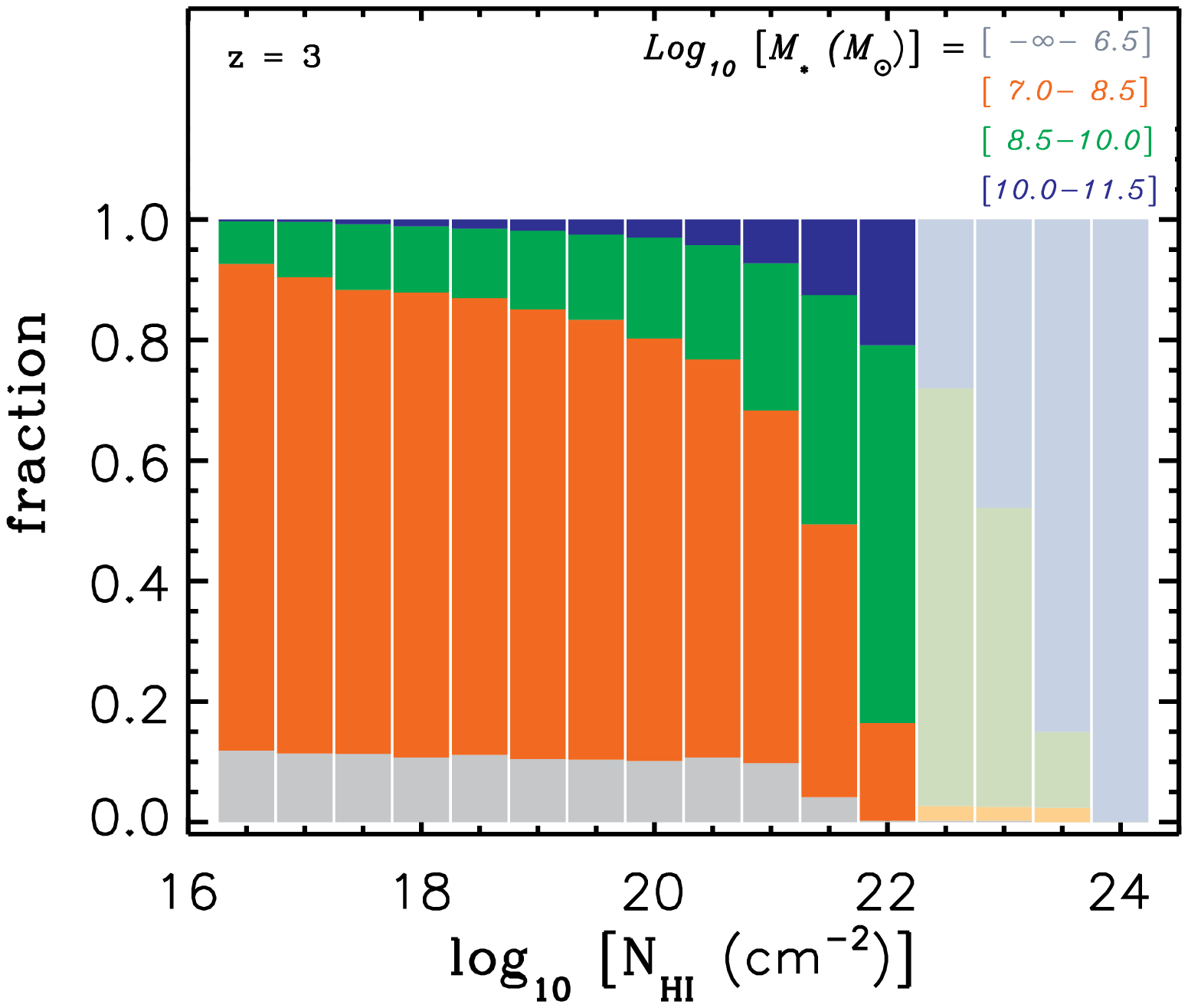}}}
\caption{The fraction of absorbers that are associated with galaxies in different star formation rate bins (left) and stellar mass bins (right) as a function of $\NHI$, at $z = 3$. The star formation rate bins and stellar mass bins are identical to that of Figure \ref{figDLA:prof-Ms-sfrI}. Because of efficient conversion of hydrogen atoms into molecules, absorbers with $\NHI \gtrsim 10^{22}\cmsq$ (indicated with the gray areas) are expected to be very rare.}
\label{figDLA:prof-Ms-sfrII}
\end{figure*}
\begin{figure*}
\centerline{\hbox{\includegraphics[height=0.39\textwidth]
             {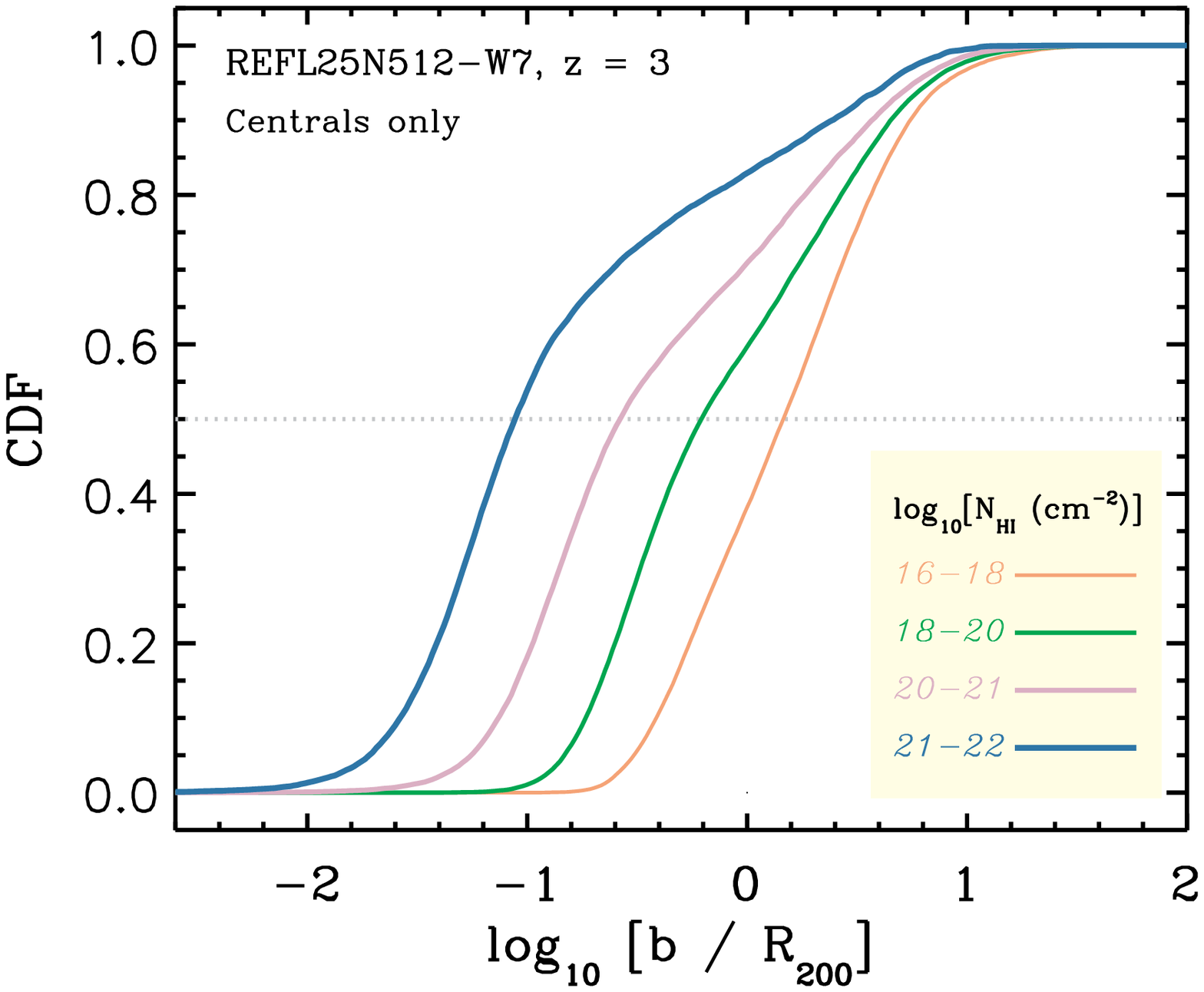}} 
             \hbox{\includegraphics[height=0.39\textwidth]
              {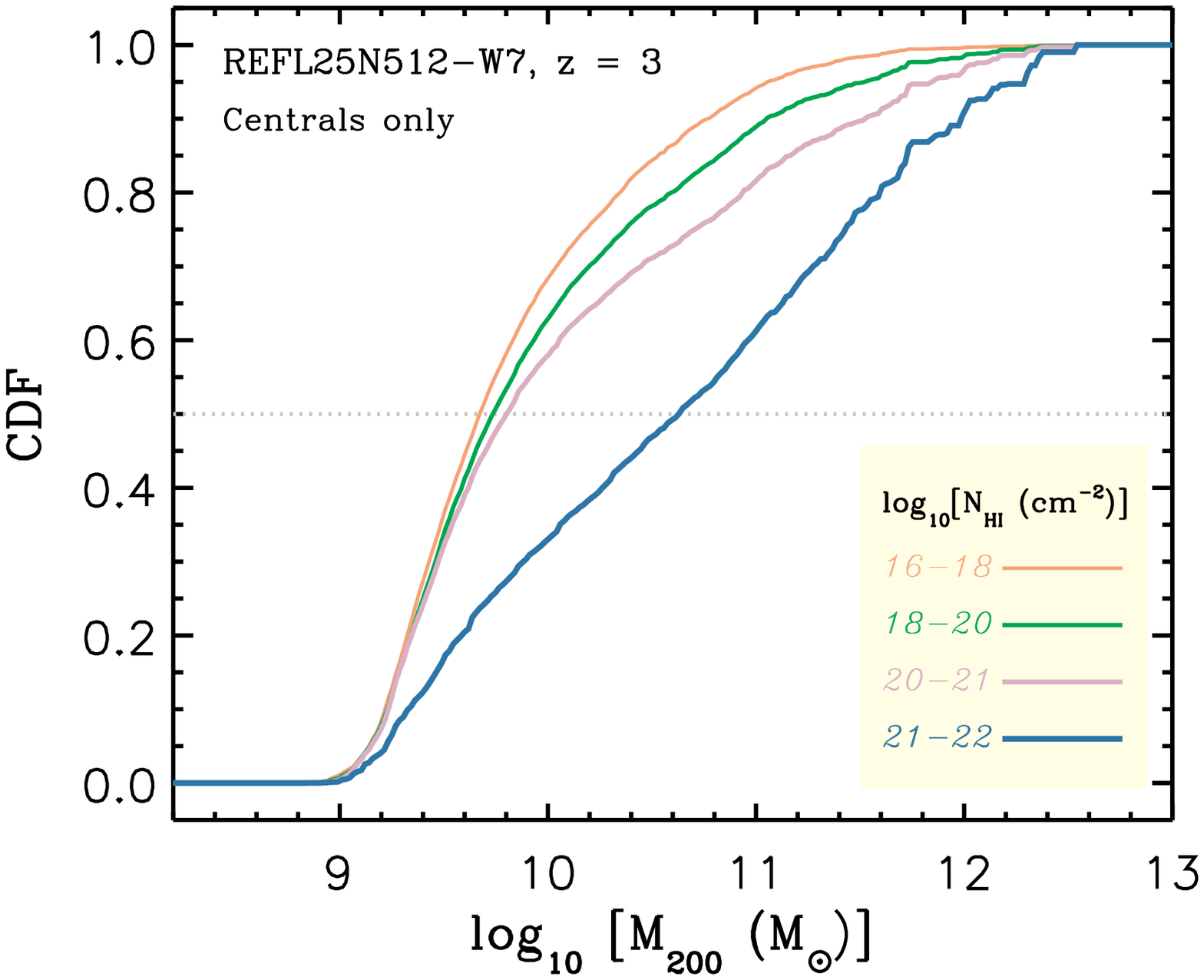}}}
\centerline{\hbox{\includegraphics[height=0.39\textwidth]
             {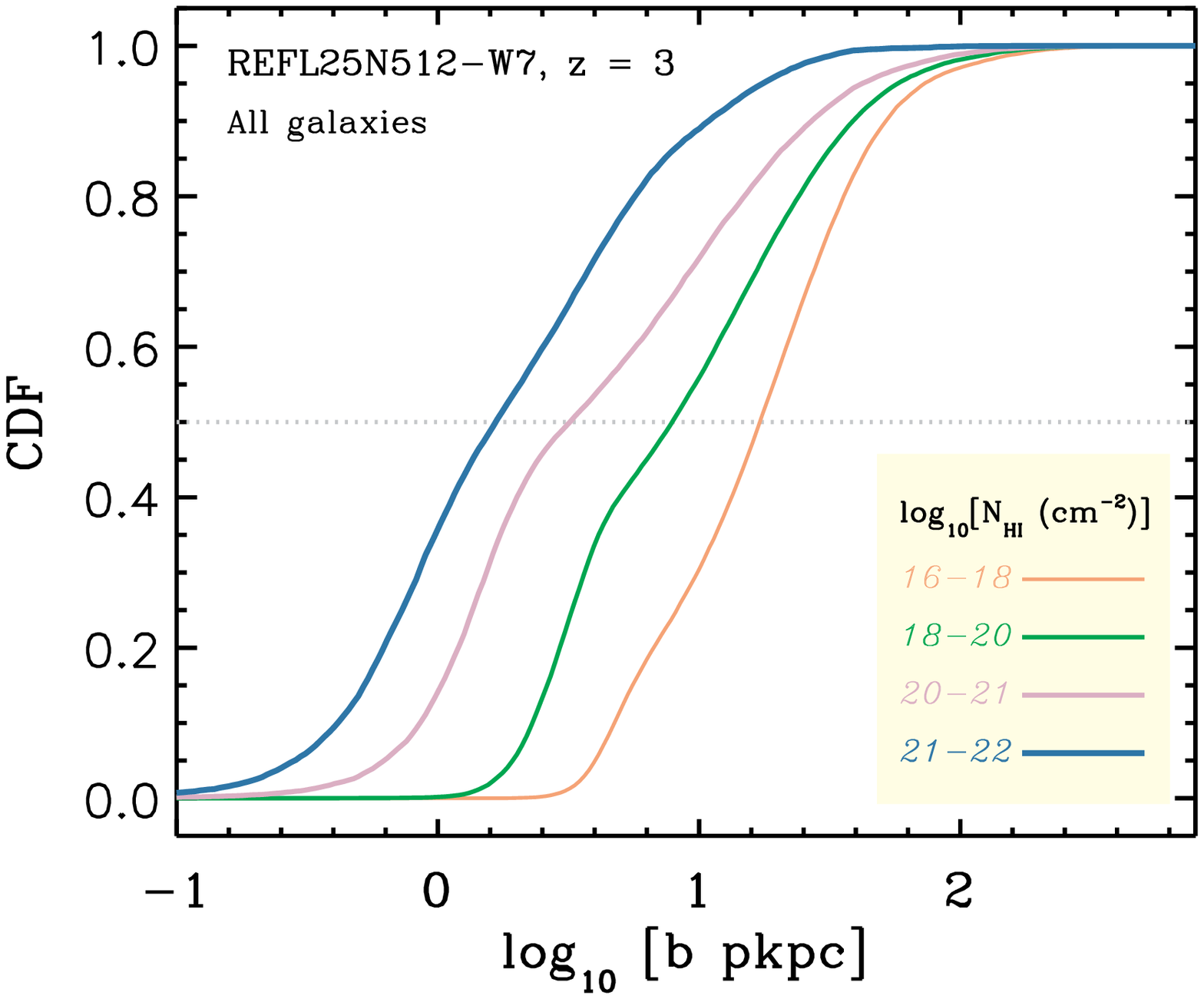}} 
             \hbox{\includegraphics[height=0.39\textwidth]
             {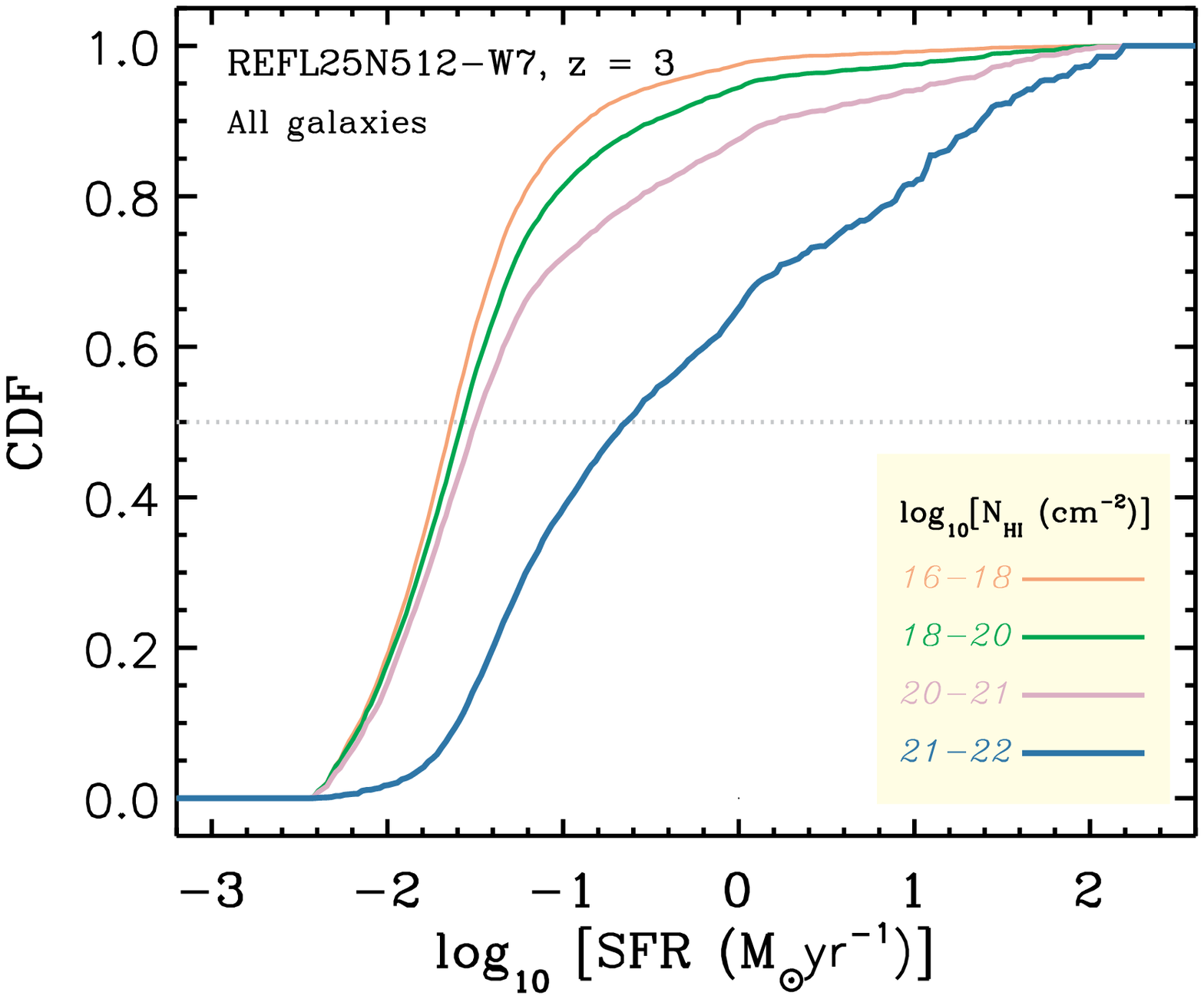}}}
\centerline{\hbox{\includegraphics[height=0.39\textwidth]
             {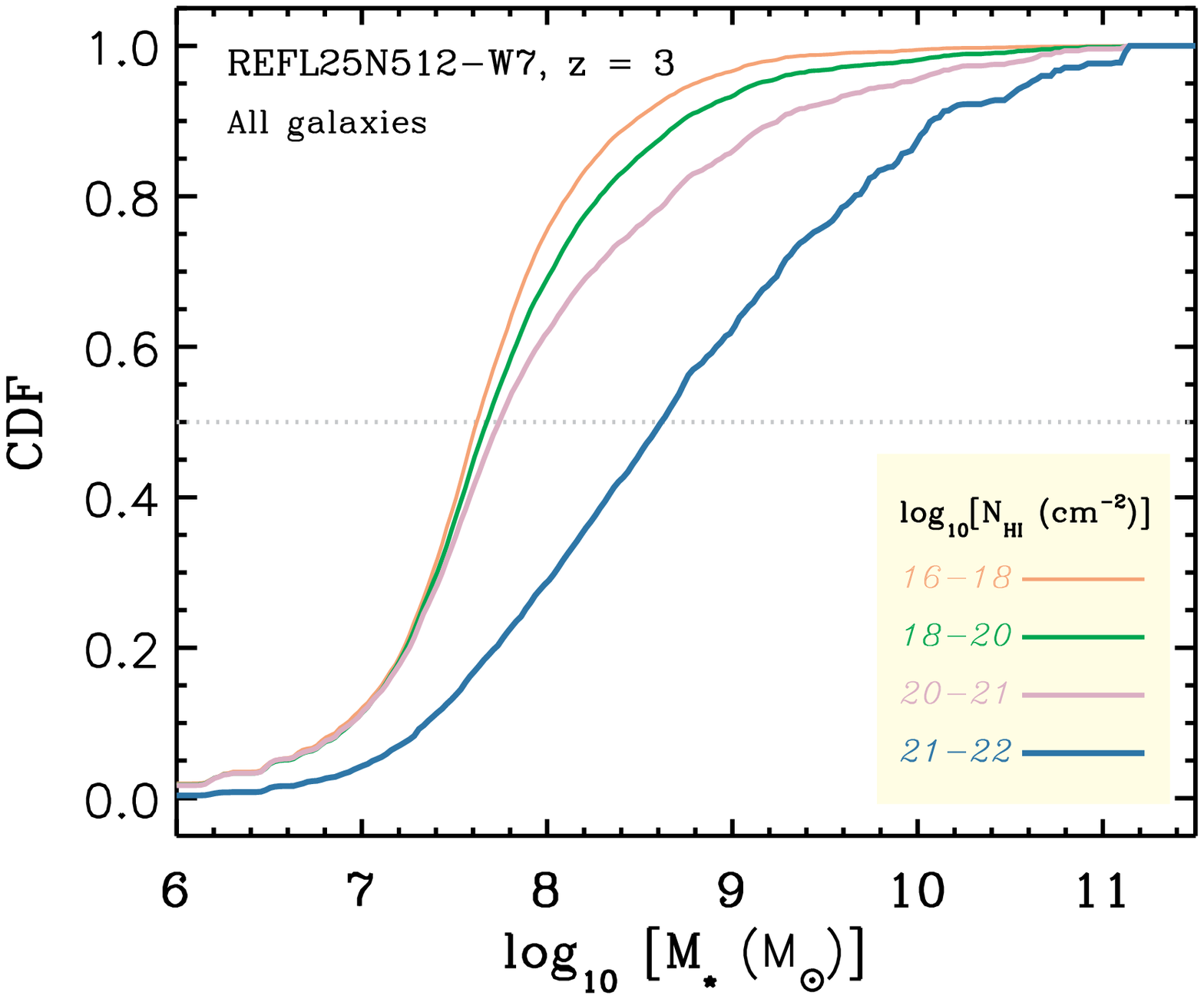}}
             \hbox{\includegraphics[height=0.39\textwidth]
             {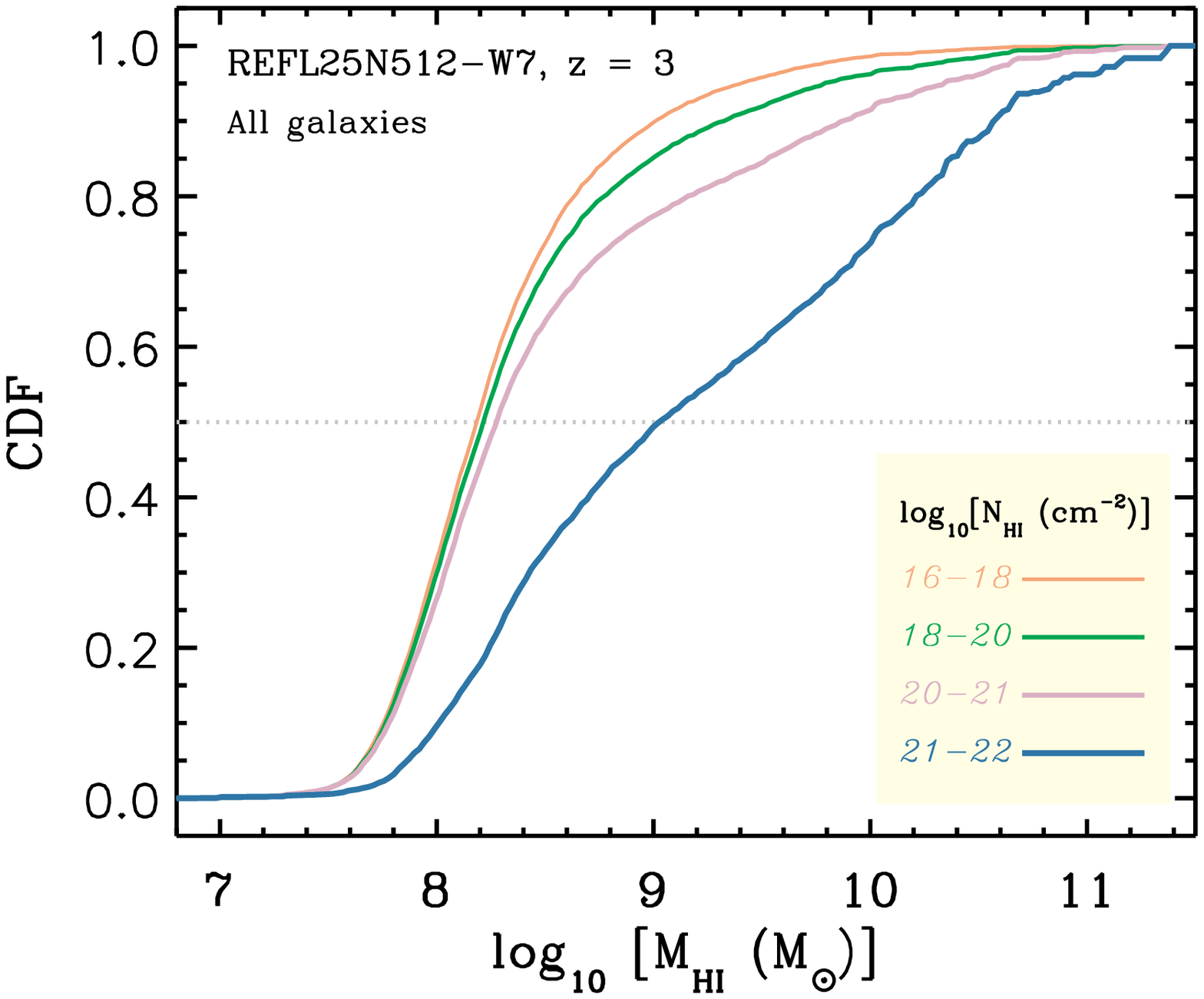}}}
\caption{Cumulative distributions of the properties of galaxies associated with strong z = 3 $\HI$ absorbers for different column density bins. The top-left and top-right panels respectively show normalized impact parameter and halo mass, taking only central galaxies into account. The other four panels take all galaxies (with $\rm{SFR} > 0.004~\Msun~\rm{yr^{-1}}$) into account. Panels from middle-left to bottom right show impact parameters, star formation rates, stellar masses and $\HI$ masses, respectively. Except for strong DLAs ($10^{21} < \NHI < 10^{22}\cmsq$), the properties of the associated galaxies are insensitive to the $\HI$ column density. However, the (normalized) impact parameters do depend strongly on the column density.}
\label{figDLA:cumdist}
\end{figure*}
\begin{figure*}
\centerline{\hbox{\includegraphics[height=0.39\textwidth]
             {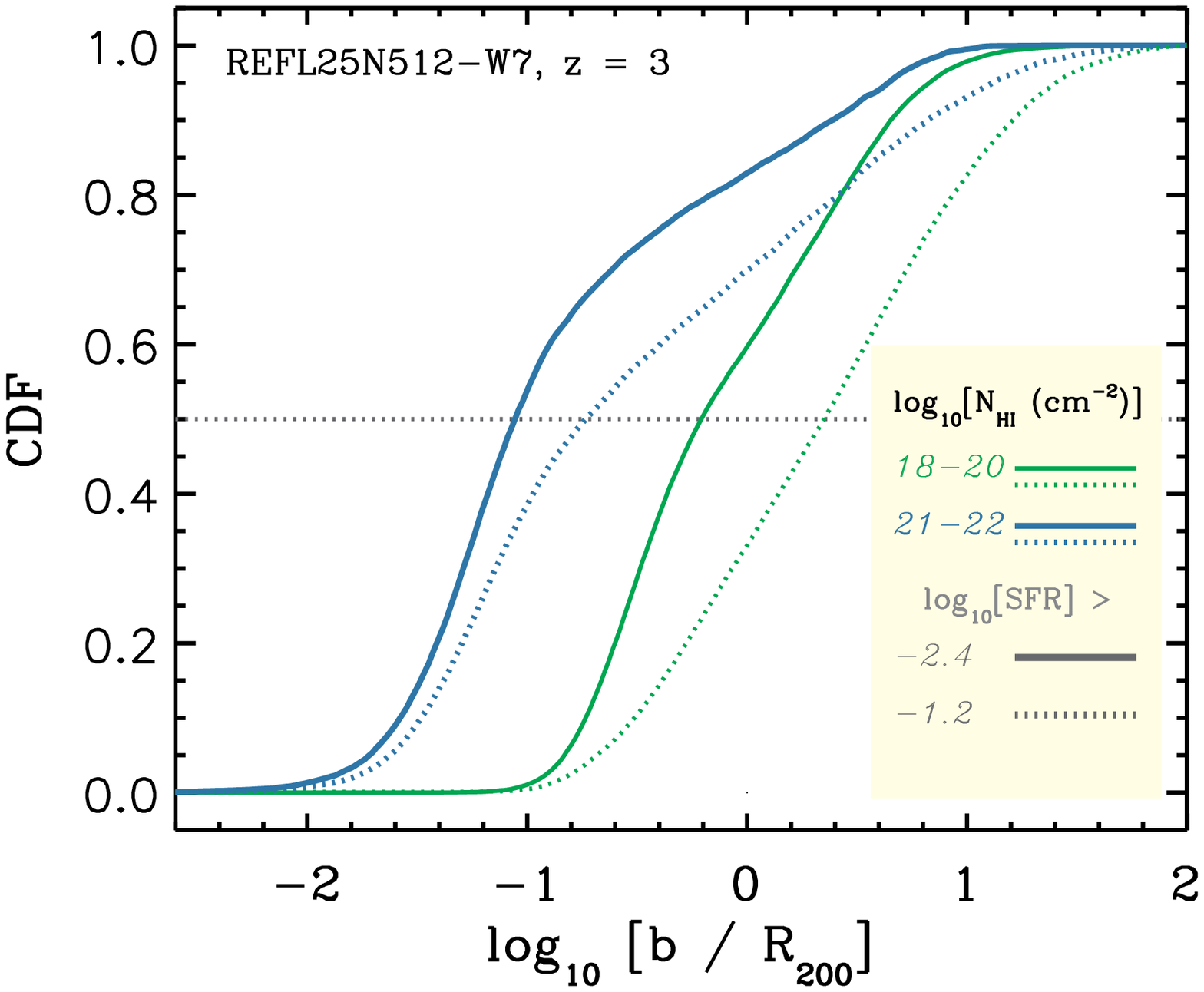}} 
             \hbox{\includegraphics[height=0.39\textwidth]
             {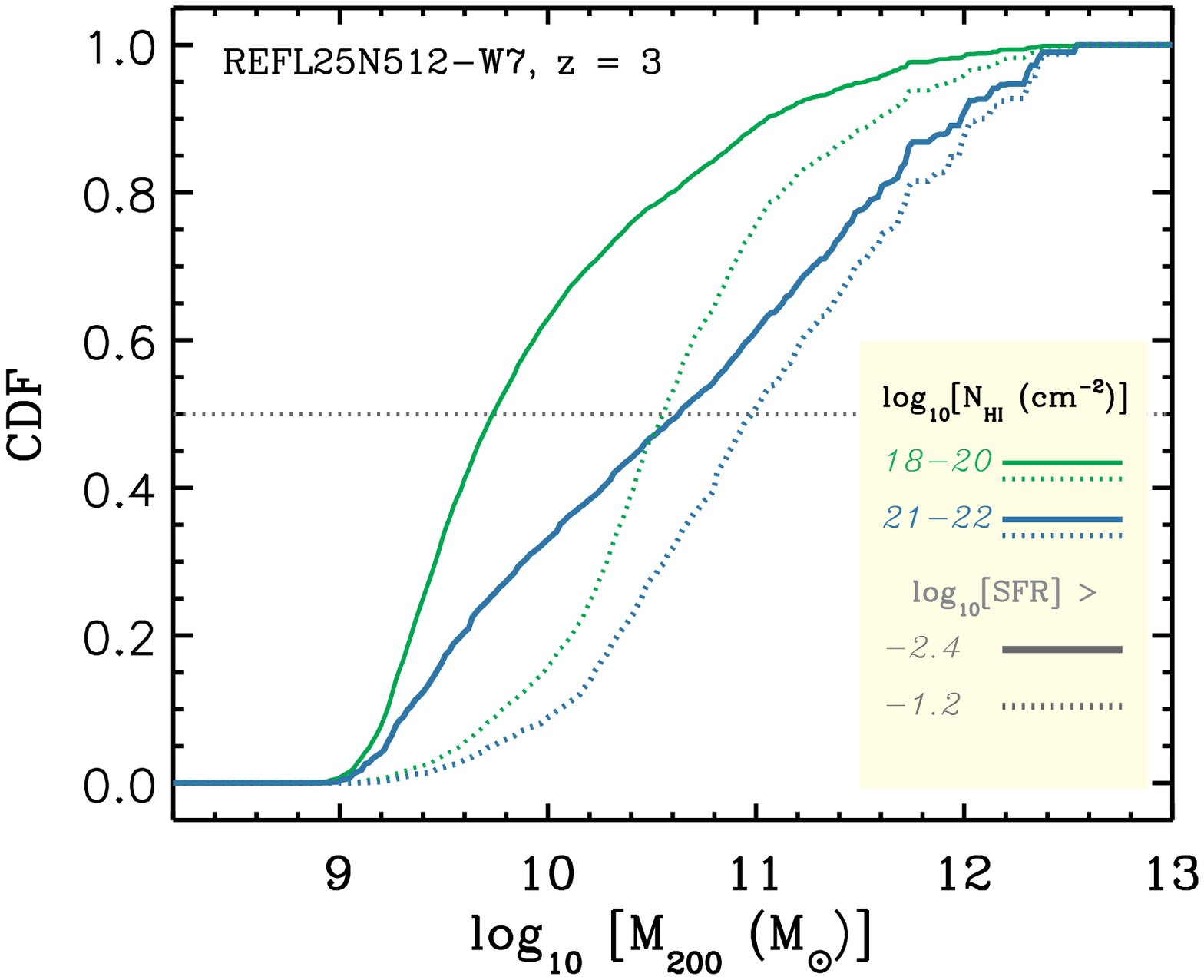}}}
\centerline{\hbox{\includegraphics[height=0.39\textwidth]
             {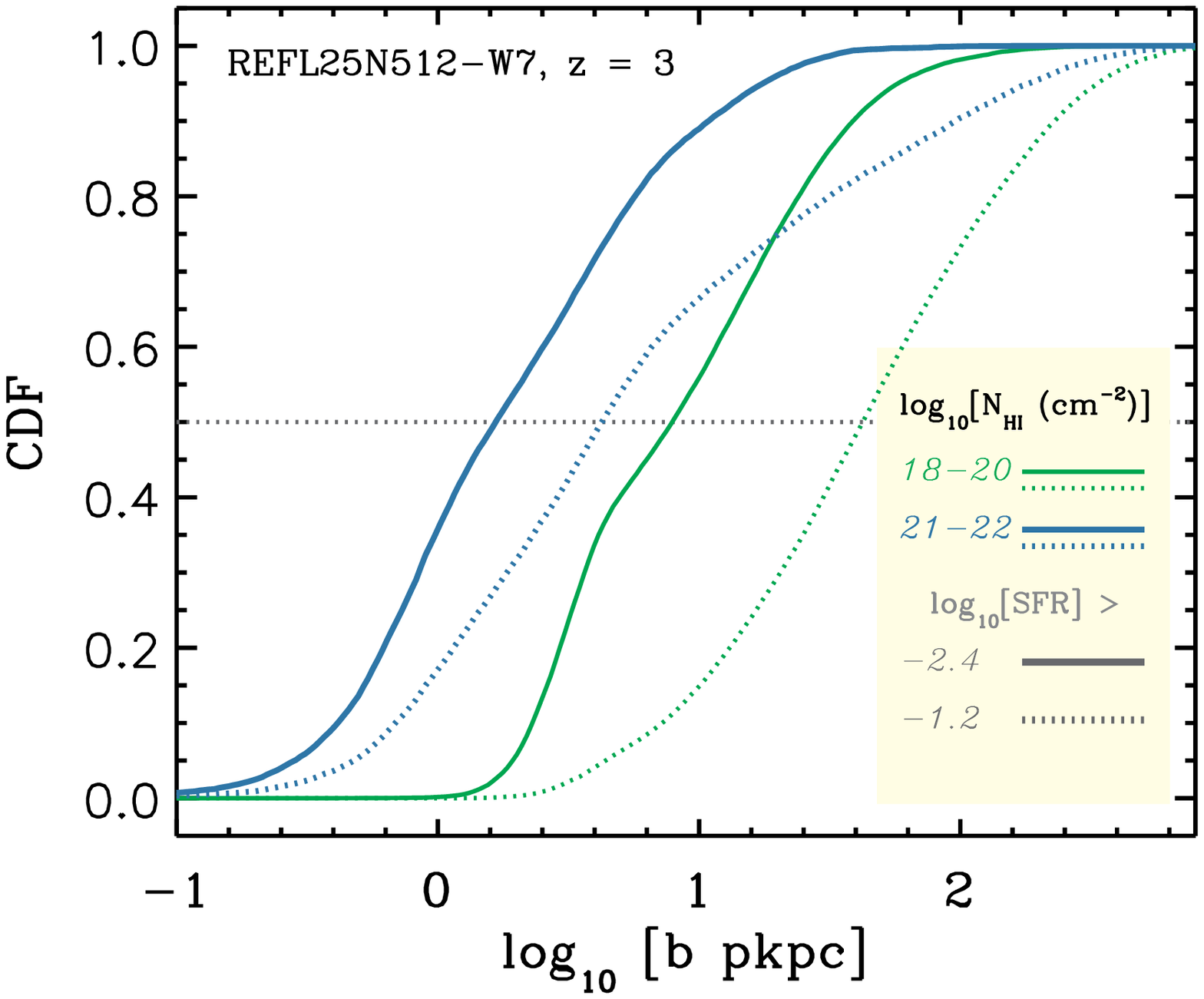}} 
             \hbox{\includegraphics[height=0.39\textwidth]
             {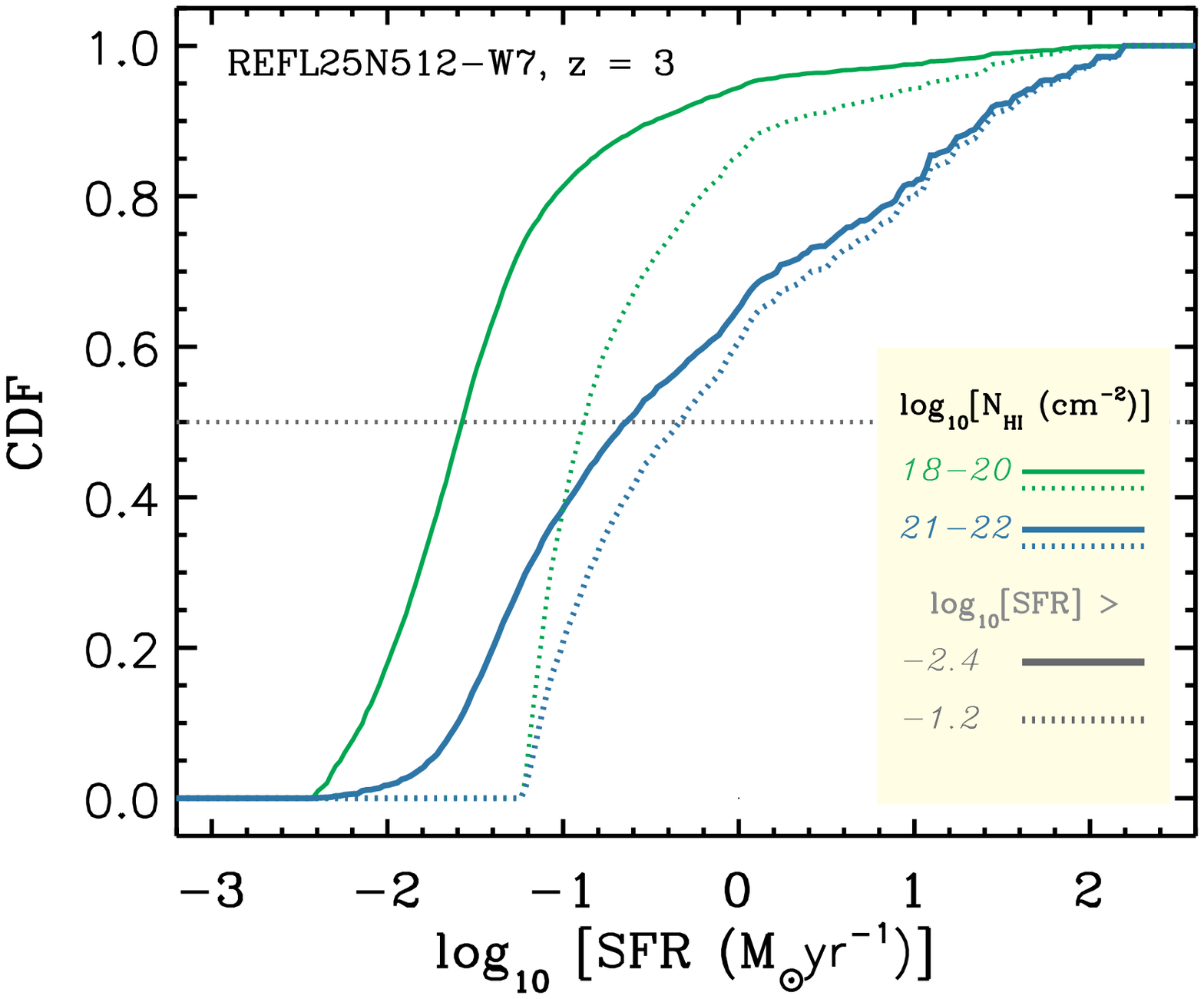}}}
\centerline{\hbox{\includegraphics[height=0.39\textwidth]
             {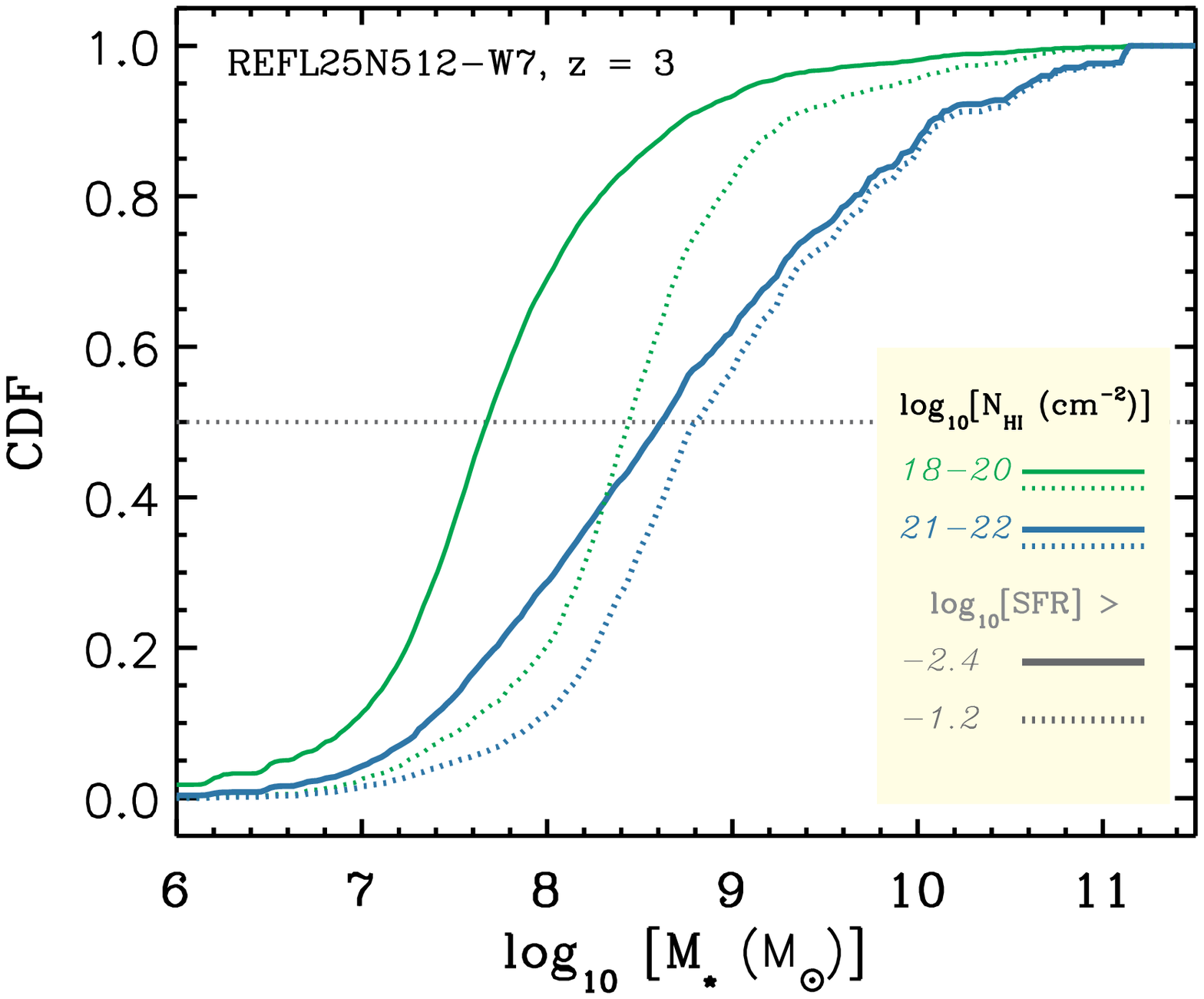}} 
             \hbox{\includegraphics[height=0.39\textwidth]
             {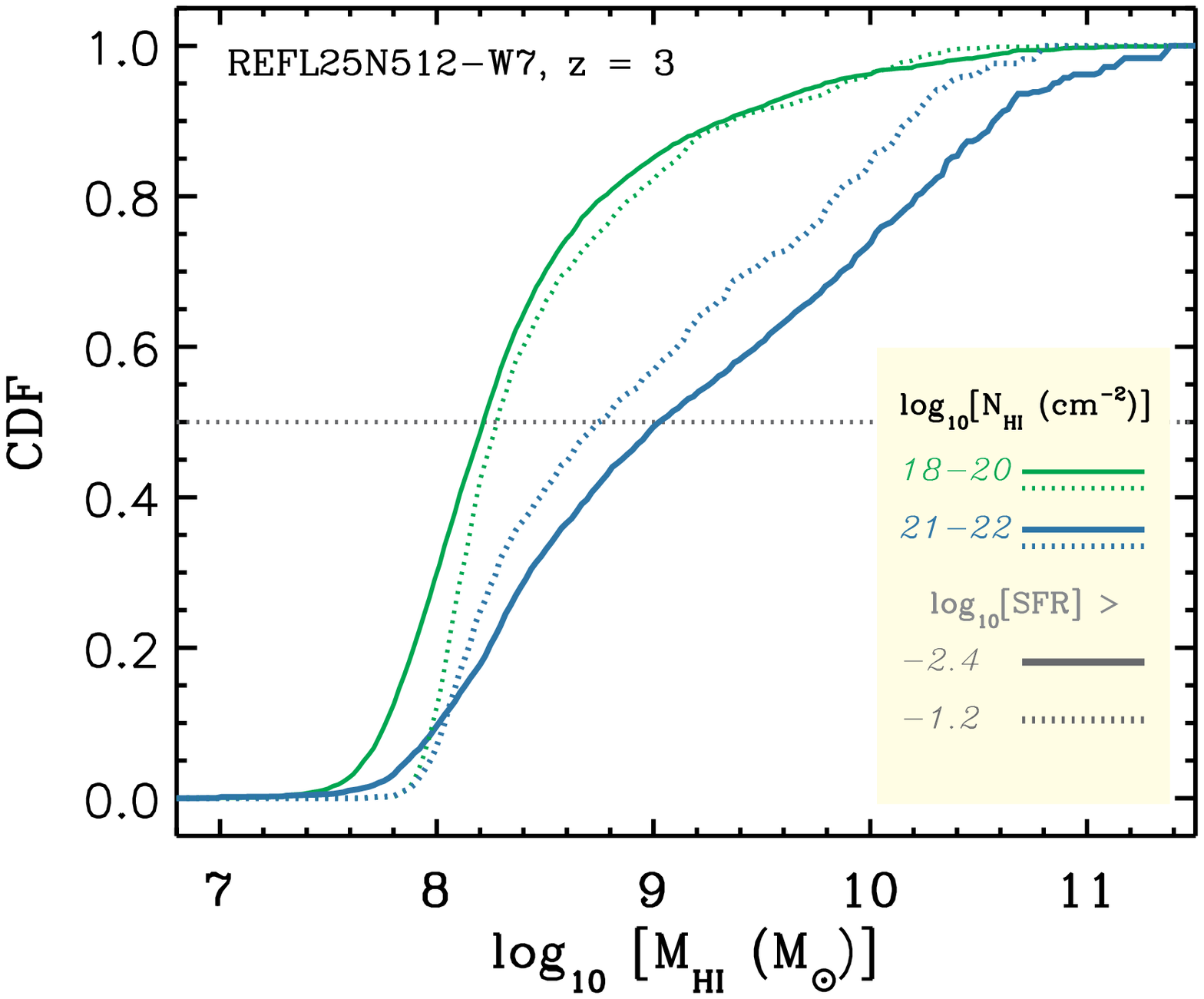}}}
\caption{The effect of the SFR threshold on the cumulative distributions that were shown in Figure \ref{figDLA:cumdist}. In each panel the blue and green solid curves correspond to LLSs (i.e.,  $10^{18} < \NHI < 10^{20}\cmsq$) and strong DLAs (i.e., $10^{21} < \NHI < 10^{22}\cmsq$), respectively. The solid curves, which are identical to those shown in Figure \ref{figDLA:cumdist}, correspond to a SFR threshold of $4 \times 10^{-3}~\Msun~\rm{yr^{-1}}$ and the dotted curves show the results obtained by imposing a SFR threshold of $6 \times 10^{-2}~\Msun~\rm{yr^{-1}}$. From the top-left to bottom-right, the panels show the cumulative distribution of normalized impact parameters, halo masses, impact parameters, SFRs, stellar masses and $\HI$ masses, respectively. The detection of galaxies with SFRs as low as $\rm{SFR} \sim 10^{-2}\Msun~\rm{yr^{-1}}$ may become possible with future instruments such as MUSE.}
\label{figDLA:cumdist-res}
\end{figure*}

\subsection{Correlations between absorbers and various properties of their associated galaxies}
\label{secDLA:phys-prop}
The gas content of galaxies is correlated with their other properties like stellar mass, size and SFR. This implies the existence of correlations between the abundance and distribution of strong $\HI$ absorbers and the properties of the galaxies that are associated with them. For instance, as mentioned earlier, at a given impact parameter, the typical stellar mass of host galaxies increases with increasing $\HI$ column density. Similarly, at a fixed $\NHI$, the typical host stellar mass increases with increasing impact parameter (see Figure \ref{figDLA:b-NHIall-2D}).

These trends are clearly visible in Figure \ref{figDLA:prof-Ms-sfrI}, which show the $b-\NHI$ relation for subsets of $\HI$ absorbers that are linked to galaxies with different star formation rates and stellar masses (shown in the left and right panels respectively). The colored curves in the left (right) panel, which show different bins of SFR (stellar mass), indicate that the impact parameters of absorbers associated with galaxies with a higher SFR or stellar mass are typically larger than the impact parameters of absorbers associated with galaxies with a lower SFR or stellar mass, in agreement with the color gradient in Figure \ref{figDLA:b-NHIall-2D}. Note that the left panel of Figure \ref{figDLA:prof-Ms-sfrI} is not directly comparable to the left panel of Figure \ref{figDLA:b-NHI}. For Figure \ref{figDLA:b-NHI} we matched absorbers to galaxies with SFR greater than three different values. For Figure \ref{figDLA:prof-Ms-sfrI}, on the other hand, all galaxies with $\rm{SFR} > 0.004~\Msun~\rm{yr^{-1}}$ were eligible to be associated with absorbers, but we only show results for absorbers that are associated with galaxies with certain properties. For reference, the dashed curve in Figure \ref{figDLA:prof-Ms-sfrI} shows the result for all absorbers and is thus identical to the blue solid curve in the left panel of Figure \ref{figDLA:b-NHI}. Note also that there is considerable scatter around the median impact parameter of each subset of absorbers, as illustrated by the shaded area around the green curves. The data points in Figure \ref{figDLA:prof-Ms-sfrI} show the confirmed, observed DLA-galaxy pairs listed in Table \ref{tbl:obs}. Pairs with reliable SFR or mass estimates are shown with the green circles and blue cubes, respectively. The simulation results agree well with the available observations. 

Figure \ref{figDLA:prof-Ms-sfrII} show the fraction of $\HI$ absorbers as a function of their column density, that are associated with galaxies with a particular SFR or mass. The fraction of $\HI$ absorbers associated with massive galaxies (which also have high SFRs), decreases rapidly with decreasing $\HI$ column density. Most absorbers with $\NHI < 10^{21}\cmsq$ are linked to galaxies with $\rm{SFR} < 6 \times 10^{-2} ~\rm{\Msun~yr^{-1}}$, or $\rm{M_{\star}} < 3\times10^8~\Msun$. While only about $20-30\%$ of those systems are associated with more massive galaxies with higher SFRs (i.e., $\rm{M_{\star}} > 3\times10^8~\Msun$ and $\rm{SFR} > 6 \times 10^{-2} ~\rm{\Msun~yr^{-1}}$), such galaxies are associated with a large fraction of strong DLAs ($\NHI > 10^{21}\cmsq$). We reiterate that these results are not expected to change if we increase the resolution except that the lowest SFRs are expected to be reduced at higher resolution (shown with the red regions in the left panel of Figure \ref{figDLA:prof-Ms-sfrII}).

Our results are in agreement with \citet{Tescari09} and \citet{Voort12a}, who found that most $\HI$ absorbers with $\NHI < 10^{21}\cmsq$ are associated with very low-mass halos with $\rm{M_{200}} < 10^{10}~\Msun$ (i.e., $\rm{M_{\star}} \lesssim 10^8~\Msun$; see the orange regions in the right panel of Figure \ref{figDLA:prof-Ms-sfrII}). Moreover, \citet{Voort12a} found that at higher $\HI$ column densities the contribution of halos with $\rm{M_{200}} > 10^{11}~\Msun$ (i.e., $\rm{M_{\star}} \gtrsim 10^9~\Msun$; see the green regions in the right panel of Figure \ref{figDLA:prof-Ms-sfrII}) increases rapidly.

We also note that \citet{Moller13} extrapolated the observed mass-metallicity relation of galaxies to the very low metallicities that are typical of observed DLAs at $z \approx 2.5$ and concluded that those metallicities are consistent with what is expected from very low-mass galaxies (i.e., $\rm{M_{\star}} \approx 10^8~\Msun$). While this conclusion is consistent with our findings, it is not clear whether one can directly compare the metallicity of the mostly neutral gas seen as DLAs with that of the $\HII$ regions on which the observed mass-metallicity relation is based, given that the $\HII$ regions are expected to be much closer to actively star-forming regions and therefore preferentially more enriched. 

To investigate further the distribution of galaxy properties for absorbers with different $\HI$ column densities, we split absorbers in different $\NHI$ bins and show the cumulative distribution of different properties of the associated galaxies in Figure \ref{figDLA:cumdist}. The top-left and top-right panels of this figure show the cumulative distribution of normalized impact parameters and halo masses, where in contrast to the other four panels, only central galaxies are taken into account (because the $\sim 30\%$ of our galaxies that are satellites do not have well-defined viral radii or halos of their own). The remaining four panels, from middle-left to the bottom-right, show the cumulative distribution of impact parameters, star formation rates, stellar masses and $\HI$ masses respectively.

Comparing the cumulative distribution of (normalized and absolute) impact parameters with the other four panels in this figure indicates that for $\NHI < 10^{21}\cmsq$ the $\HI$ column density of absorbers is much more sensitive to their projected distance from their associated galaxies (i.e., the impact parameter) than to the properties of the associated galaxies such as the stellar mass, SFR, $\HI$ mass or halo mass. As shown in the top-left panel, while more than $50\%$ of strong DLAs are within $R \lesssim 0.1~R_{200}$, most weak LLSs ($\NHI \lesssim 10^{18}\cmsq$) reside beyond the virial radius of their host galaxies.

However, as the difference between strong DLAs with $\NHI > 10^{21} \cmsq$ (blue solid curves) and lower $\HI$ column densities shows, galaxies associated with strong DLAs have distinct distributions in SFR, $\HI$ mass, stellar mass and halo mass. For instance, the median SFR of galaxies that are associated with strong DLAs is $\approx 10$ times higher than that of galaxies associated with weak DLAs or LLSs. Similar trends also hold for $\HI$ masses, stellar and halo masses. Only about $10\%$ of strong DLAs are linked to the galaxies that are typically associated with other $\HI$ absorbers. Strong DLAs, i.e., $\NHI > 10^{21} \cmsq$, are thus preferentially linked to relatively massive galaxies ($\rm{M_{\star}} \gtrsim 10^9~\Msun$) while other $\HI$ absorbers are distributed among more abundant galaxies with extremely low masses.

The middle-right panel of Figure \ref{figDLA:cumdist} is particularly useful to understand the difficulty of detecting the galaxy counterparts of LLSs and DLAs. The distribution of SFRs shown in this plot can be used to predict the probability of detecting the galaxy counterpart as a function of the detection threshold. For instance, if the detection threshold is equivalent to a SFR of $ 1 ~\rm{\Msun~yr^{-1}}$, the chance of detecting the galaxy counterpart of a strong DLA with $\NHI \approx 10^{22} \cmsq$ is 3 in 10. For weak DLAs (i.e.,  $10^{20} < \NHI < 10^{21} \cmsq$) the non-detection rate would be even higher at $\approx 90\%$. This explains the large number of non-detections in observational studies \citep[e.g.,][]{Foltz86,Smith89,Lowenthal95,Bunker99,Prochaska02,Kulkarni06,Rahmani10,Bouche12}.

As mentioned before, with the detection thresholds typical of current observations, only galaxies that have $\rm{SFR} > 1-10 ~\rm{\Msun~yr^{-1}}$ can be identified at $z \sim 3$. Because galaxies are resolved to much lower SFRs in our simulation, it is not straightforward to compare current observations to the results shown in Figure \ref{figDLA:cumdist}. With the advent of future instruments like MUSE \citep{Bacon10}, the accessible Ly$\alpha$ detection thresholds can be pushed to lower SFRs which allows the identification of galaxies (Ly$\alpha$ emitters) with SFRs as low as $10^{-2}-10^{-1} ~\rm{\Msun~yr^{-1}}$. These deep observations can be used to identify faint galaxies associated with strong $\HI$ absorbers and analyze the cumulative distribution of their properties. However, the results would still depend on the accessible detection threshold and might be different from what is shown in Figure \ref{figDLA:cumdist}. To address this issue, we compare the cumulative distributions for two different detection thresholds in Figure \ref{figDLA:cumdist-res}. The panels in this figure are identical to Figure \ref{figDLA:cumdist}, but show the result for only two $\HI$ column density bins. The green and blue curves, respectively, represent the $\HI$ absorbers (or the galaxies associated with them) that have $10^{18} < \NHI < 10^{20}\cmsq$ (i.e., LLSs) and $10^{21} < \NHI < 10^{22}\cmsq$ (i.e., strong DLAs). The solid curves show our fiducial detection threshold of $\rm{SFR} > 4 \times 10^{-3}~\rm{\Msun~yr^{-1}}$, which was also used in Figure \ref{figDLA:cumdist}. The dotted curves indicate a higher detection threshold of $\rm{SFR} > 6 \times 10^{-2}~\rm{\Msun~yr^{-1}}$, which is comparable to what will be accessible using deep MUSE observations. The substantial differences between the solid and dotted curves imply that the distributions will still be sensitive to the detection threshold. In other words, the bias introduced by the finite detection threshold should be taken into account when interpreting/modeling the observed distributions.  
\begin{figure*}
\centerline{\hbox{\includegraphics[width=0.53\textwidth]
             {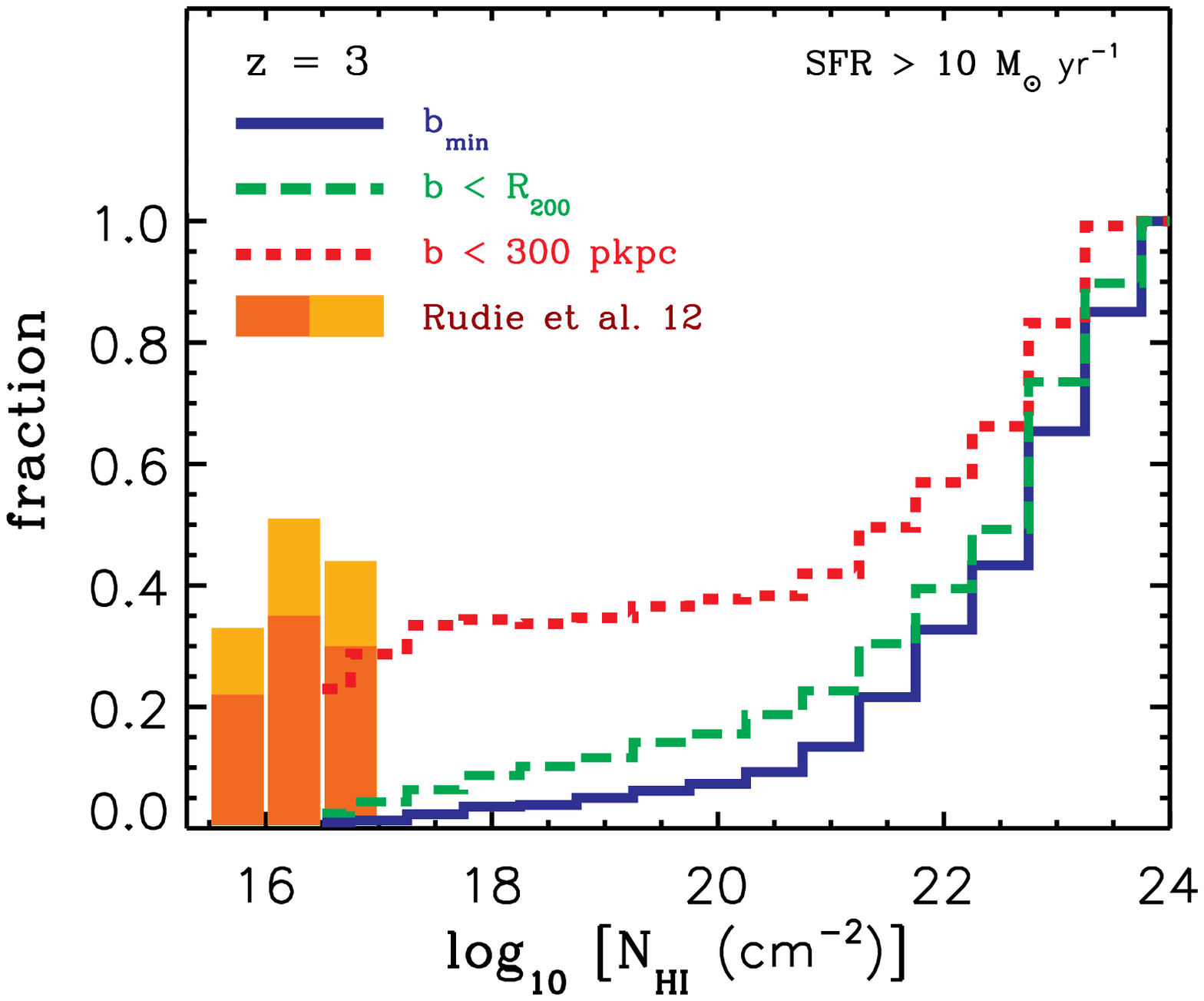}}
             \hbox{\includegraphics[width=0.53\textwidth]
             {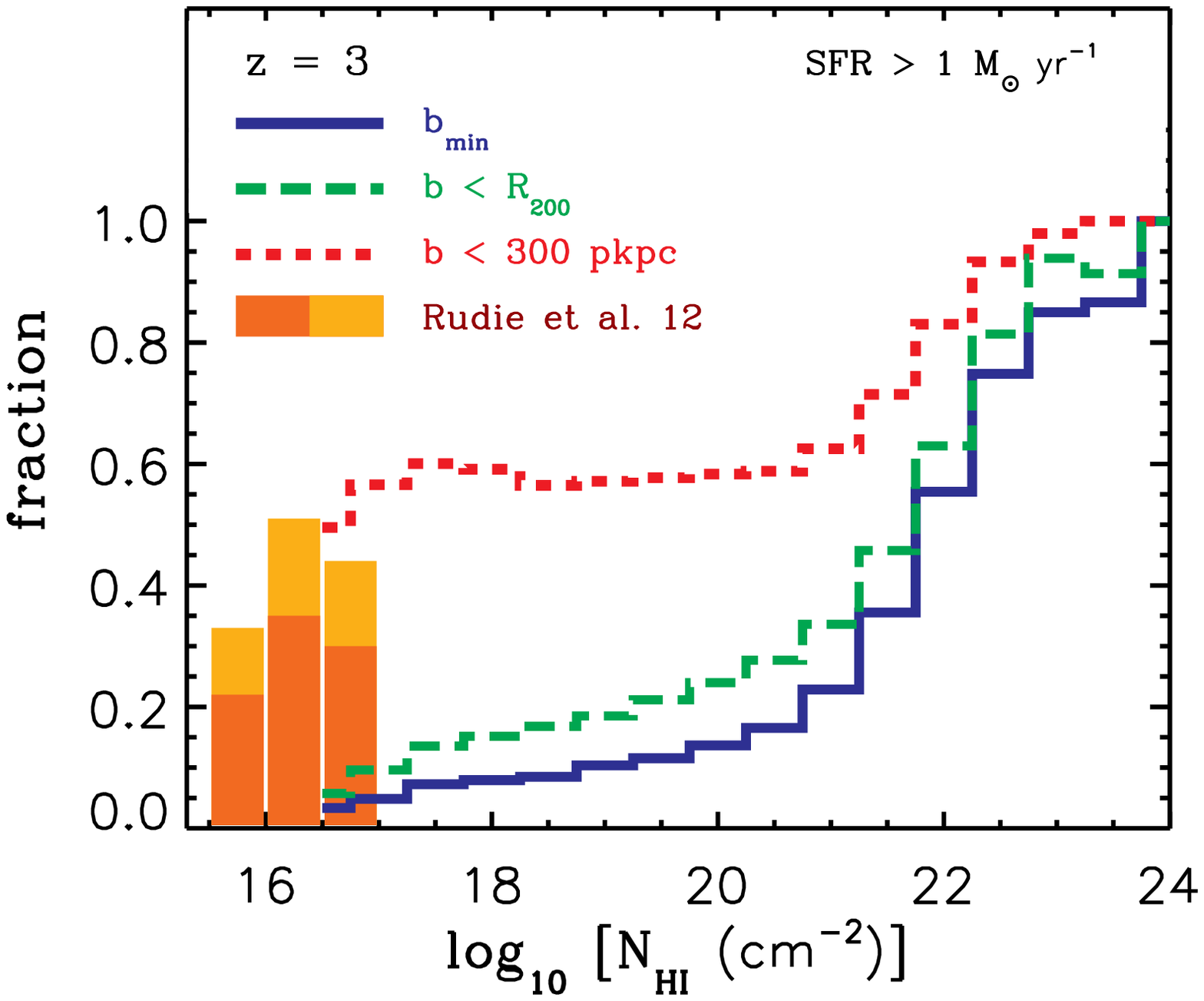}}}

\caption{Fraction of strong $\HI$ systems that are associated to galaxies with  $\rm{SFR} > 10 ~\rm{\Msun~yr^{-1}}$ (left) and  $\rm{SFR} > 1 ~\rm{\Msun~yr^{-1}}$ (right) in our simulation at $z = 3$. Curves with different line styles and colors show results that are obtained using different methods for associating $\HI$ absorbers to galaxies that are above the imposed SFR threshold: blue solid curves are obtained by associating $\HI$ absorbers to their closest galaxies, where all the simulated galaxies are taken into account (see Figures \ref{figDLA:prof-Ms-sfrI} and \ref{figDLA:prof-Ms-sfrII}); green long-dashed curves show the fraction of absorbers that reside within the virial radius of the selected galaxies, and finally, red short-dashed curves show the fraction of absorbers that are within $300$ proper kpc from the selected galaxies. For comparison, the colored bars show \citet{Rudie12} findings for the fraction of absorbers within $300$ proper kpc from galaxies with similar SFRs are shown here. The dark orange parts of the bars indicate the observed fractions and the light orange parts show the correction for missing galaxies in their spectroscopic sample. The predicted fractions are in excellent agreement with observations and show that a large fraction of strong $\HI$ absorbers are less than $300$ proper kpc away from galaxies with $\rm{SFR} > 1 ~\rm{\Msun~yr^{-1}}$. However, most of those systems are far beyond the virial radius of those galaxies and are closely associated with less massive objects with lower SFRs that fall below the detection limits of current observations.}
\label{figDLA:sel-thresh}
\end{figure*}

\subsection{Are most strong $\HI$ absorbers at $z \sim 3$ associated with Lyman-Break galaxies?}
\label{secDLA:LBG}
In this work, we adopted the absorber-centered point of view in which each absorber is associated with the galaxy with the smallest projected separation and a LOS velocity difference smaller than 300 km/s. As we showed in the previous sections, our simulation predicts that most strong $\HI$ absorbers are associated with very low-mass galaxies (i.e., $\rm{M_{\star}} < 10^{8}~\Msun$) with low SFRs ($ < 10^{-1} ~\rm{\Msun~yr^{-1}}$). We showed that this prediction is consistent with the impact parameters of observed DLA-galaxy counterparts and, more importantly, the high incident rate of finding no detectable galaxy close to DLAs. 

An alternative approach is to search for $\HI$ absorbers around galaxies (i.e., the galaxy-centered point of view) \citep[e.g.,][]{Steidel95,Adelberger03,Adelberger05,HP07,Steidel10,Rakic12,Rudie12,Prochaska13}. While the two approaches are complementary, their results may appear contradictory. Using the galaxy-centered approach, \citet{Rudie12} found that at $z\approx 2-3$ most Lyman Limit absorbers (i.e., $\NHI \approx 10^{17}\cmsq$) have an impact parameter $b < 300$ proper kpc, and a LOS velocity difference $ < 300$ km/s with respect to rest-frame UV-selected star-forming galaxies (see their Figure 30). Given that the typical galaxy mass in their sample is $\rm{M_{\star}} \approx 10^{10}~\Msun$, one might conclude that their result is in conflict with our finding that most strong $\HI$ absorbers (i.e., LLSs and DLAs) are closely associated with galaxies with $\rm{M_{\star}} < 10^{8}~\Msun$.

To understand the source of this apparent discrepancy, we first note that the transverse distance of 300 proper kpc and the 300 km/s LOS velocity difference that were adopted by \citet{Rudie12}, are, respectively, 300 and 50 per cent larger than the virial radius and the circular velocity of the halos expected to host their galaxies \citep[e.g.,][]{Trainor12,Rakic12,Rakic13}. In other words, the region \citet{Rudie12} define as the ``circumgalactic medium'' lies well beyond the virial radius of the halos that are thought to host their galaxies. 

In addition, it is important to note that galaxies are strongly clustered and that low-mass and high-mass galaxies broadly trace the same underlying large-scale structures. This implies that many strong $\HI$ absorbers are likely to be found close to massive galaxies, even if they are physically much more closely associated with low-mass galaxies. As a result, searching for $\HI$ absorbers within a reasonably large radius around massive, and hence easily observable galaxies, one recovers a large fraction of the existing strong $\HI$ absorbers. This effect is illustrated in the bottom-right panel of Figure \ref{figDLA:stamp}, which shows the $\HI$ column density distribution around a galaxy with $\rm{M_{\star}} = 10^{10}~\Msun$. While the maximum projected distance between the massive galaxy and all the $\HI$ absorbers that are shown is less than 300 proper kpc, nearly all of those absorbers have low-mass galaxies very close to them.

The aforementioned arguments are illustrated more quantitatively in Figure \ref{figDLA:sel-thresh}, which shows the fraction of strong $\HI$ absorbers in our simulation at $z = 3$ that are in the vicinity of galaxies with SFRs comparable to those of LBGs at similar redshifts\footnote{The SFRs of the galaxies used in \citet{Rudie12} vary from several to a few hundreds $\rm{\Msun~yr^{-1}}$.}. The red dashed curve in the left (right) panels shows the fraction of $\HI$ absorbers with impact parameter $< 300$ proper kpc from galaxies with $\rm{SFR} >  10 ~\rm{\Msun~yr^{-1}}$ ($\rm{SFR} >  1 ~\rm{\Msun~yr^{-1}}$), as a function of $\NHI$. These results are in excellent agreement with the fractions measured by \citet{Rudie12} (shown with the colored bars). The fraction of absorbers that are within 300 proper kpc from these galaxies is $\approx 0.4-0.6$ for $10^{16}<\NHI < 10^{21} \cmsq$. However, only a very small fraction of LLSs are within the virial radii (shown with the green long-dashed curves). As the blue solid curves show, even smaller fractions of LLSs remain associated to such massive galaxies if we account for galaxies with lower SFRs, which are typically most closely associated with LLSs but are too faint to be observed. 

\citet{Font-Ribera12} have recently estimated the host halo mass of $z\sim 2-3$ DLAs to be $\sim 10^{12}~\rm{\Msun}$ by cross-correlating DLAs with the $\HI$ Ly$\alpha$ forest on scales of tens of Mpc and by estimating the Ly$\alpha$ forest bias factors from independent Ly$\alpha$ forest correlation measurements. As was the case for \citet{Rudie12}, at first sight these observations appear to contradict our results, as well as the large number of non-detections in searches for the counterparts of DLAs. We intend to investigate whether we can reproduce the measurements of \citet{Font-Ribera12} by mimicking their analysis and testing their assumptions. However, as their method relies on measurements of clustering on scales that are of the order or larger than our simulation box, this test will unfortunately only become possible when we simulate much larger volumes with a resolution similar to that employed here.

\section{Summary and conclusions}
\label{secDLA:conclusions}
We have used cosmological simulations that have been post-processed using accurate radiative transfer corrections that account for photoionization by the UVB and recombination radiation as well as collisional ionization, to investigate the relation between strong $\HI$ absorbers (i.e., LLSs and DLAs) and galaxies at $z = 3$. The simulation we used for our study has been shown to closely reproduce the observed $\HI$ column density distribution function \citep{Altay11,Rahmati13a}. After identifying sight-lines with high $\HI$ column densities (i.e., $\NHI > 3 \times 10^{16}~\cmsq$) and calculating the line-of-sight velocities of absorbers, we used a procedure similar to that used in observational studies to associate absorbers with nearby galaxies: we associated each strong $\HI$ absorber to the galaxy which has the shortest transverse distance to the absorber and a line-of-sight velocity difference within $\pm 300~\rm{km~s^{-1}}$. Having associated all strong $\HI$ absorbers in the simulation to galaxies, we investigated statistical trends between the strength of the absorbers and the distance to, and properties of, their host galaxies. 

Among the various dependencies we studied in this work, we found that the anti-correlation between the $\HI$ column density of absorbers and the transverse distance that connects them to their host galaxies (i.e., the impact parameter) to be the strongest. While LLSs have impact parameters $\gtrsim10$ proper kpc, DLAs are typically within a few proper kpc from the nearest galaxies. Relative to the virial radius of the halo that hosts the nearest central galaxy, LLSs have typical impact parameters $\gtrsim {R_{200}}$, while DLAs are typically $\sim 10$ times closer to the center of the nearest halo (i.e. $\lesssim 0.1~{R_{200}}$). The predicted strong anti-correlation between the impact parameter of strong $\HI$ absorbers and their $\HI$ column densities agrees with observations and previous work. We also found a relatively large scatter around the median impact parameter of absorbers, at a given $\NHI$, due to the complex geometry of gas distribution around galaxies, and also the variation in the size and gas content of the galaxies that host the absorbers. 

We predict that most strong $\HI$ absorbers are closely associated with very low-mass galaxies, $\rm{M_{\star}} \lesssim 10^{8}~\Msun$, but that the fraction of strong $\HI$ absorbers that are linked to more massive galaxies increases with the $\HI$ column density. This correlation between column density and galaxy mass is particularly pronounced for strong DLAs, i.e., $\NHI > 10^{21} \cmsq$, the majority of which are associated with galaxies with $\rm{M_{\star}} \gtrsim 10^{9}~\Msun$. We analyzed different properties of galaxies that are linked to strong $\HI$ absorbers with different $\HI$ column densities and found similar trends as we found for stellar mass: most LLSs and DLAs are closely associated with galaxies that have low halo masses ($\rm{M_{200}} \lesssim 10^{10}~\Msun$), low SFRs ($ \lesssim 10^{-2} ~\rm{\Msun~yr^{-1}}$) and low $\HI$ masses ($\rm{M_{\HI}} \lesssim 10^{8}~\Msun$), but strong DLAs (i.e., $\NHI > 10^{21} \cmsq$) are typically linked to more massive galaxies with significantly higher halo masses ($\rm{M_{200}} \gtrsim 10^{10}~\Msun$), SFRs ($ \gtrsim 10^{-1} ~\rm{\Msun~yr^{-1}}$) and $\HI$ masses ($\rm{M_{\HI}} \gtrsim 10^{9}~\Msun$). 

By analyzing subsets of strong $\HI$ absorbers for which the associated galaxies have specific properties, we found that observationally confirmed DLA-galaxy pairs that have measured masses or SFRs, have impact parameters that are in good agreement with our predictions. We stress, however, that the majority of DLAs are predicted to be more closely associated with galaxies that are at smaller impact parameters, but are too faint to be detected with current surveys. Hence, the masses and impact parameters of the observed galaxy counterparts of DLAs are both biased high. This is consistent with the large number of non-detections in observational campaigns that searched for galaxies close to DLAs \citep[e.g.,][]{Foltz86,Smith89,Lowenthal95,Bunker99,Prochaska02,Kulkarni06,Bouche12} and also the relatively large number density of extremely faint Ly$\alpha$ emitters \citep{Rauch08,Rauch11}. 

In order to facilitate the comparison between cosmological simulations and observations, we provided statistics on DLA-galaxy pairs for a few different SFR thresholds. However, a proper comparison requires observational studies aiming to find galaxies close to DLAs, to report their detection limit, the maximum allowed velocity separation, and either the impact parameter of the nearest detected galaxy or, in the case of non-detections (which must always be reported), the maximum impact parameter that has been searched. For the few studies that report such information \citep[e.g.,][]{Teplitz98,Mannucci98}, we found good agreement with our simulation.

Interestingly, some recent observational studies indicate that strong $\HI$ absorbers at $z\sim 2-3$ are associated with surprisingly massive galaxies. In particular, \citet{Rudie12} studied the distribution of $\HI$ absorbers around a sample of rest-frame UV-selected Lyman-break galaxies (LBGs) with typical masses of $\rm{M_{\star}} \sim 10^{10}~\Msun$ at $z\sim2-3$ and found that nearly half of the absorbers with $10^{16} < \NHI < 10^{17}\cmsq$ reside within a line-of-sight velocity difference of $300~\rm{km~s^{-1}}$ and a transverse separation of 300 proper kpc from a LBG, a region they labelled the circumgalactic medium. This result appears to contradict our finding that most strong $\HI$ absorbers are associated with galaxies with $\rm{M_{\star}} \lesssim 10^{8}~\Msun$. We demonstrated, however, that even though the absorbers are physically most closely associated with low-mass galaxies, these galaxies cluster sufficiently strongly around galaxies as massive as LBGs to reproduce the observations of \citet{Rudie12}. Moreover, we noted that the required clustering is not even that strong: since 300 proper kpc and $300~\rm{km~s^{-1}}$ exceed the virial radius and the circular velocity of the halos thought to host LBGs by more than 300 and 50 per cent, respectively \citep[e.g.,][]{Trainor12,Rakic12,Rakic13}, nearly all the volume of the ``circumgalactic medium'' lies beyond the virial radius if we employ the definition of \citet{Rudie12}. Our results suggest that it is more sensible to define the circumgalactic medium to be the region within the virial radius.

Future deep observational surveys using new instruments (e.g., MUSE; \citealp{Bacon10}) will be able to detect fainter galaxies near $\HI$ absorbers. However, missing faint galaxies is a generic feature for any survey that has a finite detection limit and that takes an absorber-centered point of view. The incompleteness problem can be overcome by taking a galaxy-centered point of view, but this approach is inefficient for rare absorbers such as the interesting strong $\HI$ systems we studied here. Moreover, while galaxy-centered surveys can measure the statistical distribution of absorbers (such as their covering factor), we still need to avoid interpreting the selected galaxy as the counterpart to any absorber that is detected. Absorber-centered surveys will probably remain the most efficient way to build up large numbers of galaxy-DLA pairs. Even with modest detection limits, such surveys provide highly valuable constraints on the relation between absorbers and galaxies, provided all non-detections are reported and that the detection limits and the maximum possible impact parameter are clearly specified.

\section*{Acknowledgments}
\addcontentsline{toc}{section}{Acknowledgments}
We would like to thank the anonymous referee for a timely response. We also thank A. Pawlik, X. Prochaska, M. Rai\v{c}evi\`{c} and M. Shirazi for useful discussion, reading an earlier version of this paper and providing us with comments that improved the text. The simulations presented here were run on the Cosmology Machine at the Institute for Computational Cosmology in Durham (which is part of the DiRAC Facility jointly funded by STFC, the Large Facilities Capital Fund of BIS, and Durham University) as part of the Virgo Consortium research programme. This work was sponsored with financial support from the Netherlands Organization for Scientific Research (NWO), also through a VIDI grant and an NWO open competition grant. We also benefited from funding from NOVA, from the European Research Council under the European Union's Seventh Framework Programme (FP7/2007-2013) / ERC Grant agreement 278594-GasAroundGalaxies and from the Marie Curie Training Network CosmoComp (PITN-GA-2009-238356).

\appendix
\section{Choosing the maximum allowed LOS Velocity difference}
\label{ap:max-vel-dif}
\begin{figure*}
\centerline{\hbox{\includegraphics[width=0.45\textwidth]
             {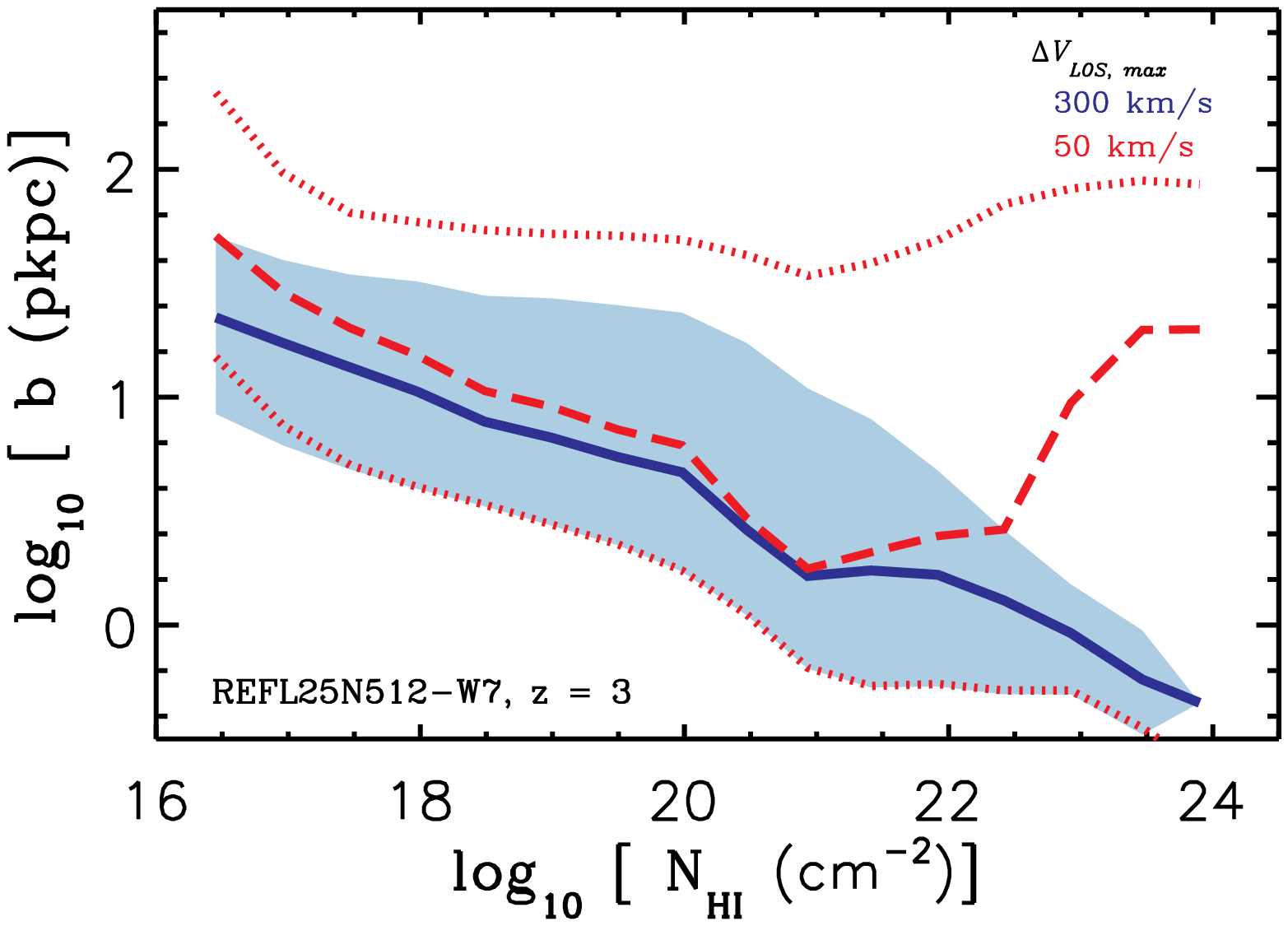}}
             \hbox{\includegraphics[width=0.45\textwidth]
             {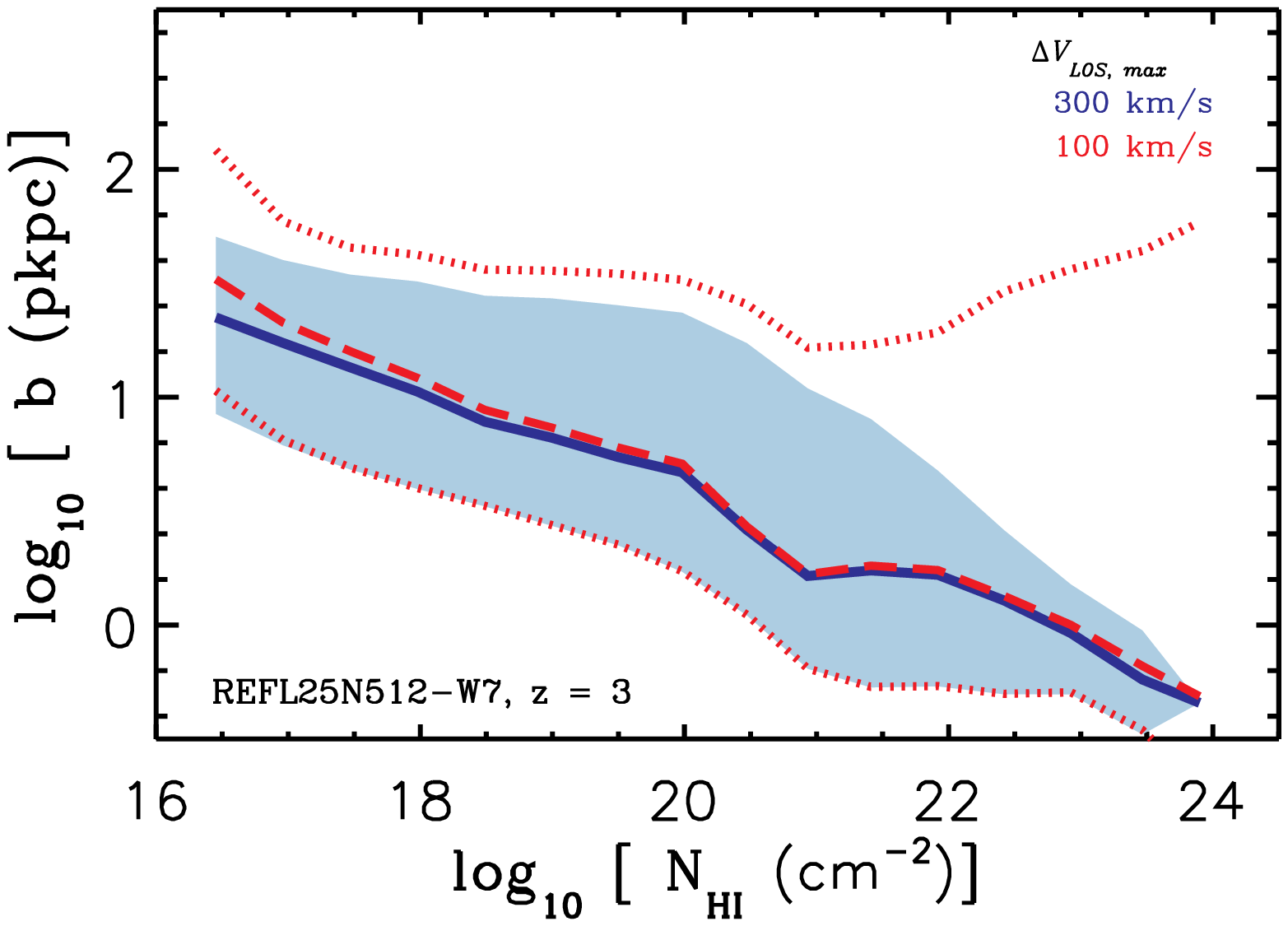}}}
\centerline{\hbox{\includegraphics[width=0.45\textwidth]
             {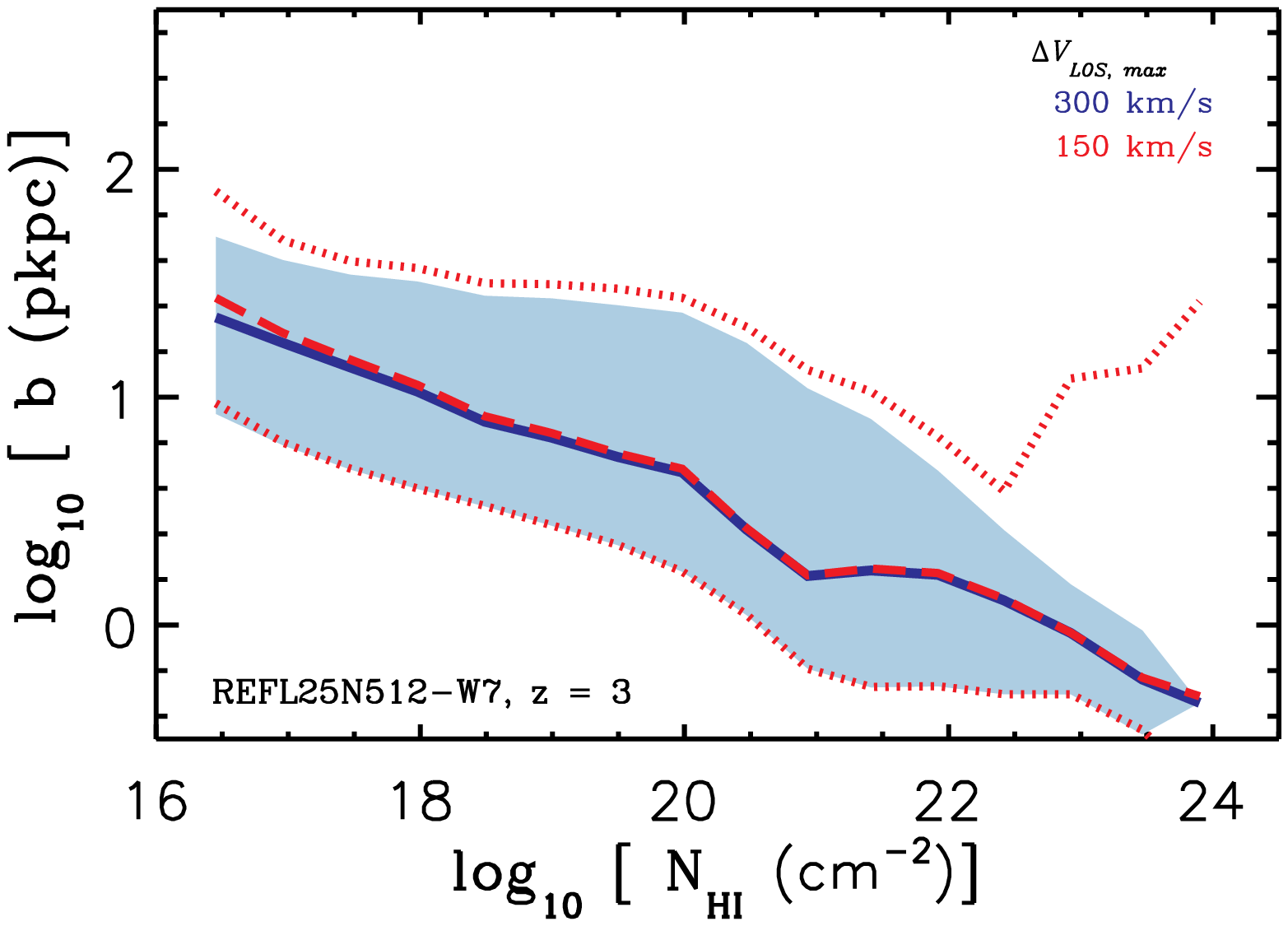}}
             \hbox{\includegraphics[width=0.45\textwidth]
             {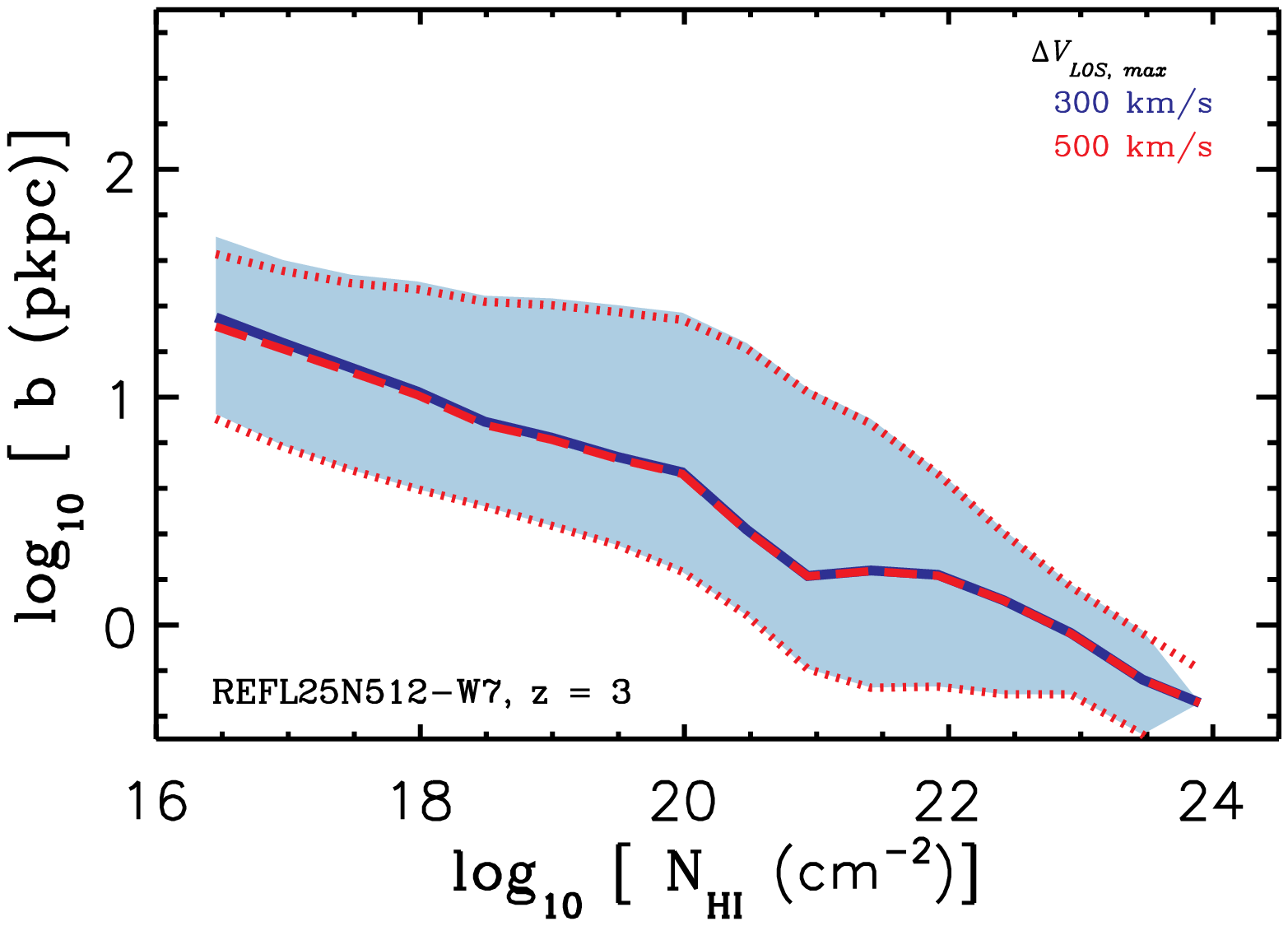}}}
\caption{The impact of changing the maximum allowed LOS velocity difference between $\HI$ absorbers and galaxies associated with them. The blue solid curve in all panels shows the median impact parameter of absorbers in our simulation at $z =3$ for our fiducial value of $\Delta V_{\rm{LOS,~max}}= 300~\rm{km/s}$ and the shaded area around the blue solid curve shows the $15\%-85\%$ percentiles. In each panel, the same result is shown for a different $\Delta V_{\rm{LOS,~max}}$ by red dashed and dotted curves which, respectively, indicate the median and the $15\%-85\%$ percentiles. The $\Delta V_{\rm{LOS,~max}}$ that is compared with our fiducial choice of $300~\rm{km/s}$ varies from $50~\rm{km/s}$ in the top-left panel to $500~\rm{km/s}$ in the bottom-right panel. The median impact parameters are converged for $\Delta V_{\rm{LOS,~max}}> 100~\rm{km/s}$, and the scatter around it is converged for $\Delta V_{\rm{LOS,~max}}> 300~\rm{km/s}$.} 
\label{figDLA:max-vel-dif}
\end{figure*}

As discussed in $\S$\ref{secDLA:matching}, when we associate $\HI$ absorbers with galaxies we take into account their relative LOS velocities. This allows us to take out systems that appear to be close in projection but are separated by large distances. In analogy to observational studies, if the difference between the LOS velocities of $\HI$ absorbers and galaxies is larger than some minimum value, then we do not associate them as counterparts even if they are very close in projection. If the allowed LOS velocity differences are too small, peculiar velocities of $\HI$ absorbers around galaxies would prevent associations of objects that are separated by small LOS distances. As shown in Figure \ref{figDLA:max-vel-dif}, the median impact parameter of absorbers as a function of $\NHI$ is converged for maximum LOS velocity differences of $\Delta V_{\rm{LOS,~max}}> 100~\rm{km/s}$ and the scatter around the median impact parameters is converged for $\Delta V_{\rm{LOS,~max}}> 300~\rm{km/s}$. Therefore, we adopt $\Delta V_{\rm{LOS,~max}}= 300~\rm{km/s}$ which is also consistent with recent observational studies \citep[e.g.,][]{Rakic12,Rudie12}.
\section{Impact of feedback}
\label{ap:feedback}
The evolution of gas and stars is determined by complex baryonic interactions that are modeled in cosmological simulations by combining various physically motivated and empirical ingredients. In this context, different feedback mechanisms can change both the distribution of gas around galaxies with a given mass and the abundance of galaxies with different masses \citep[e.g.,][]{Voort12b,Haas12}. As a result, the strength and details of various feedback mechanisms can change the distribution of $\HI$ absorbers \citep{Altay13} and may also affect the relative distribution of $\HI$ absorbers and galaxies. 

\begin{figure*}
\centerline{\hbox{\includegraphics[width=0.45\textwidth]
             {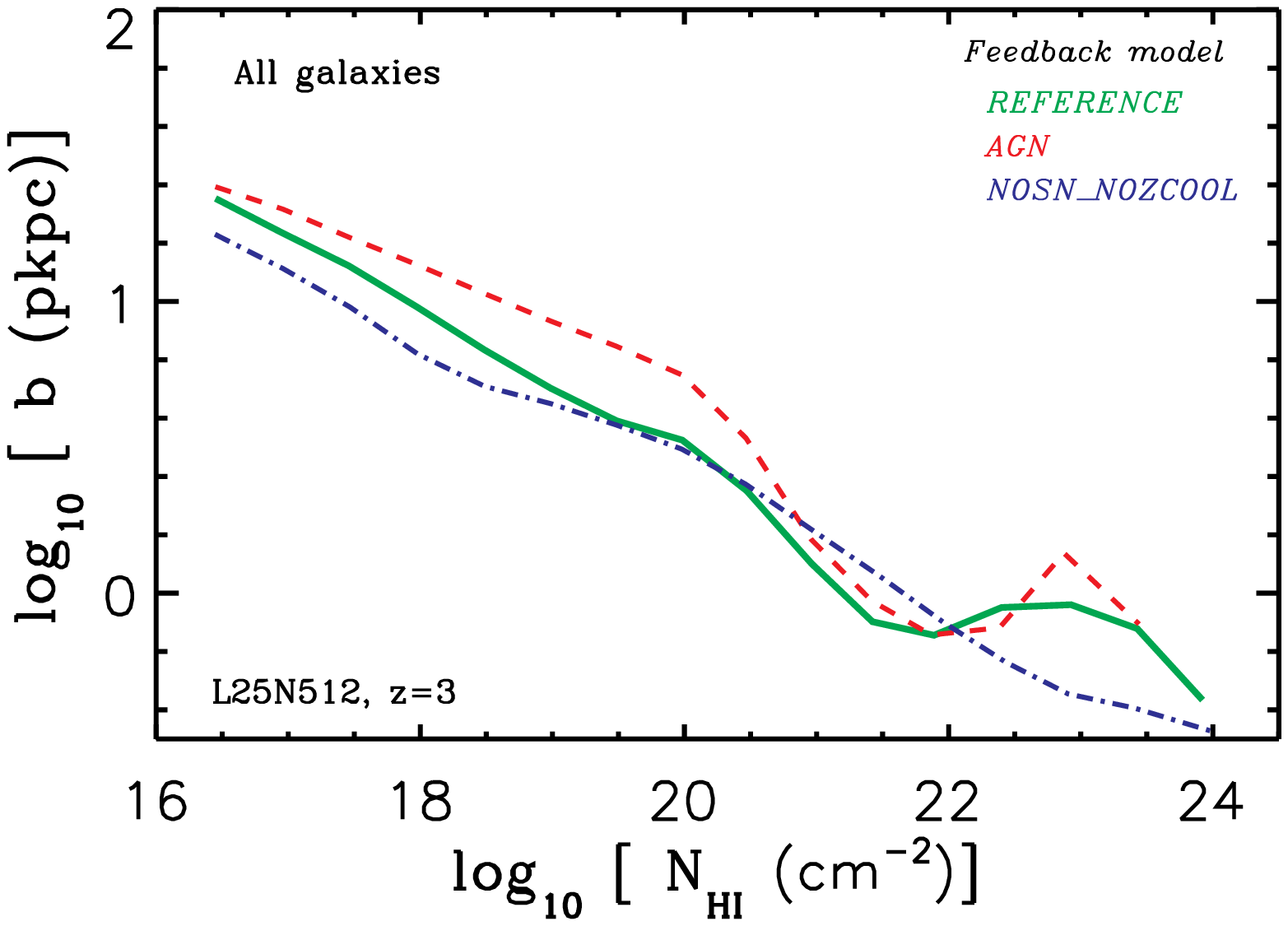}}
             \hbox{\includegraphics[width=0.45\textwidth]
             {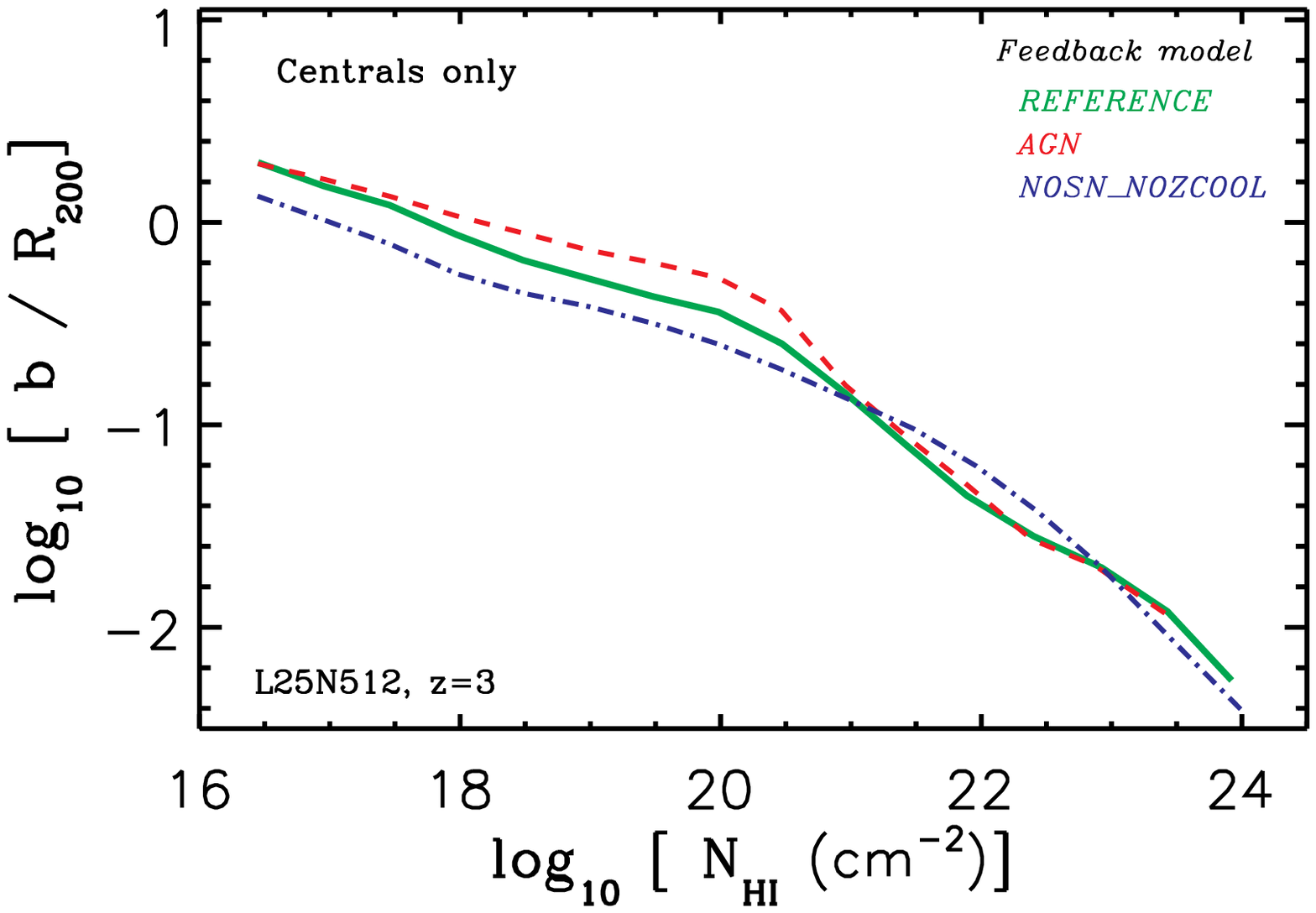}}}
\caption{The (normalized) impact parameter as a function of $\NHI$ at $z = 3$ for simulations with different feedback models is shown in the (right) left panel. The green solid, red dashed and blue dot-dashed curves respectively show the reference simulation, the impact of adding AGN feedback and the result of turning off SNe feedback and metal cooling. While the impact parameters are sensitive to the adopted feedback prescription, the differences are much smaller than the intrinsic scatter caused by the complex geometry of gas distribution around galaxies with a wide range of sizes.}
\label{figDLA:Models-prof}
\end{figure*}

To quantify the impact of feedback on our results, we compare the relation we found for our reference model between the impact parameter of absorbers and their $\NHI$ to the same relation in similar simulations taken from the OWLS suite that use different feedback prescriptions. Figure \ref{figDLA:Models-prof} shows this comparison between the reference simulation at $z = 3$ and a model that includes very efficient AGN feedback and a model with neither SNe feedback nor metal cooling. The solid green curve in the left panel of Figure \ref{figDLA:Models-prof} shows the reference model while the red dashed and blue dot-dashed curves respectively indicate the simulation with AGN and the simulation without SNe feedback and metal cooling (NOSN\_NOZCOOL). The relation between the normalized impact parameter and $\NHI$ is shown in the right panel of Figure \ref{figDLA:Models-prof}, where only central galaxies are taken into account for the matching process. 

Absorbers with $\NHI \lesssim 10^{21}\cmsq$ in the AGN simulation have typical impact parameters that are slightly larger than in the reference model. At very high $\HI$ column densities, however, the two models are similar. We also found that the contribution of galaxies with different stellar masses and SFRs as a function of $\NHI$ in the AGN simulation is very similar to the reference model (not shown). The only difference between the two simulations in this context is that the number of strong $\HI$ absorbers associated with very massive galaxies ($\rm{M_{\star}}\gtrsim 10^{10}~\Msun$, $\rm{SFR} \gtrsim  1 ~\rm{\Msun~yr^{-1}}$) becomes smaller. Turning off both SNe feedback and metal cooling slightly reduces the typical impact parameters of $\HI$ absorbers at almost all $\HI$ column densities. 

The differences due to variations in feedback are much smaller than the intrinsic scatter in the expected impact parameter at a given $\NHI$. The strong anti-correlation between the impact parameter of absorbers and their $\HI$ column densities is present and similar in all simulations despite the large variations in feedback mechanisms. Therefore, we conclude that feedback variation has only a minor impact on our main results.
\begin{figure*}
\centerline{\hbox{\includegraphics[width=0.45\textwidth]
             {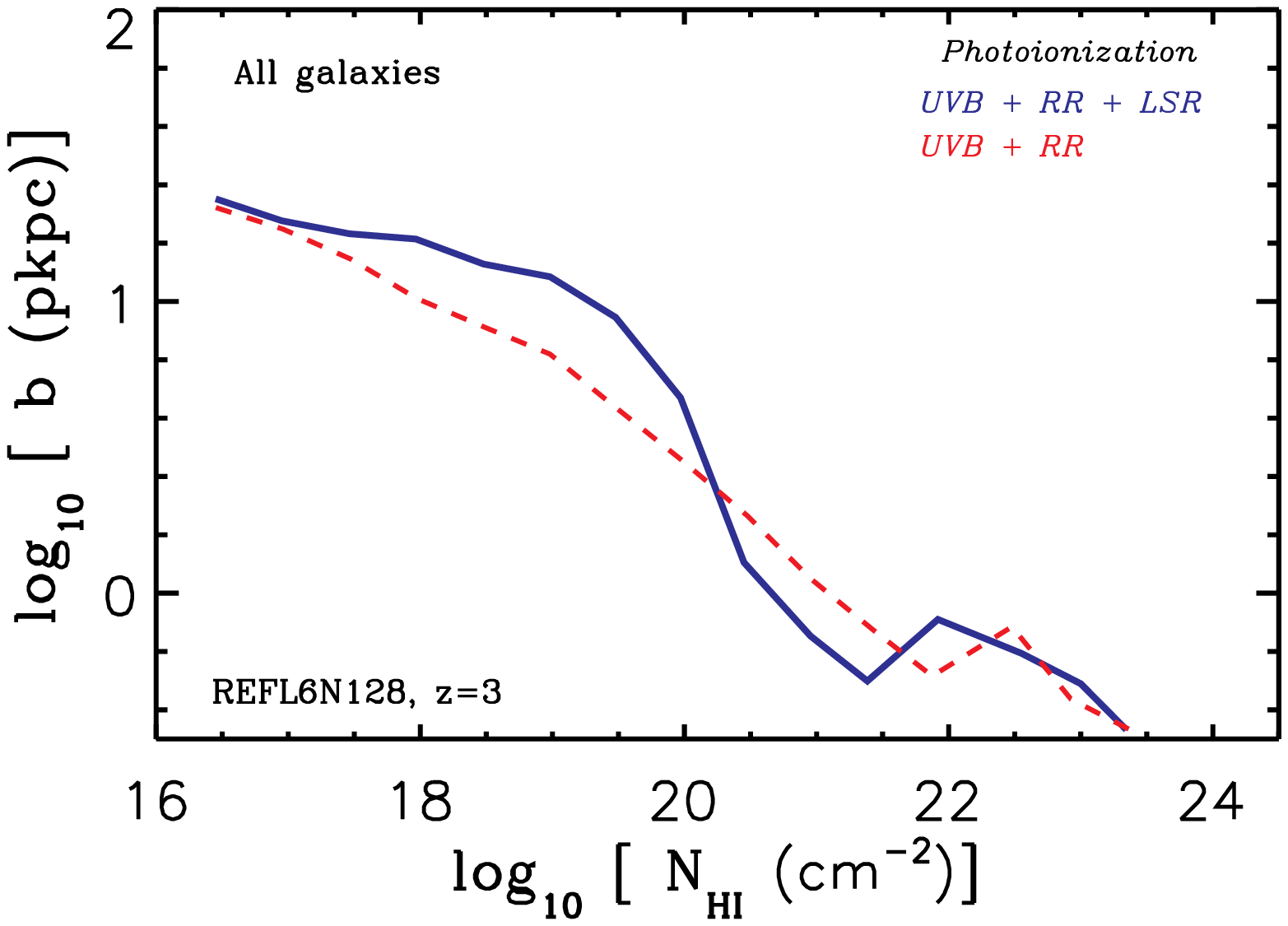}}
             \hbox{\includegraphics[width=0.45\textwidth]
             {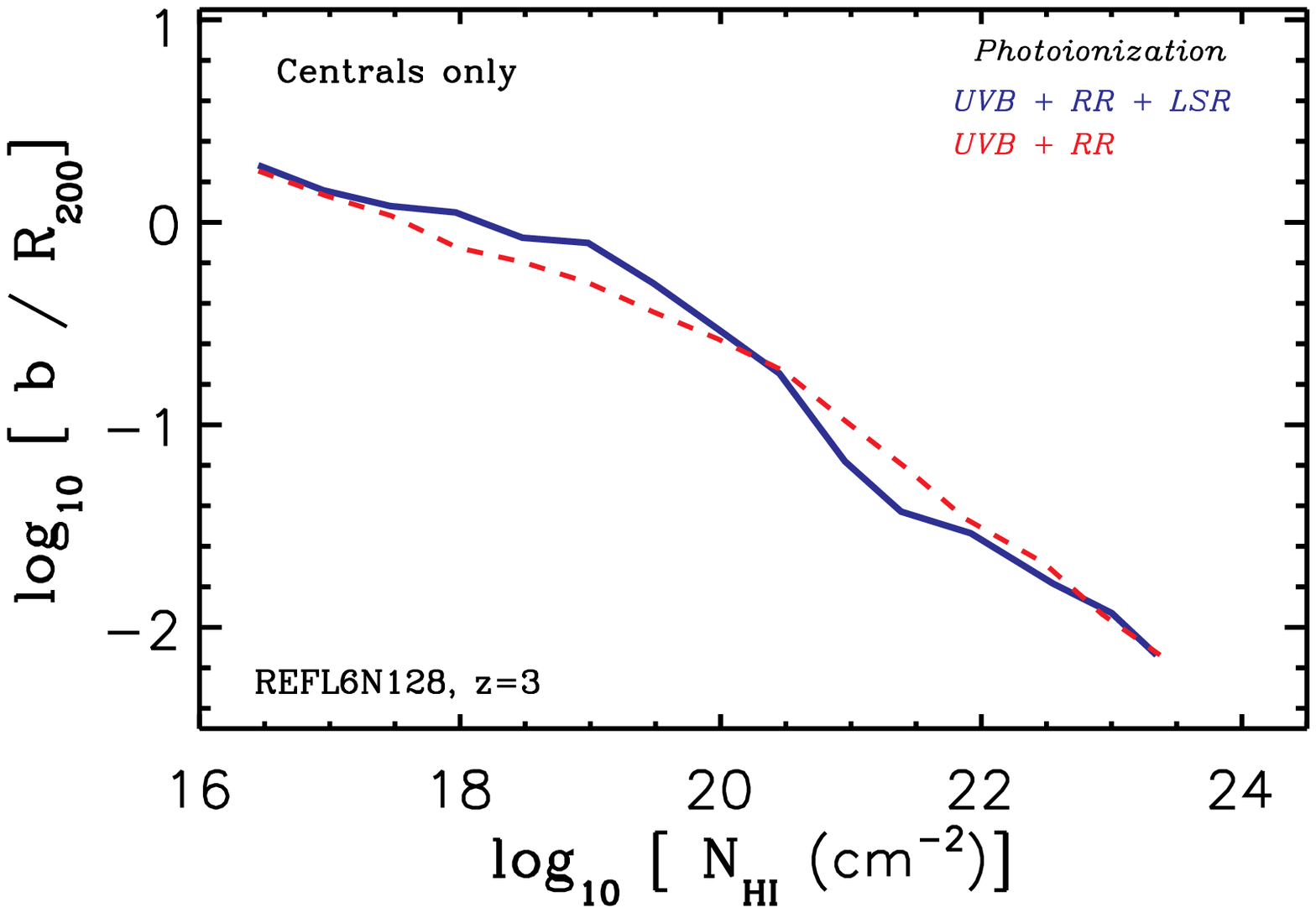}}}
\caption{The (normalized) impact parameter as a function of $\NHI$ at $z = 3$ for the \emph{REFL06N128} simulation with different photoionization models is shown in the (right) left panel. The red dashed curve shows the result when only the UVB and recombination radiations (RR) are present while the blue solid curve indicates the result of including local stellar radiation (LSR). Photoionization from local stellar radiation reduces the impact parameter of DLAs by up to $50\%$ and increases the typical impact parameter of Lyman Limit systems by similar amount.} 
\label{figDLA:LSR}
\end{figure*}
\section{Impact of local stellar radiation}
\label{ap:LSR}
After the reionization of the Universe, the background radiation produced by stars and quasars keeps hydrogen atoms mostly ionized. While the ionizing background is close to uniform in the intergalactic medium, it becomes highly non-uniform close to radiation sources. In particular, as was shown in \citet{Rahmati13b} and \citet{Schaye06}, for absorbers with $\NHI \gtrsim 10^{17}\cmsq$ local stellar radiation becomes important. Since we study the distribution of such strong $\HI$ absorbers, it is important to investigate the impact of local stellar radiation on the results of this work, where we neglect it.

The impact of local stellar radiation on the spatial distribution of strong $\HI$ absorbers at $z= 3$ is shown in Figure \ref{figDLA:LSR} for the \emph{REFL06N128} simulation. The blue solid curve shows the result of a radiative transfer calculation that accounts for the photoionization by the ultraviolet background (UVB) radiation, recombination radiation and local stellar radiation as explained in \citet{Rahmati13b}, while the red dashed curve indicates the result of including only photoionization by the UVB and recombination radiation in the same simulation. Local stellar radiation can change the median impact parameters of strong $\HI$ absorbers by up to $50\%$. The median impact parameters of DLAs with $\NHI \sim 10^{21}\cmsq$ is reduced because their $\HI$ column densities decreases due to additional photoionization by local stellar radiation. For LLSs on the other hand, local stellar radiation mainly affects systems that are closer to the galaxies. This results in an increase in the median impact parameter of absorbers at a given $\NHI$ by decreasing the $\HI$ column density of absorbers at shorter impact parameters. At lower $\HI$ column densities (i.e., $\NHI \lesssim 10^{17}\cmsq$) where the effect of local stellar radiation is negligible \citep{Rahmati13b}, the impact parameters remain unchanged\footnote{We note that using a small simulation box results in underproducing the strong $\HI$ absorbers due to missing very massive galaxies \citep{Rahmati13a}. Given that the contribution of very massive galaxies to the total abundance of absorbers increases with increasing the $\NHI$ (see Figure \ref{figDLA:prof-Ms-sfrI} and  \ref{figDLA:prof-Ms-sfrII}), missing them in the small simulation box allows the smaller galaxies to be the main DLA counterparts and hence decreases the typical impact parameters of strong $\HI$ absorbers.}. We conclude that while the effect of local stellar radiation is not negligible, it does not change the conclusions we present in this work.

\section{Resolution tests}
\label{ap:res}
We use three different cosmological simulations that have identical box sizes, but different resolutions. These simulations are part of the OWLS project \citep{Schaye10} and have cosmological parameters that are consistent with WMAP year-3 values (i.e., $\{\Om=0.238,\ \Ob=0.0418,\ \Ol=0.762,\ \sigeight=0.74,\ \ns=0.951,\ h=0.73\} $), slightly different from our fiducial cosmology. The simulation with the highest resolution (\emph{REFL25N512}) has identical mass resolution and box size as the simulation we use in this work and the other two simulations have 8 times (\emph{REFL25N256}) and 64 time (\emph{REFL25N128}) lower resolutions (see Table \ref{tbl:sims} for more details).
\begin{table*}%{sidewaystable}%
\caption{List of cosmological simulations used in this work. The details of the model ingredients are discussed in \citet{Schaye10}. From left to right the columns show: simulation identifier; comoving box size; number of dark matter particles (there are equally many baryonic particles); initial baryonic particle mass; dark matter particle mass; comoving (Plummer-equivalent) gravitational softening; maximum physical softening; final redshift; remarks about the used model, cosmology and the use of explicit radiative transfer calculations instead of a fitting function for the $\HI$ calculations (RT).} 
\begin{tabular}{lccccccccc}
\hline
Simulation & $L$ & $N$ & $m_{\rm b}$ & $m_{\rm dm}$ & $\epsilon_{\rm com}$ & $\epsilon_{\rm prop}$ & $z_{\rm end}$ & Model \\  
& $(\Mpch)$ & & $(\Msunh)$ & $(\Msunh)$ & $(\kpch)$  & $(\kpch)$ & &\\
\hline 
\emph{REFL06N128} &  6.00 & $128^3$ & $1.4 \times 10^6$
& $ 6.3 \times 10^6$ & 1.95 & 0.50 & 0 & REF, WMAP7 cosmology, RT \\
\emph{REFL25N512-W7} &  25.00 & $512^3$ & $1.4 \times 10^6$
& $ 6.3 \times 10^6$ & 1.95 & 0.50 & 2 & REF, WMAP7 cosmology \\
\emph{REFL25N512} &  25.00 & $512^3$ & $1.4 \times 10^6$
& $ 6.3 \times 10^6$ & 1.95 & 0.50 & 1 &  REF\\
\emph{AGN} &  25.00 & $512^3$ & $1.4 \times 10^6$
& $ 6.3 \times 10^6$ & 1.95 & 0.50 & 2 & with AGN \\
\emph{NOSN\_NOZCOOL} &  25.00 & $512^3$ & $1.4 \times 10^6$
& $ 6.3 \times 10^6$ & 1.95 & 0.50 & 2 & w/o SN, w/o metal cooling \\
\emph{REFL25N256} &  25.00 & $256^3$ & $1.1 \times 10^7$
& $ 5.1 \times 10^7$ & 3.91 & 1.00 & 2 & REF\\
\emph{REFL25N128} &  25.00 & $128^3$ & $8.7 \times 10^7$
& $ 4.1 \times 10^8$ & 7.81 & 2.00 & 0 & REF\\
%\emph{REFL50N512} &  50.00 & $512^3$ & $1.1 \times 10^7$
%& $ 5.1 \times 10^7$ & 3.91 & 1.00 & 0 & REF\\
\hline
\end{tabular}
\label{tbl:sims}
\end{table*}%{sidewaystable}%

\begin{figure}
\centerline{\hbox{\includegraphics[width=0.5\textwidth]
             {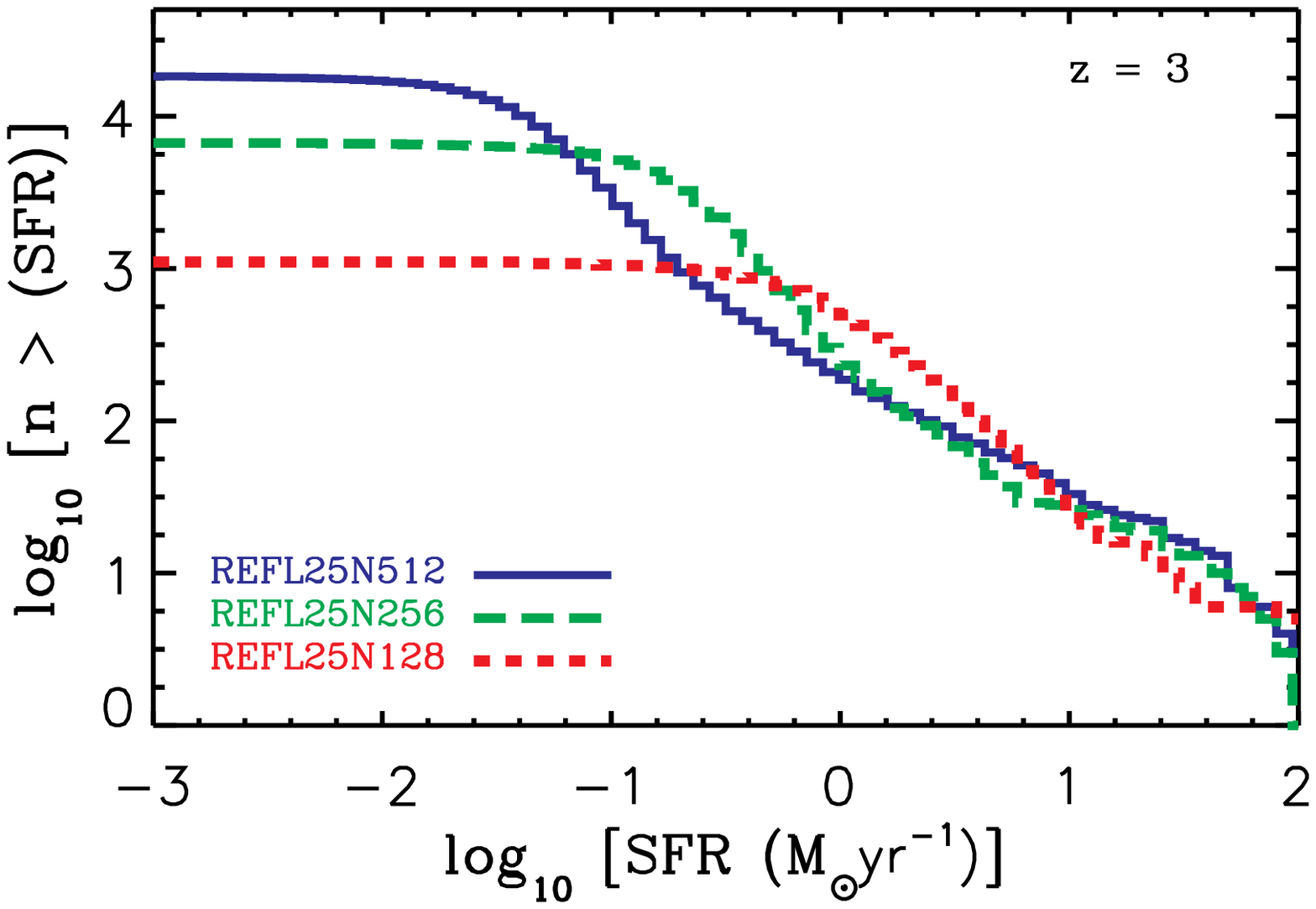}}}
\caption{Cumulative number of galaxies that are resolved in simulations at $z = 3$ with different resolutions as a function of their SFR. Blue solid, green long-dashed and red dashed curves show the \emph{REFL25N512}, \emph{REFL25N256} and \emph{REFL25N128} simulations, respectively.} 
\label{figDLA:n-sfr}
\end{figure}
\begin{figure*}
\centerline{\hbox{\includegraphics[width=0.45\textwidth]
             {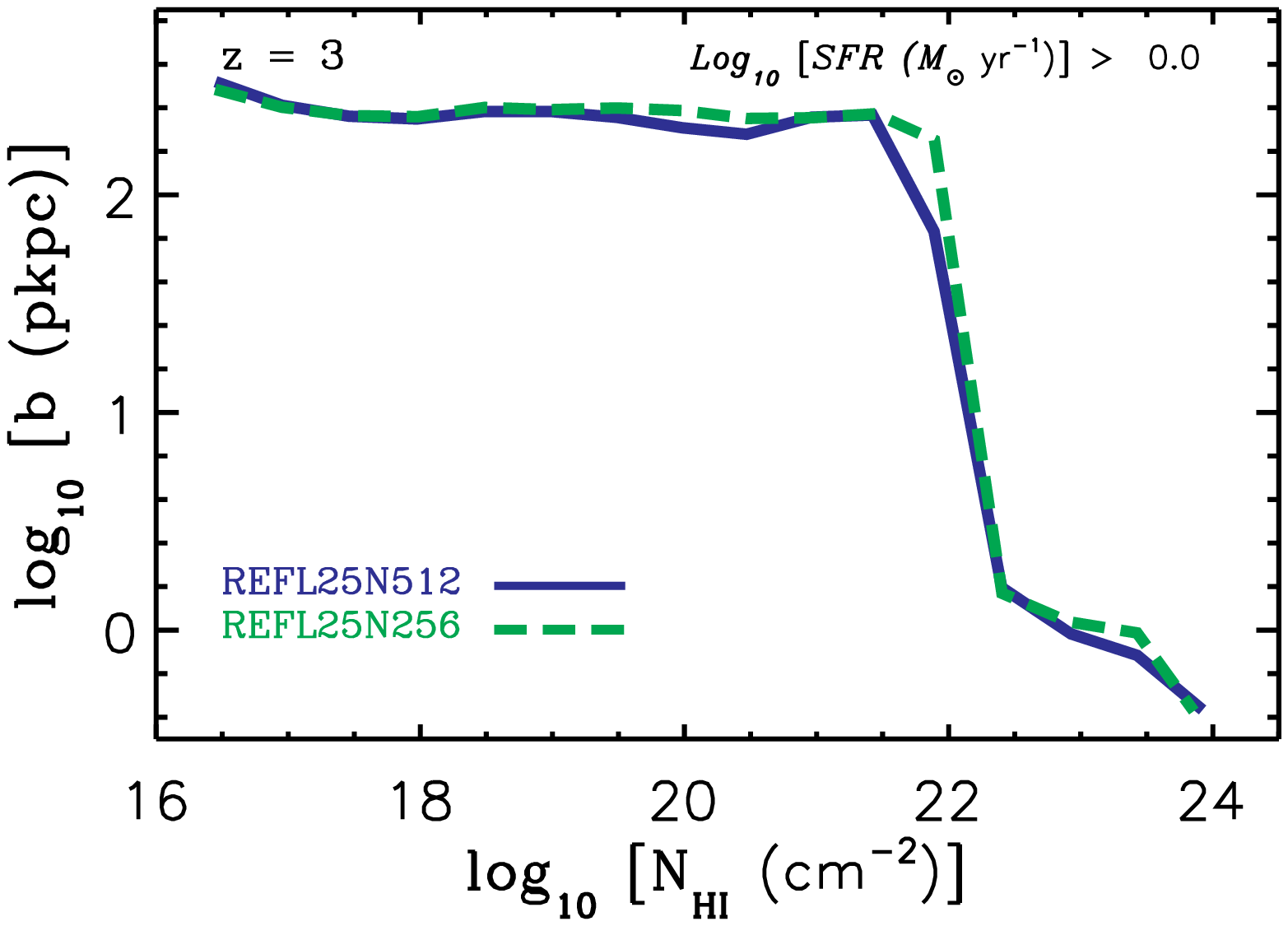}}
             \hbox{\includegraphics[width=0.45\textwidth]
             {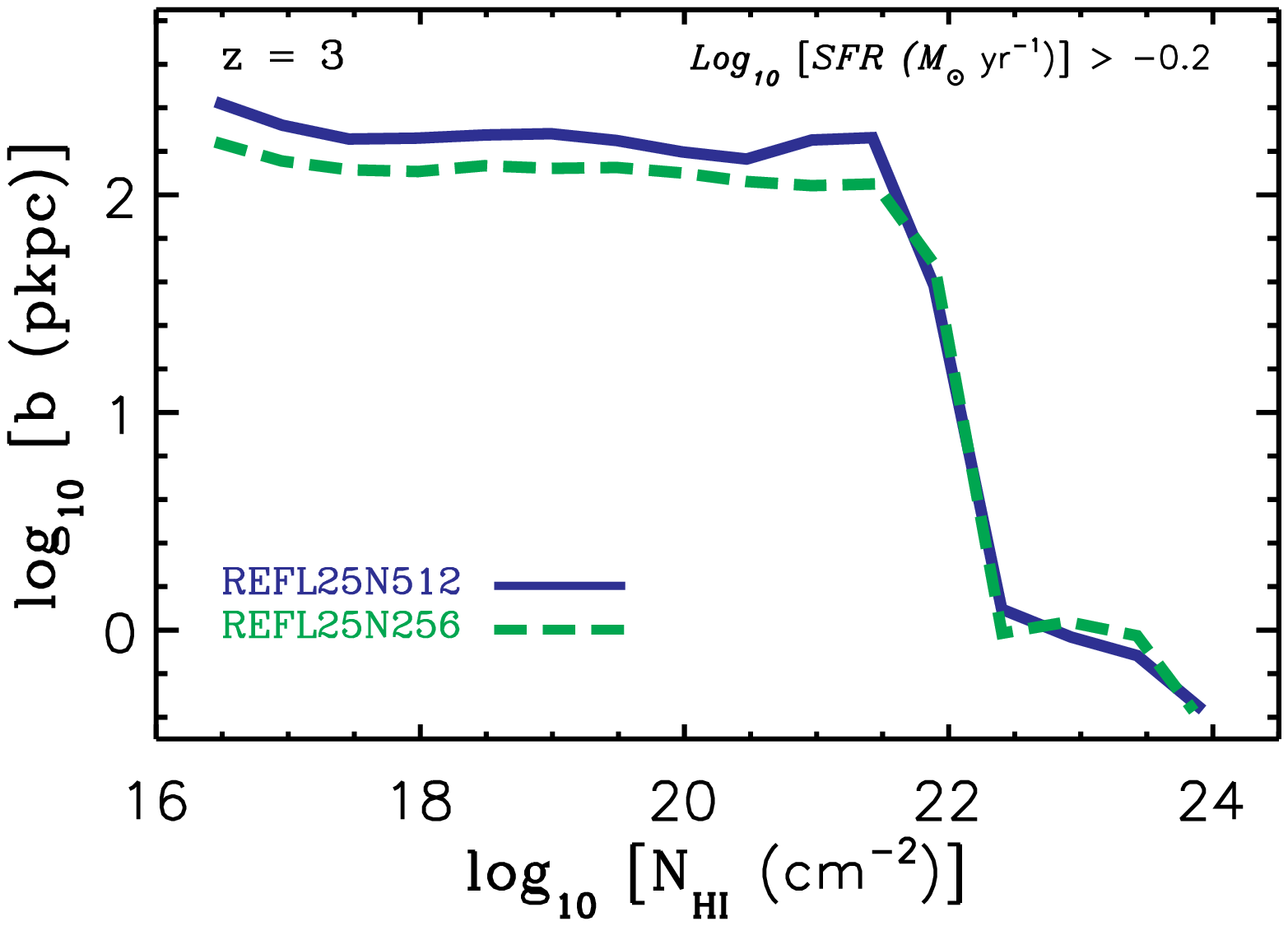}}}
\centerline{\hbox{\includegraphics[width=0.45\textwidth]
             {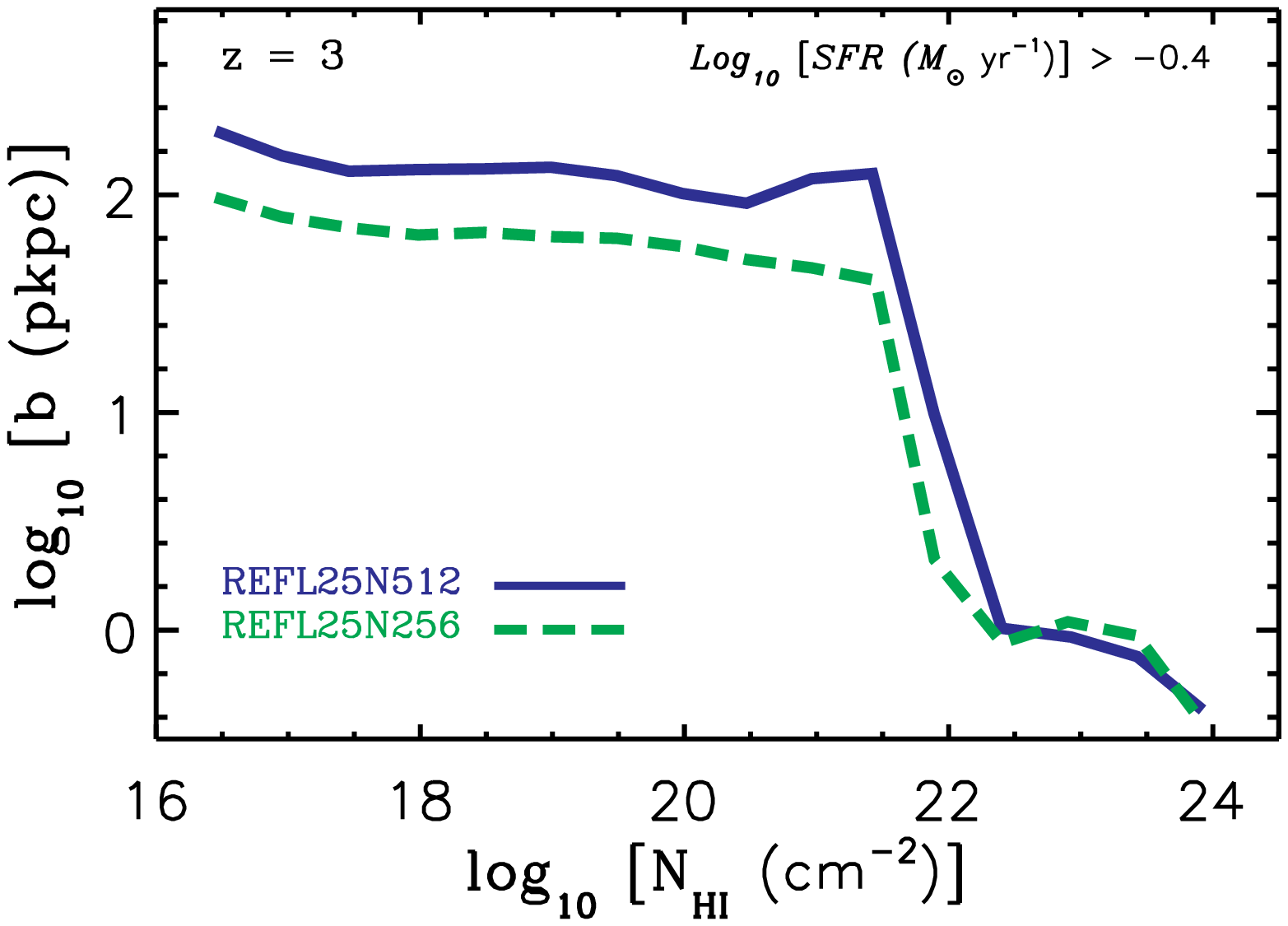}}
             \hbox{\includegraphics[width=0.45\textwidth]
             {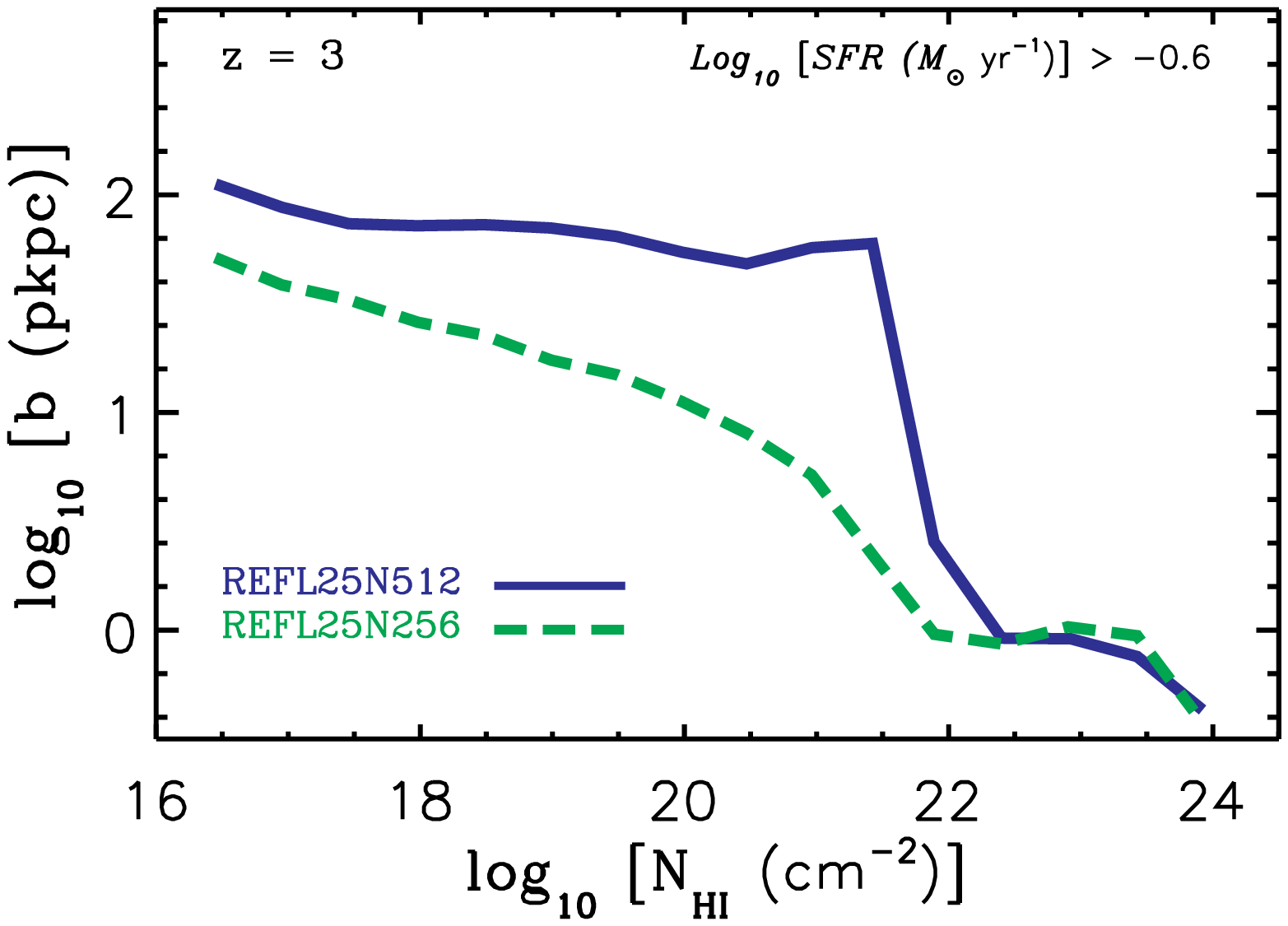}}}
\caption{The resolution dependence of the $b-\NHI$ relation at different SFR thresholds. The blue solid and green dashed curves represent the \emph{REFL25N512} and the \emph{REFL25N256} simulations, respectively. From top-left to bottom-right the SFR thresholds are $1$, $0.63$, $0.4$ and $0.2 ~\rm{\Msun~yr^{-1}}$, respectively. The results are similar in the two simulations for $\rm{SFR} > 0.4 ~\rm{\Msun~yr^{-1}}$ ($Log_{10} [SFR ({\Msun~yr^{-1}})] > -0.4$).} 
\label{figDLA:res-bNHI}
\end{figure*}
As we showed in \citet{Rahmati13a}, the $\HI$ column density distribution function is converged for LLS and most DLAs at the resolution that we use in this work. The positions of galaxies in cosmological simulations are determined by the distribution of overdensities and are therefore not expected to be highly sensitive to the resolution except perhaps on very small scales. This is, however, not true for the number of galaxies that are resolved in simulations. As the resolution increases, the number of structures that are resolved also increases. This can be seen from Figure \ref{figDLA:n-sfr} which shows the cumulative number of galaxies that are identified in our simulations as a function of the adopted SFR threshold\footnote{We note that contrary to the \emph{REFL25N512} simulation in which the cosmological parameters were set to WMAP year-3 values, our reference simulation in this work assumes the WMAP year-7 cosmology. As a result its cumulative distribution of galaxies flattens at lower SFR than what is shown by the blue solid curve in Figure \ref{figDLA:n-sfr}.}. Comparing the blue solid and green long-dashed curves shows that the number of galaxies that have $\rm{SFR} >  1 ~\rm{\Msun~yr^{-1}}$ is nearly identical in the two simulations with the highest resolutions. Together with the converged $\HI$ column density distribution function, this result implies that the relation between the impact parameter of $\HI$ absorbers and galaxies with $\rm{SFR} \gtrsim 1 ~\rm{\Msun~yr^{-1}}$ is also converged in the \emph{REFL25N256} simulation. As Figure \ref{figDLA:res-bNHI} shows, this is indeed the case. For $\rm{SFR} > 0.4 ~\rm{\Msun~yr^{-1}}$ the differences are small and the two simulations are nearly identical for $\rm{SFR} > 1 ~\rm{\Msun~yr^{-1}}$. This suggests that the $b-\NHI$ relation is close to converged in the \emph{REFL25N512} simulation for $\rm{SFR} > 0.4/8 = 5\times 10^{-2} ~\rm{\Msun~yr^{-1}}$.
\begin{figure*}
\centerline{\hbox{\includegraphics[width=0.45\textwidth]
             {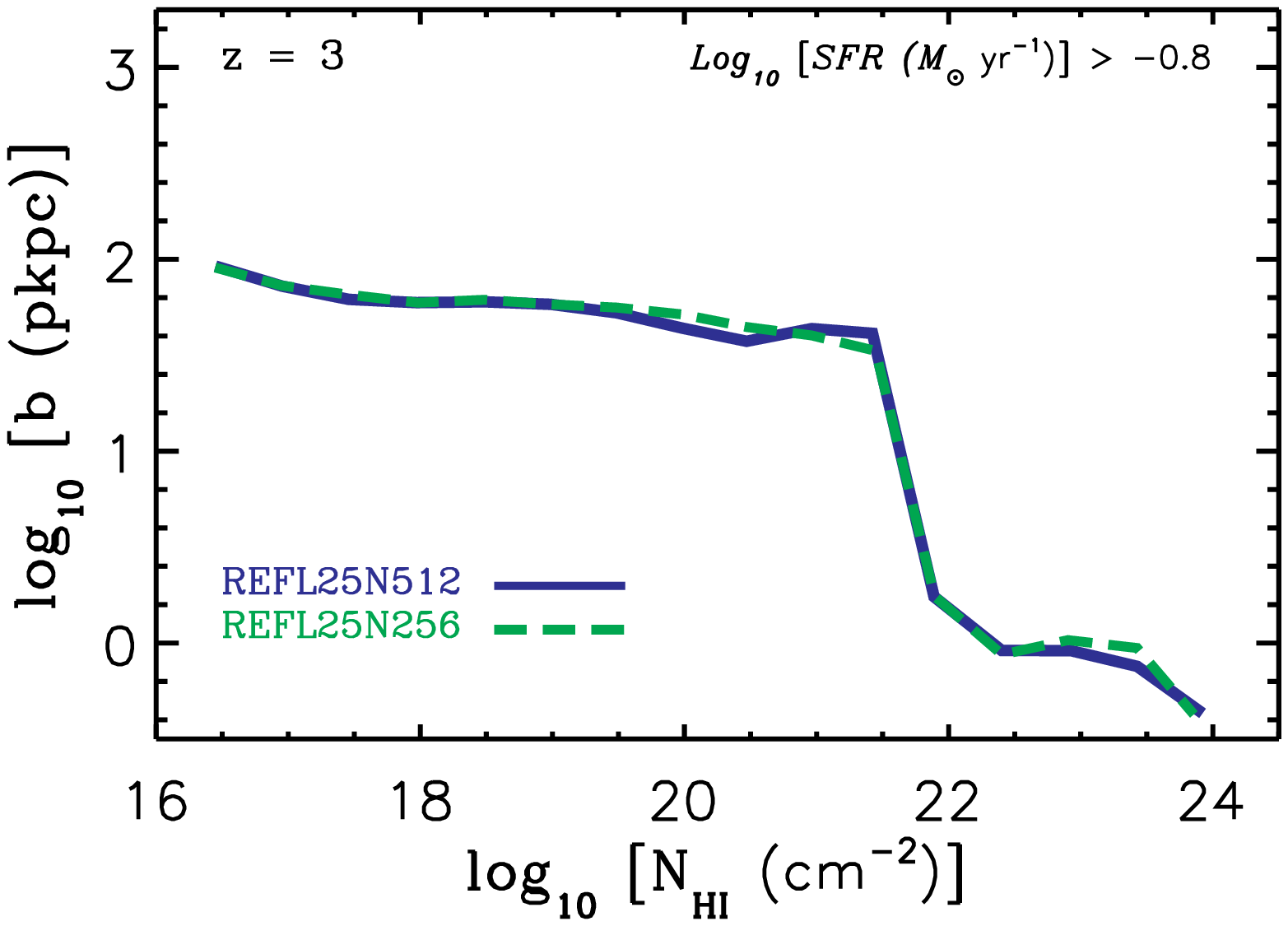}}
             \hbox{\includegraphics[width=0.45\textwidth]
             {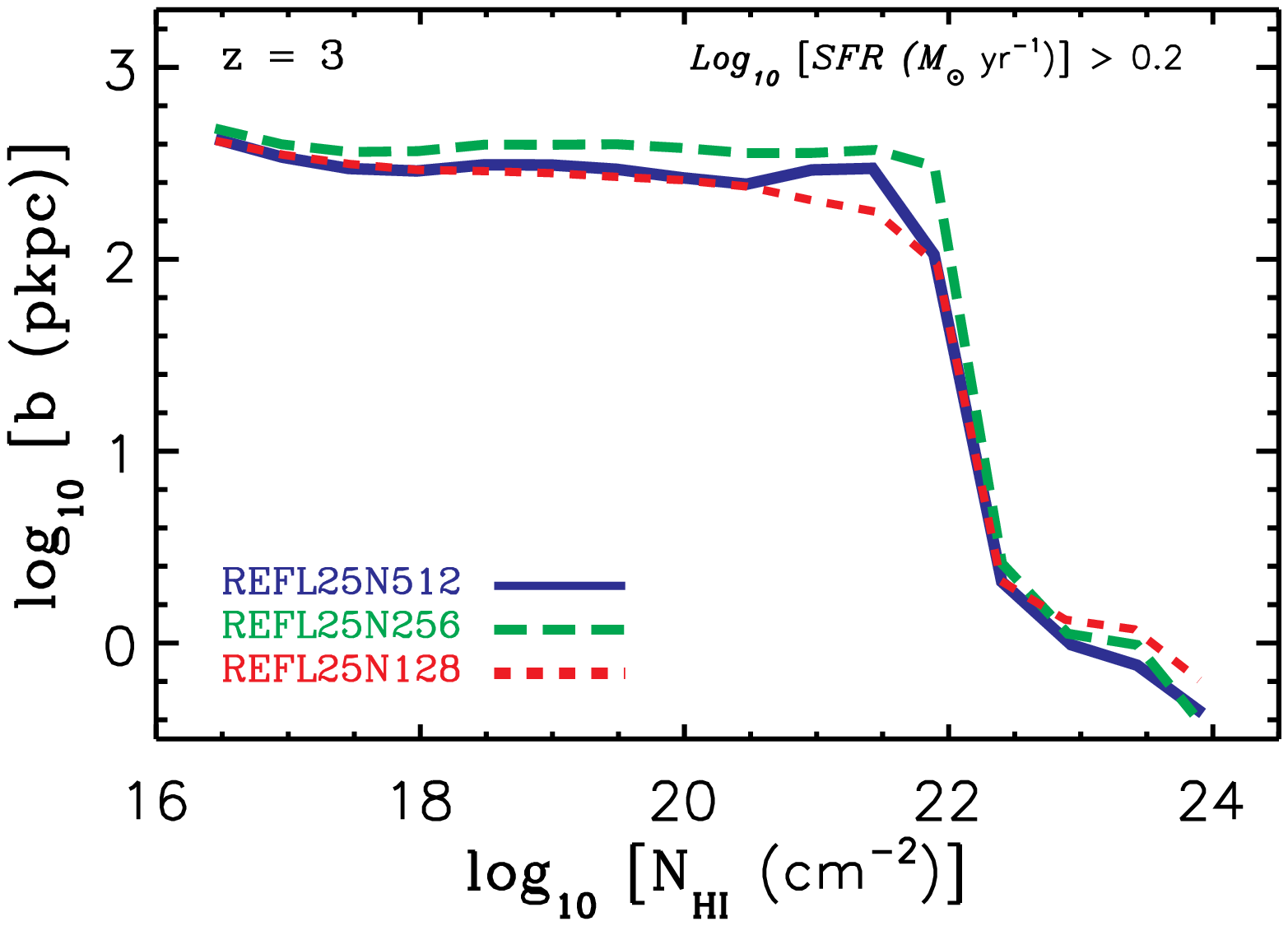}}}
\caption{Impact parameter of $\HI$ absorbers as a function of their $\NHI$ for different SFR thresholds chosen to correspond to the same galaxy number density. In the left panel, the blue solid curve shows the result for the \emph{REFL25N512} simulation if only galaxies with $\rm{SFR} > 0.16 ~\rm{\Msun~yr^{-1}}$ are taken into account. The green long-dashed curve shows the result for \emph{REFL25N256} if the SFR threshold is such that the cumulative number density of galaxies is matched to that of galaxies with $\rm{SFR} > 0.16 ~\rm{\Msun~yr^{-1}}$ in \emph{REFL25N512} simulation. In the right panel, the blue solid curve shows the $b-\NHI$ relation for \emph{REFL25N512} if only galaxies with $\rm{SFR} > 1.6 ~\rm{\Msun~yr^{-1}}$ are taken into account. The green long-dashed and red dashed curves show, respectively, the results for \emph{REFL25N256} and  \emph{REFL25N128} if the SFR thresholds are chosen such that the total number density of galaxies that are taken into account is matched to that of galaxies with $\rm{SFR} > 1.6 ~\rm{\Msun~yr^{-1}}$ in the \emph{REFL25N512} simulation (i.e., $2\times 10^{-3}$ galaxies per comoving $\rm{Mpc}^{-3}$). For a fixed total number density of galaxies the relation between impact parameters and the $\HI$ column density of absorbers is insensitive to the resolution.} 
\label{figDLA:res-nfix}
\end{figure*}
By increasing the number density of galaxies in a simulation, one would expect a decrease in the average distance between $\HI$ absorbers and galaxies. On the other hand, one might expect to retrieve the same relation between impact parameters and $\NHI$ of absorbers if the total number density of galaxies were the same in simulations with different resolutions. As Figure \ref{figDLA:res-nfix} shows, this is indeed true for our simulations. In each panel we choose different SFR thresholds for different resolutions in order to match the total number density of galaxies above the SFR threshold for all the simulations. This result suggests that if one keeps the total number density of galaxies fixed to the value that corresponds to the SFR threshold that we use in our study, increasing the resolution is not expected to change the $b-\NHI$ relation and other conclusions we derived in this work. We used the relatively low SFR threshold of $\rm{SFR} > 4\times 10^{-3} ~\rm{\Msun~yr^{-1}}$ to include as many bound (sub) structure as possible in our analysis. The total number density of galaxies that are selected in our reference simulation (i.e., \emph{REFL25N512-W7}) by this criterion is $0.5$ galaxy per comoving $\rm{Mpc}^3$ (i.e., equivalent to $31.5$ galaxy per proper $\rm{Mpc}^3$). We note, however, that the SFR threshold that corresponds to this total number density is expected to be lower than $4\times 10^{-3} ~\rm{\Msun~yr^{-1}}$ at higher resolutions. This can be seen in Figure \ref{figDLA:n-sfr} which shows that at a fixed cumulative number density, the SFRs of galaxies in the \emph{REFL25N256} simulation that have $\rm{SFR} <  1 ~\rm{\Msun~yr^{-1}}$ decrease with increasing resolution. 

\end{document}